\documentclass[letterpaper,11pt]{article}
\pdfoutput=1

\usepackage{jheppub}

\usepackage{subfig}
\usepackage{xspace}
\usepackage[countmax]{subfloat}
\usepackage{slashed}
\usepackage{color}
\usepackage{graphicx}                   
\usepackage{amsmath}
\usepackage{amssymb}
\usepackage{bbold}
\usepackage{verbatim}
\usepackage{bm}

\usepackage{bbm}

\newcommand{\nbar}{{\bar n}}

\newcommand{\ecf}[2]{e_{#1}^{(#2)}} 
\newcommand{\ecflp}[2]{\tilde e_{#1}^{(#2)}} 
\newcommand{\ecfop}[2]{\mathbf{e_{#1}^{(#2)}}} 
\newcommand{\ecfnobeta}[1]{e_{#1}} 

\newcommand{\ecfLa}{e_{2}^{(\alpha)}}
\newcommand{\ecfLb}{e_{2}^{(\beta)}}
\newcommand{\ecfres}{e_{3}^{(\alpha)}}
\newcommand{\ecfresgen}{e_\text{res}}
\newcommand{\ecfresgenLP}{\tilde e_\text{res}}
\newcommand{\ecfreslp}{\tilde e_{3}^{(\alpha)}}

\newcommand{\eec}[2]{e_{#1}^{(#2)}}
\newcommand{\eeclp}[2]{\tilde e_{#1}^{(#2)}}

\newcommand{\anglediff}{\tan ^2\frac{R}{2}-\tan ^2\frac{\theta_{sj}}{2}}
\newcommand{\sja}{n_{sj}}
\newcommand{\sjabar}{\bar{n}_{sj}}

\newcommand{\sje}{z_{sj}}
\newcommand{\sjtheta}{\theta_{sj}}
\newcommand{\sjOmega}{\Omega_{sj}}

\newcommand{\outj}{B}

\def\nbar{\bar n}

\DeclareRobustCommand{\Sec}[1]{Sec.~\ref{#1}}

\DeclareRobustCommand{\App}[1]{App.~\ref{#1}}
\DeclareRobustCommand{\Tab}[1]{Table~\ref{#1}}

\DeclareRobustCommand{\Fig}[1]{Fig.~\ref{#1}}
\DeclareRobustCommand{\Figs}[2]{Figs.~\ref{#1} and \ref{#2}}
\DeclareRobustCommand{\Eq}[1]{Eq.~(\ref{#1})}
\DeclareRobustCommand{\Eqs}[2]{Eqs.~(\ref{#1}) and (\ref{#2})}
\DeclareRobustCommand{\Ref}[1]{Ref.~\cite{#1}}
\DeclareRobustCommand{\Refs}[1]{Refs.~\cite{#1}}

\newcommand{\tr}{\text{tr}}
\newcommand{\jet}{\text{jet}}

\preprint{MIT--CTP 4635}

\title{
Non-Global Logarithms, Factorization, and the Soft Substructure of Jets
}

\author{Andrew J. Larkoski,}
\author{Ian Moult,}
\author{and Duff Neill}

\affiliation{Center for Theoretical Physics, Massachusetts Institute of Technology, Cambridge, MA 02139, USA}

\emailAdd{larkoski@mit.edu}
\emailAdd{ianmoult@mit.edu}
\emailAdd{dneill@mit.edu}

\abstract{ 
An outstanding problem in QCD and jet physics is the factorization and resummation of logarithms that arise due to phase space constraints, so-called non-global logarithms (NGLs).  In this paper, we show that NGLs can be factorized and resummed down to an unresolved infrared scale by making sufficiently many measurements on a jet or other restricted phase space region.  Resummation is accomplished by renormalization group evolution of the objects in the factorization theorem and anomalous dimensions can be calculated to any perturbative accuracy and with any number of colors. To connect with the NGLs of more inclusive measurements, we present a novel perturbative expansion which is controlled by the volume of the allowed phase space for unresolved emissions. Arbitrary accuracy can be obtained by making more and more measurements so to resolve lower and lower scales.  We find that even a minimal number of measurements produces agreement with Monte Carlo methods for leading-logarithmic resummation of NGLs at the sub-percent level over the full dynamical range relevant for the Large Hadron Collider. We also discuss other applications of our factorization theorem to soft jet dynamics and how to extend to higher-order accuracy. 
}

\begin{document} 
\maketitle

\section{Introduction}\label{sec:intro}

A fundamental problem in QCD and collider physics is the identification of hierarchical scales in a system defined by some number of observations made on that system.  Generically, ratios of these scales appear in logarithms at every order of the perturbative expansion of the cross section, and can become large in the soft or collinear regions of phase space.  To tame these large logarithms and so to improve the convergence of the perturbative expansion, resummation of the logarithms to all orders is required.  This resummation requires the factorization of the physics that dominates at each scale from one another, so that one can guarantee that an all-orders description is possible in a particular region of phase space.  This program has seen enormous success with applications to predictions for deep inelastic scattering (e.g. \cite{Antonelli:1999kx,Dasgupta:2001eq,Dasgupta:2002dc,Manohar:2003vb,Becher:2006mr,Kang:2013nha}), $e^+e^-$ (e.g. \cite{Catani:1992ua,Dokshitzer:1998kz,Catani:1998sf,Becher:2008cf,Abbate:2010xh,Banfi:2014sua}), $p\bar{p}$ and $pp$ collision experiments (e.g. \cite{Collins:1985ue,Sterman:1986aj,Catani:2003zt,Banfi:2004nk,Ahrens:2008nc,Stewart:2009yx,Berger:2010xi,Banfi:2010xy}), and weak decays (e.g. \cite{Misiak:1992bc,Buchalla:1995vs,Chetyrkin:1996vx,Buras:1998raa,Misiak:2006zs}), amongst others.

Strictly speaking the picture outlined above for the resummation of large logarithms is only known to completely capture all logarithms to a given accuracy if all radiation in an event contributes to the observables.  Such observables are referred to as ``global'' if all final state particles contribute to their value.  Global observables at particle colliders include thrust \cite{Farhi:1977sg}, angularities \cite{Berger:2003iw}, or weak boson $p_T$ distributions in Drell-Yan production \cite{Collins:1984kg}, for example.  However, global observables are only a subset of observables that are interesting for studying QCD or new physics.  Jets, and observables definedå on their constituents, rapidity gaps  \cite{Marchesini:1988hj,Berger:2001ns,Oderda:1998en,Oderda:1999kr}, or any observable that is only defined by radiation in a limited region of the full phase space are referred to as ``non-global'' \cite{Dasgupta:2001sh}.  Especially with the phase space available for high $p_T$ jets at the Large Hadron Collider, there has been substantial effort in defining jet substructure observables \cite{Abdesselam:2010pt,Altheimer:2012mn,Altheimer:2013yza}, and with the discovery of the Higgs boson \cite{Chatrchyan:2012ufa,Aad:2012tfa}, identifying vector boson fusion events with forward jets is essential for determining properties of the Higgs.  Thus, non-global observables are widely used and therefore require detailed theoretical understanding.

Unlike global observables, non-global observables are sensitive to both the relevant scales within the jet or identified phase space region as well as the scale outside the jet.\footnote{For compactness, regardless of how the identified phase space region is defined, we will refer to it as a ``jet''.}  While the out-of-jet region is not directly measured, emissions originating from the outside can contaminate and affect the in-jet region of phase space on which measurements are performed.  Therefore, for an accurate description of the cross section to a given logarithmic accuracy of a non-global observable, we must resum not only logarithms of ratios of in-jet scales (global logarithms), but also logarithms of  ratios of in-jet to out-of-jet scales.  The latter logarithms are referred to as non-global logarithms (NGLs).

While the existence of NGLs has been known for some time, and NGLs have been well studied in the literature \cite{Dasgupta:2001sh,Dasgupta:2002bw,Dasgupta:2002dc,Banfi:2002hw,Appleby:2002ke,Weigert:2003mm,Rubin:2010fc,Banfi:2010pa,Kelley:2011tj,Hornig:2011iu,Hornig:2011tg,Kelley:2011aa,Kelley:2012kj,Hatta:2013iba,Schwartz:2014wha,Khelifa-Kerfa:2015mma,Caron-Huot:2015bja}, they have proved challenging to understand.  The leading NGLs in the large $N_c$ limit can be resummed by Monte Carlo methods \cite{Dasgupta:2001sh,Dasgupta:2002bw} or the Banfi-Marchesini-Smye (BMS) equation \cite{Banfi:2002hw}, but a systematic understanding of NGLs to all logarithmic orders is lacking.\footnote{While this paper was being finalized, \Ref{Caron-Huot:2015bja} appeared, which presented a novel formalism for the resummation of NGLs using a non-linear evolution equation for a ``color density matrix", and discussing its relation to reggeization \cite{Kuraev:1976ge,Lipatov:1976zz} and the BFKL \cite{Lipatov:1985uk,Kuraev:1977fs,Balitsky:1978ic} and B-JIMWLK \cite{JalilianMarian:1996xn,JalilianMarian:1997gr,Iancu:2001ad} equations. \Ref{Caron-Huot:2015bja} did not, however, demonstrate resummation of NGLs for a particular observable, nor did it prove a factorization theorem in which the evolution equation embeds. However, the procedure for incorporating the resummation for an observable was sketched. In this paper, we demonstrate the factorization for observables, hence capturing observable dependence, and show how the resummation occurs as a linear renormalization group evolution. The approach presented here also exhibits connections to reggeization and the BFKL equation, although we have chosen not to focus on these aspects in this paper. }  Other effects, such as finite $N_c$ at leading logarithmic order \cite{Weigert:2003mm,Hatta:2013iba} and at fixed-order \cite{Kelley:2011aa,Hornig:2011tg,Hornig:2011iu,Kelley:2012kj,Schwartz:2014wha,Khelifa-Kerfa:2015mma} have been studied in detail.  A systematic understanding of NGLs to all orders has so far not been possible because it has not been shown how to factorize the ratio of in-jet to out-of-jet scales from one another.  Therefore, to understand and resum NGLs to arbitrary accuracy requires proving factorization of in-jet and out-of-jet scales from one another.  In this paper, we will do this and present a procedure for systematic improvement of the accuracy to which the NGLs are computed.

Rather than understanding NGLs directly, recently it has been emphasized that the effects of NGLs can be reduced or power-suppressed in some cases \cite{Rubin:2010fc,Dasgupta:2013via,Dasgupta:2013ihk,Larkoski:2014wba}.  To do this, one can use jet grooming techniques \cite{Butterworth:2008iy,Ellis:2009su,Ellis:2009me,Krohn:2009th,Larkoski:2014wba} that remove soft, wide-angle radiation in the jet which is (potentially) likely sensitive to out-of-jet scales.  While the removal of NGLs using these grooming techniques is indeed one potential solution to the problem (although factorization and resummation for groomed observables is currently also not understood to all orders) here we will attack them directly and work toward their all-orders description.

To understand the problem of NGLs more precisely, consider measuring the mass $m$ and energy $E_\text{jet}$ of a jet and an inclusive observable on the region outside the jet, such as the out-of-jet energy, $E_\text{out}$.  Importantly, note that only soft radiation can be sensitive to out-of-jet scales by the collinear safety of the jet finding algorithm, as only soft radiation can cross phase space boundaries.  Assuming that soft and collinear physics factorizes, we introduce a soft function $S$ which encodes the pattern of soft radiation from the dipoles (pairs of eikonal Wilson lines) in the event.  For a global observable, the soft function only depends on scales set directly by the measurements on the event.  However, in this case, the soft function is sensitive to in-jet and out-of-jet scales \cite{Hornig:2011tg,Hornig:2011iu,Kelley:2011aa}:
\begin{equation}
S\equiv S\left(m,E_\jet,E_\text{out}\right) \,.
\end{equation}
No measurement has been done on the jet to determine if the emissions setting the mass come from in the jet or outside the jet, and so it would seem like this soft function cannot be further factorized. Therefore, without isolating the in-jet and out-of-jet scales in some way, the NGLs of this system cannot be resummed.

In the case just discussed, the soft function could not be further factorized because no measurement was done on the jet to isolate the region of phase space where the NGLs are important.  For the jet mass, this region of phase space is the emission of a single soft gluon from outside the jet into the jet, near the jet boundary.  A single gluon is not an infrared and collinear (IRC) safe quantity, so we should really think of this as a soft subjet located near the jet boundary.\footnote{\Refs{Marchesini:2004ne,Marchesini:2003nh} considered the production of a heavy $q\bar q$ from a soft gluon, which identifies a similar region of phase space. }   Nevertheless, just measuring the jet mass is not sufficient to uniquely specify this region of phase space.  The resolution to the problem of NGLs is therefore clear: we must measure sufficiently many observables on the jet so as to isolate the region of phase space in which the NGLs are important. This is the goal of this paper.  By measuring several observables on a jet we are able to identify the region of phase space where the NGLs live, refactorize the soft function, and resum the NGLs by renormalization group evolution of the now-factorized soft function.

\subsection{Overview of the paper}

Here, we present a detailed overview of the content and reasoning of this paper so that our logic is not lost to the details of calculation. For simplicity, in this paper we restrict ourselves to jet production in $e^+e^-$ collisions, although the approach can be extended to other situations. Our approach in this paper to accomplishing the resummation of NGLs is the following.  First, we find a jet in an $e^+e^- \to$ hadrons collision event by identifying the broadening axis \cite{Thaler:2010tr,Thaler:2011gf,Larkoski:2014uqa} of the event and including those particles that lie within a cone of fixed radius $R$ of the broadening axis.  For much of the jet's phase space relevant for NGLs, this is identical to finding jets with the anti-$k_T$ jet algorithm \cite{Cacciari:2008gp} with radius $R$ and the Winner-Take-All (WTA) recombination scheme \cite{Bertolini:2013iqa,Larkoski:2014uqa,Salambroadening}. The broadening axis is insensitive to recoil effects on the jet axis, and so the jet axis aligns with the direction of the hardest radiation in the jet.  This is necessary to eliminate back-reaction on the jet direction from wide-angle, soft radiation.  For the region of phase space outside the jet, we measure some quantity, which we refer to as $\outj$.  This measurement sets the scale of the out-of-jet radiation and we require $\outj\ll 1$ which enforces soft and collinear dynamics to dominate the out-of-jet region. Additionally, we will assume that the out-of-jet scale is much lower than the in-jet scale which is the phase space region in which the NGLs are large and must be resummed.

Within the jet, we want to guarantee that the jet contains a soft subjet approaching the jet boundary which is sensitive to the out-of-jet scale $\outj$.  We do this by measuring several IRC safe $n$-point energy correlation functions on the jet \cite{Larkoski:2013eya,Larkoski:2014gra}.  In particular, to uniquely identify a single soft subjet and determine its energy fraction and angle from the hard jet core, we measure the two- and three-point energy correlation functions $\ecf{2}{\alpha}$, $\ecf{2}{\beta}$ and $\ecf{3}{\alpha}$, for angular exponents $\alpha,\beta$.  The measurement of three energy correlation functions is required to enforce that the soft subjet is not collinear with the hard jet core.  The soft subjet region of this three-dimensional phase space is parametrically defined by \cite{Larkoski:2014gra}
\begin{equation}
\ecf{2}{\alpha}\sim \ecf{2}{\beta}\ll 1\,, \quad \text{ and }\quad \ecfres \ll \big(\ecf{2}{\alpha}\big)^{3} \,,
\end{equation}
where we have assumed that $\alpha> \beta$.  This soft subjet region is an essential component of the description of the full phase space defined by the energy correlation functions and is required to make predictions of distributions of jet discrimination observables such as $D_2$, defined in \Ref{Larkoski:2014gra}.  The phase space formed from the simultaneous measurement of the two- and three-point energy correlation functions is described in \Ref{Larkoski:2014gra}, and we will present analytic predictions for the full double-differential cross section in a future publication \cite{usD2}.

\begin{figure}
\begin{center}
\subfloat[]{\label{fig:softjet}
\includegraphics[scale = .22]{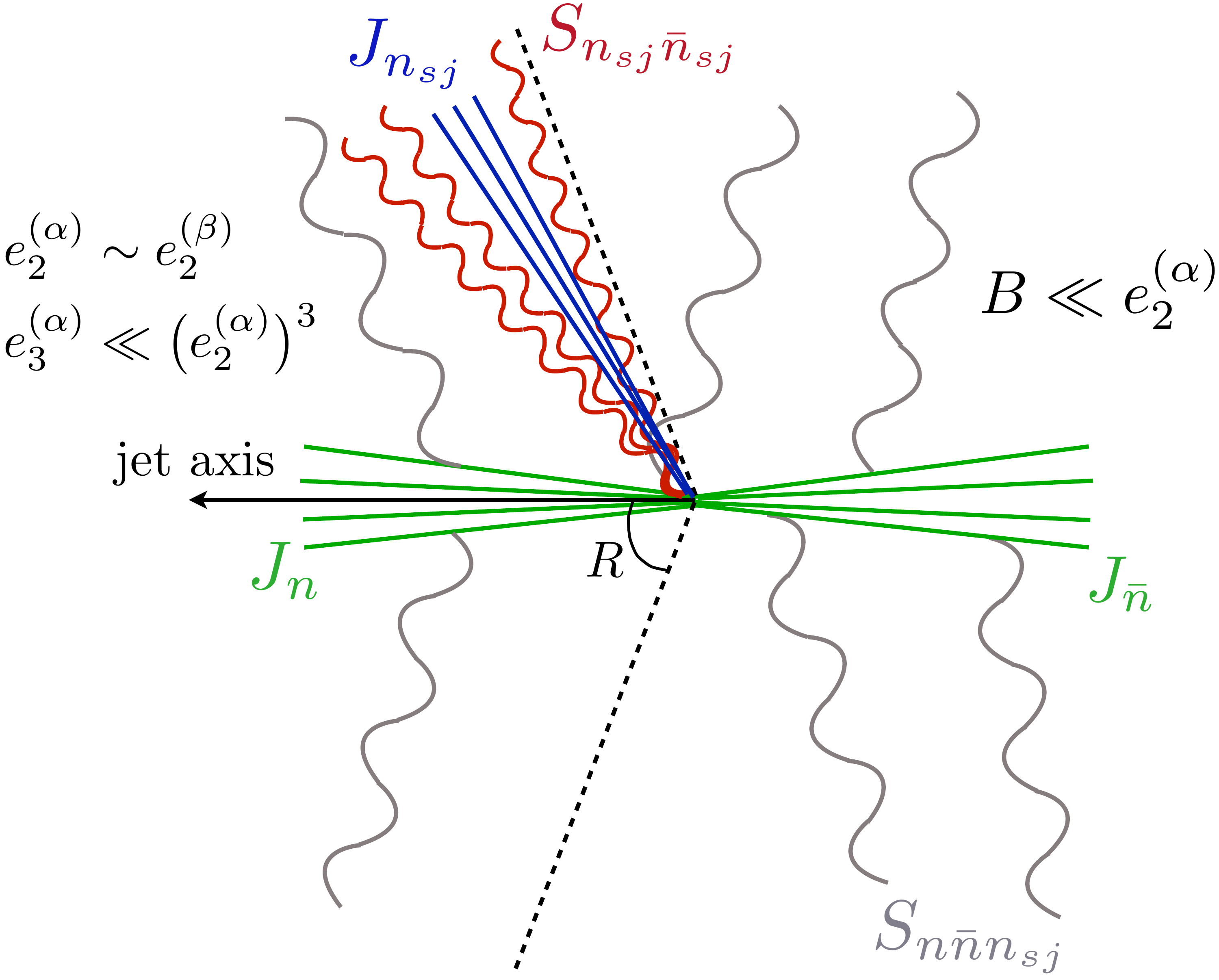}
}
$\quad$
\subfloat[]{\label{fig:ladderfact}
\includegraphics[scale = 0.2]{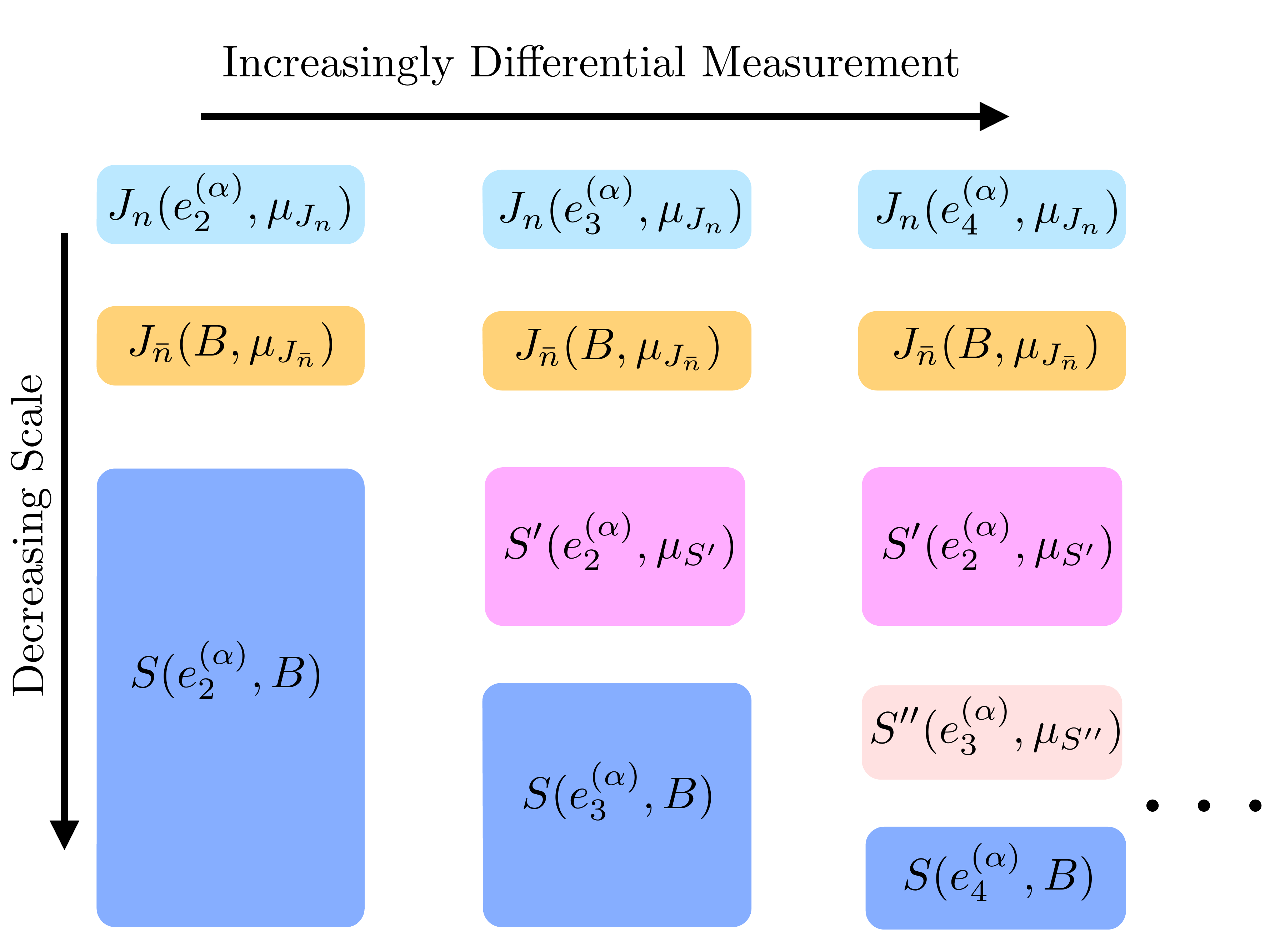}
}
\end{center}
\vspace{-0.2cm}
\caption{(a) Illustration of the phase space configuration and dominant modes for a jet containing a hard core and a soft subjet. Here the gray radiation denotes global soft radiation $S_{n\bar n n_{sj}}$, and the green radiation denotes collinear radiation along the direction of the energetic jet axes, $J_n$ and $J_{\bar n}$. The soft subjet dynamics is described by soft jet modes, $J_{n_{sj}}$ shown in blue, and boundary soft modes shown in red, $S_{n_{sj} \bar n_{sj}}$. (b)  Schematic of the ladder of factorization theorems defined by increasingly differential measurements made on the jet. With each additional measurement, the NGLs are pushed to the soft function at a lower unresolved scale. The $S'$, and $S''$ are schematic, typically being a product of multiple functions, but depend only on a single scale. 
}
\label{fig:intro}
\end{figure}

Identification of the soft subjet region of phase space enables a factorization of the cross section in this region of phase space by identification of the dominant modes that contribute to the particular values of the measured energy correlation functions.  Assuming that $B$ is an additive observable and $B\ll \ecf{2}{\alpha}$, this factorization theorem takes the form
\begin{align}\label{eq:fact_intro}
\frac{d\sigma(\outj;R)}{d\ecfLa d\ecfLb d\ecfres }&=H(Q^2) H^{sj}_{n\bar{n}}\Big(\ecfLa,\ecfLb\Big) J_{n}\Big(\ecfres\Big)\otimes J_{\bar{n}}(\outj) \nonumber\\
&\qquad\otimes S_{n\bar{n}\sja }\Big(\ecfres;\outj;R\Big)\otimes J_{\sja}\Big(\ecfres\Big)\otimes S_{\sja\sjabar}(\ecfres;R)\,,
\end{align}
where $\otimes$ denotes convolutions for any repeated observable.  Here $H(Q^2)$ and $H^{sj}_{n\bar{n}}\Big(\ecfLa,\ecfLb\Big)$ are hard functions describing the production of the dijet pair and the soft subjet, respectively. The functions $J_{n}\Big(\ecfres\Big)$ and $J_{\bar{n}}(\outj)$ are jet functions describing the dynamics of the jets along the $n$ and $\bar n$ directions. $S_{n\bar{n}\sja }\Big(\ecfres;\outj;R\Big)$ is the global soft function involving three Wilson line directions. Finally, the dynamics of the soft subjet is factorized into the two functions $J_{\sja}\Big(\ecfres\Big)$, and $S_{\sja\sjabar}(\ecfres;R)$, each of which will be discussed in more detail in \Sec{sec:Fact}.  In \Fig{fig:softjet}, we present an illustration of the modes that contribute in the soft subjet region of phase space.  Note that our factorization theorem in the soft subjet region factorizes the in-jet scales defined by $\ecf{2}{\alpha}$ and $\ecf{2}{\beta}$ from the out-of-jet scale $B$, and therefore the NGLs of ratios of the soft subjet energy to the out-of-jet scale $B$ can be resummed. In the effective field theory language, the additional measurement has converted one of the soft scales to a hard scale, allowing for the resummation of the NGLs by standard renormalization group techniques.

By studying the dynamics of the soft subjet in this region of phase space, we are led to introduce what we term the \emph {dressed gluon approximation},\footnote{The term ``dressed gluon" is also used in approaches to renormalon resummation (see e.g. \cite{Gardi:2001ny}). These approaches, although similar in the spirit of associating additional dynamics with a single gluon, attempt to describe completely distinct physical phenomena. } which captures the resummation of NGLs due to unresolved emissions associated with the soft subjet.\footnote{A similar approximation was used in calculation of jets with rapidity gaps \cite{Forshaw:2006fk,Forshaw:2008cq,Forshaw:2009fz,DuranDelgado:2011tp}, termed the ``out-of-gap'' expansion. The connection of these dressed gluon expansions to physically measurable subprocesses, as well as their realization as an expansion of the BMS equation and the role of the buffer region was not addressed in these works.}  In particular, in the region of phase space with a single soft subjet, we have one-dressed gluon. We demonstrate that the dressed gluon approximation can be used to calculate the NGLs of a more inclusive observable by marginalizing over the factorization theorem. Importantly, the one-dressed gluon approximation can be calculated to arbitrary perturbative accuracy and for any number of colors, $N_c$.

While the measurement of $\ecf{2}{\alpha},\ecf{2}{\beta},\ecf{3}{\alpha}$ has allowed us to successfully factorize the jet scales set by $\ecf{2}{\alpha}$ and $\ecf{2}{\beta}$ from $B$, because we have not resolved any further emissions in the jet, there still exist NGLs at the scale of $\ecf{3}{\alpha}$.  To resum those NGLs requires resolving two soft subjets located near the jet boundary.  To isolate this region of phase space and resum the NGLs including $\ecf{3}{\alpha}$, we can measure the two-, three- and four-point energy correlation functions, factorize the cross section and renormalize, which produces the two-dressed gluon approximation.  This then pushes the NGLs to the unresolved scale set by the four-point energy correlation function.  The procedure can then be repeated, by measuring higher-point energy correlation functions to resolve more and more soft subjets in the jet, resumming the NGLs down to some unresolved scale below which no soft subjets are identified. We illustrate this increasingly differential factorization theorem ladder for resumming NGLs in \Fig{fig:ladderfact}.  We discuss the convergence of the reorganization of the traditional perturbative expansion in terms of the number of dressed gluons, and show that the contribution from higher numbers of dressed gluons is highly suppressed by the available phase space volume. We also relate the dressed gluon expansion to an expansion of the BMS equation. We stress that these factorization theorems can be calculated to any perturbative accuracy and for arbitrary numbers of colors $N_c$, allowing for the extension to the resummation of subleading logarithmic corrections.

To justify that this step-by-step resummation procedure of the NGLs accurately captures those NGLs known to exist, we compare our dressed gluon approximation of hemisphere jet masses in the large-$N_c$ limit to Monte Carlo resummation, and the fixed-order expansion of the BMS equation.  By only including the one- and two-dressed gluon approximations, we find agreement with solutions of the BMS equation at the sub-percent level for phenomenological values of the NGLs.  This demonstrates that the one- and two-dressed gluon approximations capture the dominant contributions to the leading NGLs in the large-$N_c$ limit, with small corrections due to the presence of NGLs at lower resolution scales. The dressed gluon approximation is easily incorporated analytically into existing factorization theorems for multi-jet processes.\footnote{This assumes that the jet definitions in those factorization theorems are robust to soft subjets approaching their boundaries.} The dressed gluon approximation also provides analytical understanding of many features of jet physics and NGLs; for example, dressed gluons manifest the ``buffer region'' \cite{Dasgupta:2002bw} near the jet boundary in which emissions are forbidden.

The outline of this paper is as follows.  In \Sec{sec:phase_space} we define the soft subjet region of phase space via measurements of $\ecfnobeta{2}$ and $\ecfnobeta{3}$.  In \Sec{sec:Fact}, we present the factorization of the cross section in the soft subjet phase space region within the context of soft-collinear effective theory, calculating anomalous dimensions and renormalizing the functions appearing in the cross section.  In \Sec{sec:dressed_gluon}, we introduce the dressed gluon approximation, which follows from our factorization theorem in the soft subjet region of phase space. We demonstrate how the dressed gluon approximation can be used in the calculation of NGLs by computing the NGLs for the hemisphere invariant mass with both one and two dressed gluons, and provide a numerical comparison to the BMS equation, and various approximations found in the literature. We also discuss analytic insights into the dynamics of NGLs which are realized in the dressed gluon approximation. In \Sec{sec:nglimp}, we discuss how our approach can be extended beyond leading logarithmic accuracy and discuss the necessity of capturing NGLs arising from collinear splitting along the boundary of phase space.  We conclude in \Sec{sec:conc} and discuss directions for further understanding of NGLs and soft subjet dynamics to all-orders.  Calculational details are presented in appendices.

\section{Observables and Phase Space}
\label{sec:phase_space}

As discussed in the introduction, our strategy for resumming NGLs is to isolate the region of jet phase space which is sensitive to both in-jet and out-of-jet scales, using IRC safe measurements for which we can prove a factorization theorem.  We will consider the process $e^+e^-\to$ hadrons on which we apply a broadening axis cone algorithm.\footnote{We thank Jesse Thaler and Daniele Bertolini for discussions about such algorithms.}  For our purposes, the implementation of this algorithm is to identify the broadening axes in the event, which can be done in an inclusive fashion by minimizing 2-jettiness \cite{Stewart:2010tn,Thaler:2010tr,Thaler:2011gf} with angular exponent $\beta=1$, draw fixed cones of radius $R$ around the axes, and study the largest energy jet, defining the out-of-jet region to be the complement of the cone of radius $R$. We will often take $R=\pi/2$, in which case this divides the event into hemispheres.  The choice of jet algorithm is vital for simplifying the analysis of NGLs; the use of the broadening axes and geometric cones guarantees the jet axis lies along the direction of the hardest radiation in the jet with a circular shape.  This remains true for the complete set of possible soft subjets, even when the soft subjet lies near or on the jet boundary. This scheme is largely equivalent to finding the jets with the anti-$k_T$ jet algorithm \cite{Cacciari:2008gp} with the Winner-Take-All (WTA) recombination scheme \cite{Bertolini:2013iqa,Larkoski:2014uqa,Salambroadening}. For jets containing soft subjets, the two algorithms will give identical factorizations at leading power for much of the soft subjet phase space. However, because it is a sequential recombination algorithm, the boundary of anti-$k_T$ jets will distort in the presence of soft subjets located sufficiently near the boundary.  The power counting of the factorization theorems give a precise definition of the phase space boundaries between these regimes.

At lowest order, the region of phase space sensitive to both in-jet and out-of-jet scales consists of a single soft gluon in the jet, located near the jet boundary.  A single gluon is not an IRC safe object, and so the natural IRC safe generalization of a soft gluon is a soft subjet.  Therefore, we wish to isolate jets which have a hard core of radiation and a soft subjet at large angle from the jet core.  Because we use a recoil-free jet algorithm, this soft, wide angle subjet does not displace the jet axis from the hard jet core.

To identify the soft subjet region of the jet's phase space, we will measure a number of the $n$-point energy correlation functions \cite{Larkoski:2013eya,Larkoski:2014gra} on the jet.  The $n$-point energy correlation function is an IRC safe observable and is sensitive to $n$-prong structure in a jet.  To identify the soft subjet region of phase space we will need the two- and three-point energy correlation functions which we define for $e^+e^-$ collisions as \cite{Larkoski:2013eya}
\begin{align}
\ecf{2}{\beta}&= \frac{1}{E_J^2} \sum_{i<j\in J} E_i E_j \left(
\frac{2p_i \cdot p_j}{E_i E_j}
\right)^{\beta/2} \,, \\
\ecf{3}{\beta}&= \frac{1}{E_J^3} \sum_{i<j<k\in J} E_i E_j E_k \left(
\frac{2p_i \cdot p_j}{E_i E_j}
\frac{2p_i \cdot p_k}{E_i E_k}
\frac{2p_j \cdot p_k}{E_j E_k}
\right)^{\beta/2} \,, \nonumber
\end{align}
where $J$ represents the jet, $E_i$ and $p_i$ are the energy and four momentum of particle $i$ in the jet $J$ and $\beta$ is an angular exponent that is required to be greater than 0 for IRC safety.  The four-point and higher energy correlation functions are defined as the natural generalization.  The $n$-point energy correlation function vanishes in all soft and collinear limits of an $n$ particle configuration.  In the soft or collinear limit, the $\ecf{2}{\beta}$ are equivalent to the (recoil-free) angularities \cite{Berger:2003iw,Almeida:2008yp,Ellis:2010rwa,Larkoski:2014uqa} and when measured event-wide in $e^+e^-$ collisions, $\ecf{3}{2}$ is equivalent to the $C$-parameter \cite{Parisi:1978eg,Donoghue:1979vi} computed to ${\cal O}(\alpha_s)$.

\begin{figure}
\begin{center}
\subfloat[]{\label{fig:2pt3ptps}
\includegraphics[scale = 0.6]{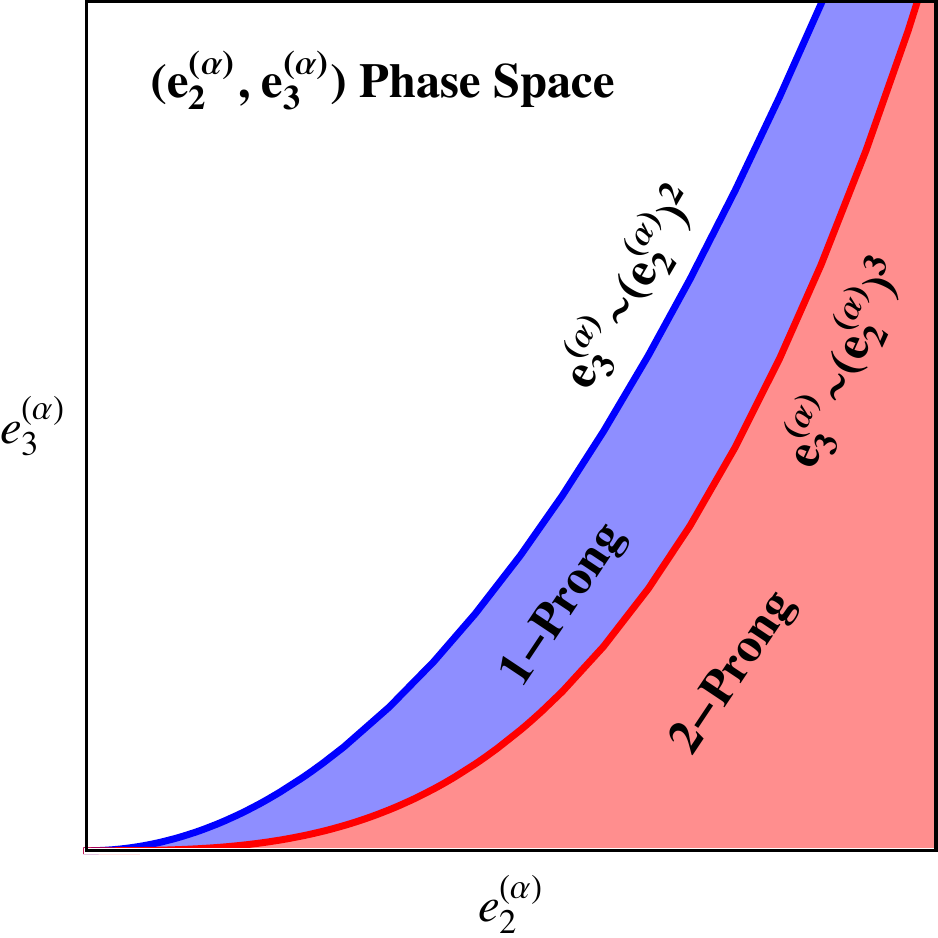}
}
$\qquad$
\subfloat[]{\label{fig:2ptps}
\includegraphics[scale = 0.6]{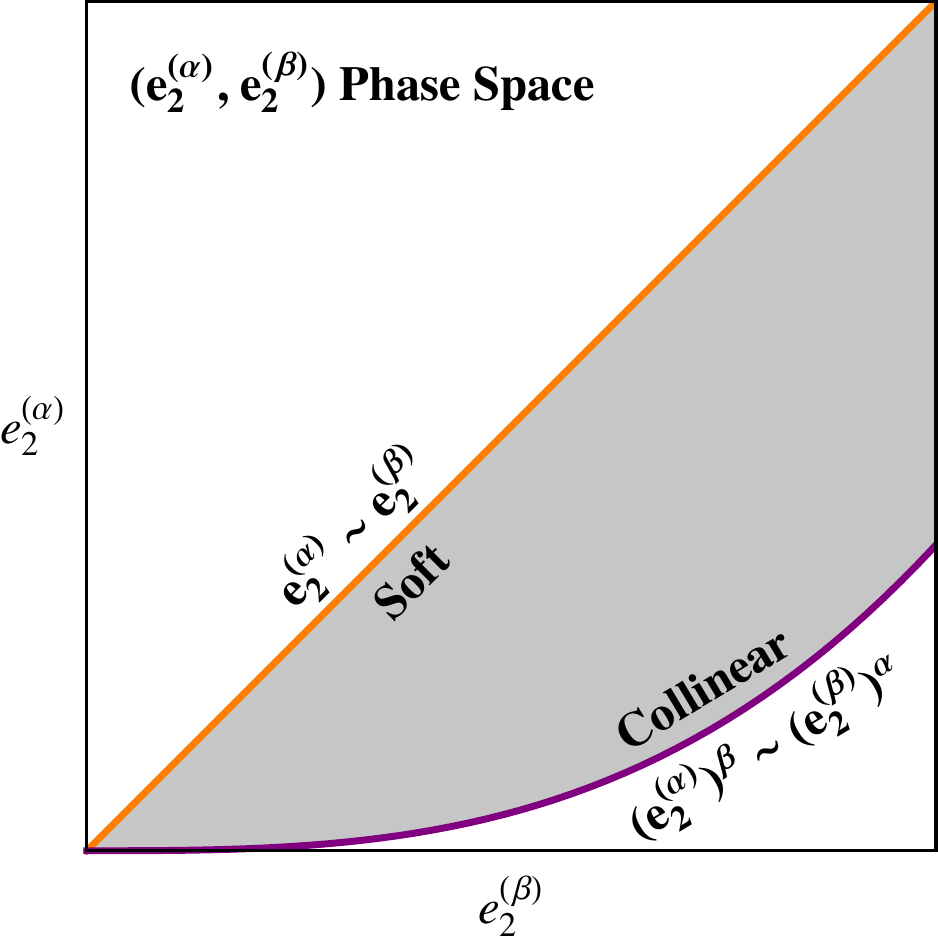}
}
\end{center}
\vspace{-0.2cm}
\caption{(a)  Illustration of the phase space for a jet on which $\ecf{2}{\alpha}$ and $\ecf{3}{\alpha}$ have been measured. Jets with a two-prong structure lie in the lower (red) region of phase space, where $\ecf{3}{\alpha} \ll (\ecf{2}{\alpha})^3$. The energy correlation functions parametrically separate the one-prong and two-prong regions of phase space. (b) Illustration of the phase space for a jet on which both $\ecf{2}{\alpha}$ and $\ecf{2}{\beta}$ have been measured, with $\alpha>\beta$, shown in gray.  Jets dominated by soft radiation lie in the upper region of the phase space, where $\ecf{2}{\alpha} \sim \ecf{2}{\beta}$. Jets with two energetic collinear subjets populate the region $\ecf{2}{\alpha} \sim {\ecf{2}{\beta}}^{\alpha/\beta}$.
}
\label{fig:phasespace}
\end{figure}

The measurements that we perform on these events are as follows.  First, we measure some observable $B$ on the region outside the identified jet $J$.  On the jet $J$, we measure the observables $\ecf{2}{\alpha}$, $\ecf{2}{\beta}$ and $\ecf{3}{\alpha}$ for angular exponents $\alpha, \beta$ and we will assume that $\alpha > \beta$.  To demand that the dynamics of the jet are dominated by soft and collinear radiation, we require that $\ecf{2}{\alpha} \ll 1$.  A jet with a hard core and a single soft subjet has 2-prong substructure, and so to identify 2-prong jets we require \cite{Larkoski:2014gra}
\begin{equation}\label{eq:powercount_2jet}
\ecf{3}{\alpha} \ll \big(\ecf{2}{\alpha}\big)^3 \,,
\end{equation}
which follows straightforwardly from power counting.
The measurement of the two- and three-point energy correlation functions on a jet resolve either 1- or 2-prong substructure, which is described in detail in \Ref{Larkoski:2014gra} and is displayed in \Fig{fig:2pt3ptps}.  Importantly, the energy correlation functions provide a parametric separation of the 1- and 2-prong regions of phase space, defined by the precise scaling relation of \Eq{eq:powercount_2jet}. Due to this parametric separation, well defined factorization theorems exist in both regions of phase space.   To enforce that the subjet is both soft and at a wide angle from the jet core, we therefore require
\begin{equation}
\ecf{2}{\alpha} \sim \ecf{2}{\beta} \,.
\end{equation} 
Combined with the condition $\ecf{2}{\alpha}\ll1$, this forces $\sje\ll 1$ and $\sjtheta\sim R$, where $R$ is the jet radius, which we assume to be an order 1 number.  In particular, this additional measurement allows us to distinguish the case of a soft subjet from the case of two energetic collinear subjets, which has been studied in \Ref{Bauer:2011uc}.  The  $\ecf{2}{\alpha}, \ecf{2}{\beta}$ phase space is described in detail in \Ref{Larkoski:2014tva} (see also \cite{Procura:2014cba}) and displayed in \Fig{fig:2ptps}. The two different subjet configurations, which exist on the boundaries of the allowed phase space defined by $\ecf{2}{\alpha}$ and $ \ecf{3}{\alpha}$, are shown in \Fig{fig:subjet_configs}.  Because we have identified a soft subjet and measured both $\ecf{2}{\alpha}$ and $\ecf{2}{\beta}$, the energy fraction $\sje$ and angle from the jet core $\sjtheta$ of the soft subjet are well-defined and IRC safe quantities.  Once the soft subjet region of phase space has been identified using the energy correlation functions, the observables we will consider as measured on the jet are $\sje$, $\sjtheta$ and $\ecf{3}{\alpha}$.  \Fig{fig:softjet} illustrates the structure of the event we are studying and the observables that we measure on the in-jet and out-of-jet regions.

\section{Effective Field Theory Description and Factorization}
\label{sec:Fact}

In this section we present a factorization theorem in the soft subjet region of phase space described in \Sec{sec:phase_space}. We use the formalism of soft-collinear effective theory (SCET) \cite{Bauer:2000yr,Bauer:2001ct,Bauer:2001yt,Bauer:2002nz}, an effective theory of QCD in the soft and collinear limits. Because the factorization theorem involves many novel features, we will discuss its structure in detail. We begin with a power counting analysis in \Sec{sec:modes} to determine the modes required in the low energy effective theory. The mode structure dictates the functions appearing in the factorization theorem, which is presented in \Sec{sec:fact_theorem}.  The novel feature of the factorization theorem is the presence of modes whose virtuality is set by their angle to the boundary of the jet and whose resummation is directly tied to the resummation of NGLs.

\subsection{Modes of the Factorization}\label{sec:modes}

\begin{figure}
\begin{center}   
\subfloat[]{\includegraphics[width=7.25cm]{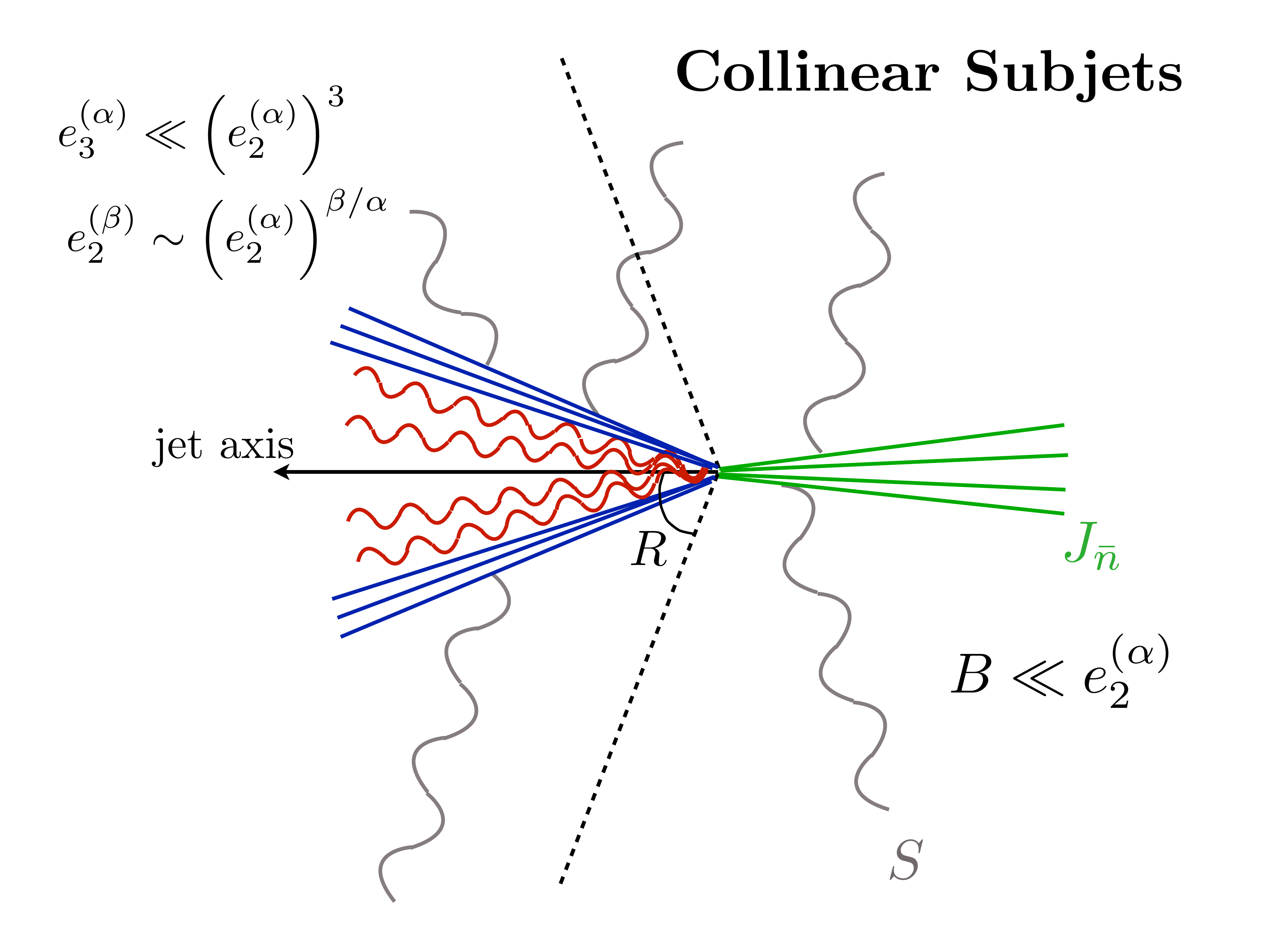} 
}\ 
\subfloat[]{
\includegraphics[width=7.25cm]{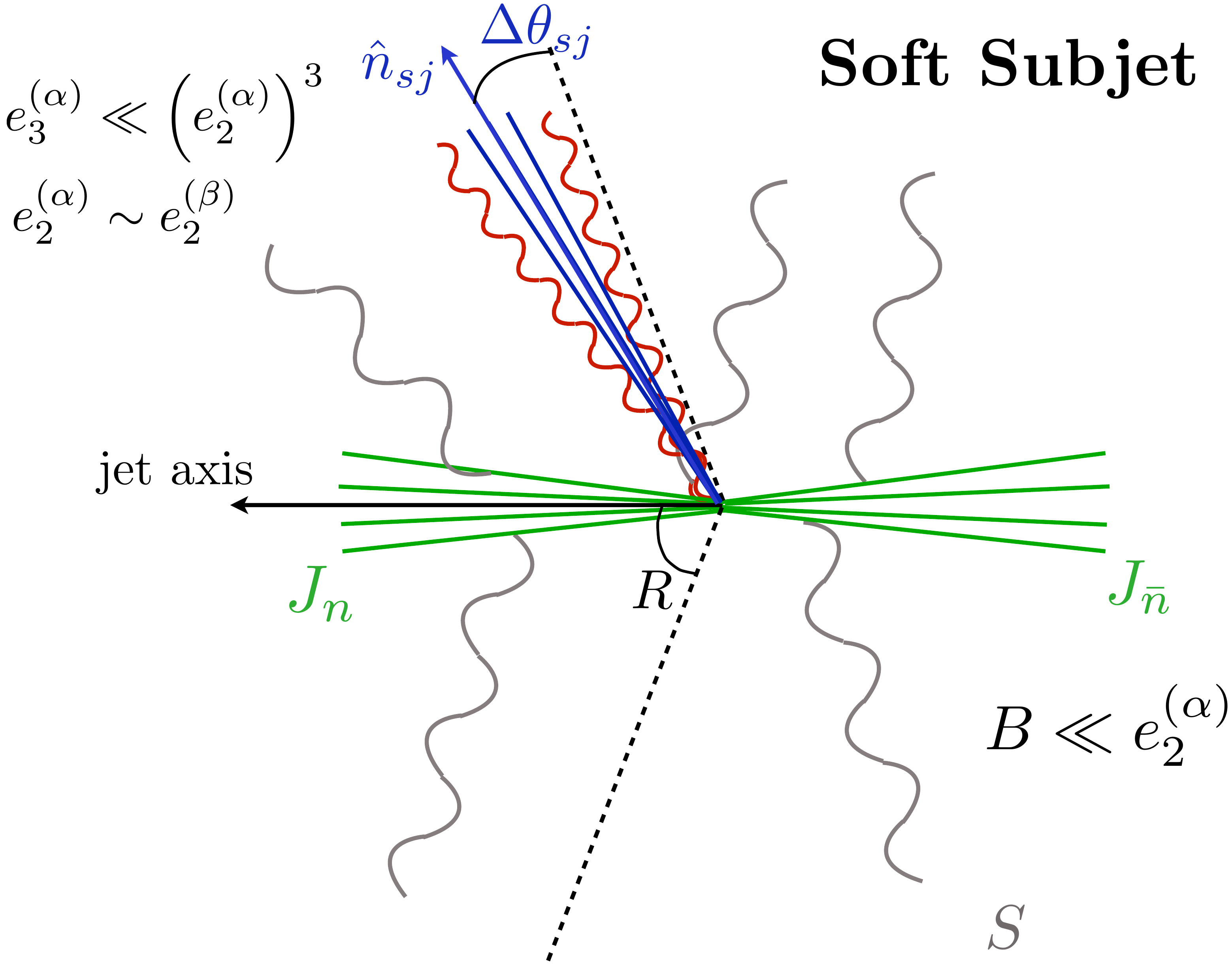}
}
\end{center}
\caption{ The two distinct subjet configurations which exist in the two prong region of phase space. (a) Two energetic collinear subjets, which has been studied in \Ref{Bauer:2011uc}, and populates the region of phase space $\ecf{2}{\beta} \sim \left ({\ecf{2}{\alpha}}\right )^{\beta/\alpha}$. (b) Wide angle soft subjet, which populates the region of phase space $\ecf{2}{\alpha} \sim \ecf{2}{\beta}$. 
}
\label{fig:subjet_configs}
\end{figure}

The simultaneous measurement of $\ecfLa $, $\ecfLb$, and $\ecfres$ defines a multi-scale structure in the low energy effective theory required for the complete factorized description of the soft subjet region of phase space. A proper understanding of the relative scalings of the modes is essential to specify the structure of the factorization theorem we will present in \Eq{fact_inclusive_form_1}, and to understand the structure of zero bin subtractions \cite{Manohar:2006nz} needed to remove overlaps. The scaling of the modes in the low energy effective theory can be determined by power counting arguments, which have been considered for multi-differential observables resolving multiple subjets in \Refs{Larkoski:2014gra,Larkoski:2014zma}. We will follow the conventions used in those papers, writing the scalings of all modes in terms of the physical observables, instead of the traditional $\lambda$, as we find this to be more transparent.  Typically, the modes of a factorization theorem are set by the observables measured. We will find in addition that the existence of the jet boundary itself will play a vital role in determining all the modes required for the complete factorized description.

Because we only measure the observable $B$ in the out-of-jet region, its mode structure is simple and so we will discuss it first.  Importantly, we require that $B$ is an additive observable, like the energy correlation functions, angularities, mass, etc., so that the factorization theorem is of a universal form.  We assume that the out-of-jet region has order-1 angular size and $B\ll 1$ and so the measured value of $B$ is dominated by collinear and global soft modes.  Given a momentum $p$, we will adopt the following notation for its components expressed in light-cone coordinates defined by the directions $a$ and $\bar a$:
\begin{equation}\label{eq:momdecomp}
(p^+,p^-,p_\perp)_{a\bar{a}}\equiv(a \cdot p,\bar{a}\cdot p,p_\perp)\,.
\end{equation}
As an example, if we assume that $B$ measures the out-of-jet broadening, the scaling of the collinear and soft modes is
\begin{align}
p_c &\sim Q \left (  1,B^2,B   \right )_{n\bar{n}}\,, \\
p_s &\sim  Q \,B \left ( 1 ,1,1  \right )_{n\bar{n}}    \,,\nonumber 
\end{align}
where $Q$ is a proxy for the energy in the out-of-jet region and is of the same order as the total scattering energy in the event. $n$ is the direction of the jet of interest, while $\bar n$ is the direction of the jet that contributes to the measured value of $B$.

Now, we turn to the in-jet modes that contribute to the measurement of the energy correlation functions.  As shown schematically in \Fig{fig:subjet_configs}, in the soft subjet region of phase space, we assign the power counting $\sjtheta\sim R\sim 1$, where $\sjtheta$ is the angle between the jet axis and the soft subjet direction.  In this region of phase space, the two-point energy correlation functions are dominated by soft, wide angle radiation and so we have
\begin{equation}
 \ecfLa \sim \ecfLb \sim \sje\,,
\end{equation}
where $\sje$ is the energy fraction of the soft subjet.  As there is no parametric difference between $\ecf{2}{\alpha}$ and $\ecf{2}{\beta}$, these observables are redundant from a power-counting perspective and in the following, we will express the scaling of all modes in terms of $\ecf{2}{\alpha}$.  Continuing, the additional measurement of $\ecf{3}{\alpha}$ on the jet resolves the hard collinear structure of the jet core, the structure of the soft subjet, and global soft radiation from the hard dipoles present in the event.\footnote{Unlike the SCET$_+$ factorization of \Ref{Bauer:2011uc}, which considered a jet with collinear subjets, a soft wide-angle subjet does not add an additional collinear-soft mode.  In the SCET$_+$ case, global soft radiation cannot resolve the hard collinear splitting, which therefore requires an additional mode to describe the dipole radiation of the subjets.}  We will denote the angular size of the hard core by $\theta_c$, the angular size of the soft subjet by $\theta_{cs}$ and the energy fraction of the global soft radiation by $z_s$.  Therefore, the parametric scaling of $\ecf{3}{\alpha}$ in the soft subjet region of phase space is set by these three contributions:
\begin{equation}\label{e3scale}
\ecf{3}{\alpha}\sim \sje (\theta_c^\alpha +\sje \theta_{cs}^\alpha +z_s)\,.
\end{equation}
This implies that only the soft subjet modes are directly sensitive to \emph{both} the $\ecfLa$ and $\ecfres$ measurements.

From the contributions to $\ecf{3}{\alpha}$ in \Eq{e3scale}, we are then able to define the momentum scaling of each contributing mode via the measured values of $\ecf{2}{\alpha}$ and $\ecf{3}{\alpha}$.  In the notation of \Eq{eq:momdecomp}, the momentum of the hard collinear and global soft radiation scales like
\begin{align}
p_c &\sim E_J \left ( \left (  \frac{\ecfres}{  \ecfLa }  \right)^{2/\alpha},1,\left (  \frac{\ecfres}{  \ecfLa }  \right)^{1/\alpha}    \right )_{n\bar{n}}\,, \\
p_s &\sim  E_J \,\frac{\ecfres}{  \ecfLa } \left ( 1 ,1,1  \right )_{n\bar{n}}    \,,\nonumber 
\end{align}
where $E_J$ is the energy of the jet and $n$ and $\bar n$ are the light-like directions of the jet of interest and the other jet in the event, respectively.  The soft subjet mode's momentum scales like 
\begin{align}
p_{sj} \sim E_J\, \ecfLa \left ( \left (  \frac{\ecfres}{  \left (\ecfLa\right)^2 }  \right)^{2/\alpha},1,\left (  \frac{\ecfres}{  \left (\ecfLa\right )^2 }  \right)^{1/\alpha}    \right )_{\sja\sjabar}   \,,
\end{align}
in the light-cone coordinates defined by the direction of the soft subjet, $n_{sj}$. These are the complete set of modes defined by the scales set by the measurements of $\ecfLa ,\ecfLb$, and $\ecfres$ alone.

If these measurements were global, that is, sensitive to all radiation in the event, this would be the complete enumeration of the modes that contribute to the measured observables.  However, because the energy correlation functions are only measured on the jet, the boundary of the jet plays an important role in defining the relevant modes as well.  In particular, since we are considering the case where the out-of-jet scale is much less than the in-jet scale, namely $B\ll \ecf{2}{\alpha}$,\footnote{Formally we take the scaling $B\sim \frac{\ecf{3}{\alpha}}{\ecf{2}{\alpha}}$, i.e., $B$ is at the global soft scale.} for the modes in the soft subjet, the angle between the soft subjet axis and the jet boundary $\Delta \sjtheta\equiv R-\sjtheta$ places additional constraints on the soft subjet dynamics, much like an additional measurement would. In the region of phase space in which the NGLs are parametrically large, and should be resummed, this must be taken into account. In \Sec{sec:nglimp} we will briefly discuss the case when the NGLs are not parametrically large, and how this factorization theorem is modified. We therefore expect that $\Delta \sjtheta$ defines a relevant scale in the effective theory, and should be included in the power counting analysis. We must consider the possibility of a mode whose angular scale with respect to the soft subjet axis is not set by the measurement of the two- and three-point energy correlation functions, but rather by the jet boundary itself.  This mode does not contribute to the two-point energy correlation functions $\ecfLa ,\ecfLb$ and its energy is set by $\ecfres$.  This therefore defines an additional soft mode which is localized around the soft subjet's direction and constrained by the jet boundary.  We therefore refer to this new mode as a {\it boundary soft mode}.  

The presence of this mode is absolutely necessary for understanding NGLs. Importantly, within the fat jet, it is the only mode that contributes to the cross section singular logarithmic terms of the form
\begin{align}
\ln\left(\frac{1}{\Delta \sjtheta}\right)\gg 1\,.
\end{align}
The necessity of this mode, justified here by power counting, also appears from explicit calculation of the functions appearing in the factorization theorem in \Eq{fact_inclusive_form_1}.  Logarithms of $\Delta \sjtheta$ can also arise from global soft radiation in the out of jet region (see \App{sec:global_soft}), and so the boundary soft mode will be required for renormalization group consistency of the factorization theorem. It is critical for the NGL resummation that the two different factorized functions at different energy scales are both sensitive to the jet boundary.

The scaling of the momentum of the boundary soft mode is determined by considering its contribution to $\ecfres$, given that its angular scale is set by $\Delta\sjtheta$.  The dominant contribution to $\ecf{3}{\alpha}$ from the boundary soft modes is
\begin{equation}
\left.\ecf{3}{\alpha}\right|_{bs}\sim z_{sj}\,z_{bs} \left(\Delta \sjtheta\right)^\alpha \,, 
\end{equation}
where the energy fraction of the boundary soft radiation is $z_{bs}$.  Therefore the boundary soft mode's momentum components scale like
\begin{align}
p_{bs} &\sim E_J\frac{  \ecfres   }{ \ecfLa   \left(\Delta \sjtheta\right)^\alpha }  \left ( \left(\Delta  \sjtheta\right)^2,1,\Delta  \sjtheta \right )_{\sja\sjabar}   \,, \nonumber 
\end{align}
written in the light-cone coordinates defined by the soft subjet axis.  For consistency of the factorization, we must enforce that the soft subjet modes cannot resolve the jet boundary and that the boundary soft modes are localized near the jet boundary.  That is, the angular size of the soft subjet modes must be parametrically smaller than that of the boundary soft modes:
\begin{equation}
\left(\Delta\sjtheta\right)^\alpha \gg  \left(\theta_{cs}\right)^\alpha\sim \frac{\ecfres}{\left (\ecfLa \right )^2}\,, \qquad \text{and}\qquad \Delta\sjtheta \ll 1\,.
\end{equation}
Therefore, the factorization theorem applies in a region of the phase space where the soft subjet is becoming pinched against the boundary of the  jet, but lies far enough away that the modes of the soft subjet do not touch the boundary.\footnote{For the anti-$k_T$ jet algorithm, if the modes of the soft subjet resolve the jet boundary (so $\Delta\sjtheta\sim \theta_{cs}$), this will in general result in distortions of the jet boundary due to clustering effects. In a strict cone algorithm, as we use here, this boundary-collinear regime is still factorizable, and is relevant for the resummation of subleading NGLs, as will be discussed in \Sec{sec:nglimp}.}

\subsection{The Factorization Theorem for a Soft Subjet}\label{sec:fact_theorem}

We will prove the following factorization theorem for the production of a soft subjet:
\begin{align}\label{fact_inclusive_form_1}
\frac{d\sigma(\outj;R)}{d\ecfLa d\ecfLb d\ecfres }&=H(Q^2) H^{sj}_{n\bar{n}}\Big(\ecfLa,\ecfLb\Big) J_{n}\Big(\ecfres\Big)\otimes J_{\bar{n}}(\outj) \nonumber\\
&\qquad\otimes S_{n\bar{n}\sja }\Big(\ecfres;\outj;R\Big)\otimes J_{\sja}\Big(\ecfres\Big)\otimes S_{\sja\sjabar}(\ecfres;R)\,,
\end{align}
valid under the assumptions on the phase space described in \Sec{sec:modes}.
Here convolutions are implicit in any variable that is twice repeated, and we have explicitly indicated the dependence on the jet boundaries with the jet radius $R$. The operator definitions of the functions are given in \App{app:Factorized_Functions}, but the physical origin of each function is straightforward to understand, with each function describing the dynamics of one of the modes described in \Sec{sec:modes}. A brief description of the functions appearing in \Eq{fact_inclusive_form_1} is as follows:
\begin{itemize}
\item $H(Q^2)$ is the hard function for the production of a dijet event at an $e^+e^-$ collider.
\item $H_{n\bar{n}}^{sj}\Big(\ecfLa,\ecfLb\Big) $ is the hard function describing the production of the soft subjet from the initial $q\bar{q}$ dipole, and describes dynamics at the scale set by $\ecfLa,\ecfLb$.
\item $J_{n}\Big(\ecfres\Big)$ is a jet function at the scale $\ecfres$ describing the hard collinear modes of the identified jet.
\item  $J_{\bar{n}}(\outj)$ is a jet function describing the collinear modes of the out-of-jet region of the event.
\item $S_{n\bar{n}\sja }\Big(\ecfres;\outj;R\Big)$ is the global soft function, involving three Wilson line directions, $n, \bar n, \sja$.
\item $J_{\sja}\Big(\ecfres\Big)$ is a jet function describing the dynamics of the soft subjet modes, which carry the bulk of the energy in the soft subjet.
\item $S_{\sja\sjabar}(\ecfres;R)$ is a soft function describing the dynamics of the boundary soft modes. It depends only on two Wilson line directions $\sja, \sjabar$.
\end{itemize}
The factorization theorem of \Eq{fact_inclusive_form_1} therefore achieves a complete factorization of the modes presented in \Sec{sec:modes}. As discussed in \Sec{sec:phase_space}, in the soft subjet region of phase space, we can relate the variables $\ecfLa\,, \ecfLb$ to the physically more transparent $\sje, \sjtheta$ variables with a simple Jacobian factor, giving the factorization theorem
\begin{align}\label{fact_inclusive_form_2}
\frac{d\sigma(\outj;R)}{d\sje\, d\sjtheta\, d\ecfres}&=H(Q^2) H^{sj}_{n\bar{n}}\Big(\sje,\sjtheta\Big) J_{n}\Big(\ecfres\Big)\otimes J_{\bar{n}}(\outj) \nonumber\\
&\qquad\otimes S_{n\bar{n}\sja }\Big(\ecfres;\outj;R\Big)\otimes J_{\sja}\Big(\ecfres\Big)\otimes S_{\sja\sjabar}(\ecfres;R)\,.
\end{align} 
The calculation to one-loop of various objects in this factorization theorem is presented in \App{app:oneloopcalcs}.

\begin{figure}
\begin{center} 
\includegraphics[width=14cm]{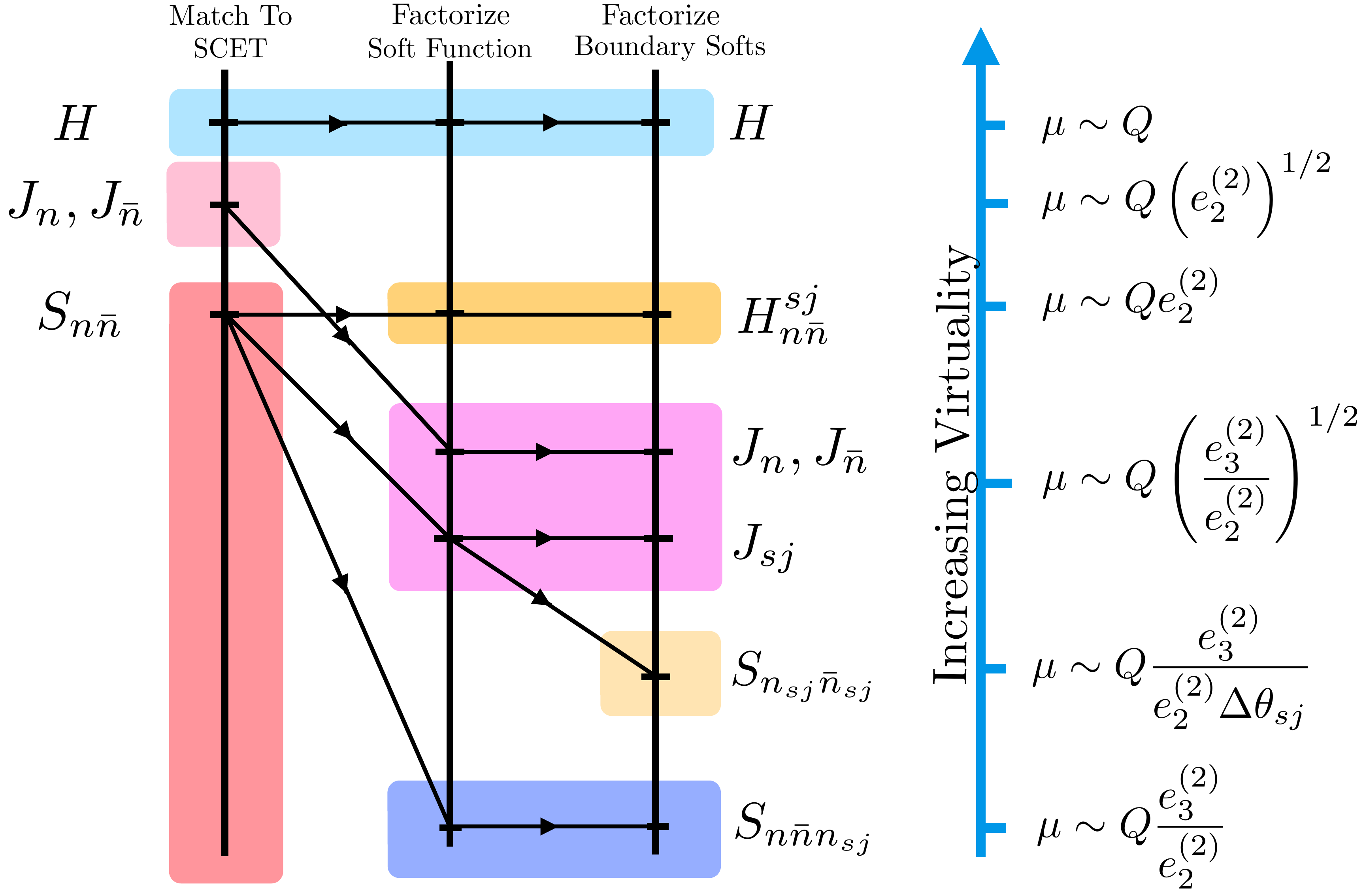} 
\end{center}
\caption{ A schematic of the multi-stage matching procedure used to prove the factorization theorem of \Eq{fact_inclusive_form_1} for the soft subjet region of phase space. As discussed in the text, the matching proceeds in three stages: matching to SCET, refactorizing the soft function to describe the soft jet production, and factorizing the boundary soft mode. The canonical scales of the modes in the final factorization theorem are shown on the right, ordered in virtuality. Here we have chosen an angular exponent $\alpha=2$ for concreteness.  
}
\label{fig:scales_factorization}
\end{figure}

We now describe how the factorization theorem of \Eq{fact_inclusive_form_1} arises in a multi-stage matching onto the effective theory involving the modes of \Sec{sec:modes}.\footnote{In the case that $\sje\gg\outj$, such a multi-stage matching is not necessary and one can construct the factorization theorem along the lines of \Refs{Stewart:2010tn,Ellis:2010rwa}. However, here we pursue the multi-stage matching since it will be necessary in a forthcoming publication \cite{us:inclusive} dealing with the factorization of soft subjets when $\sje\ll \outj$.} This multi-stage matching procedure is shown schematically in \Fig{fig:scales_factorization}. In the first step, QCD is matched onto an SCET theory at the hard scale of the $e^+e^-$ event by matching the electroweak currents of both theories \cite{Fleming:2007qr,Bauer:2008dt,Chiu:2012ir}. This theory is then evolved down to the soft scales defined by $\ecfLa\sim \ecfLb$, where these soft modes can be decoupled via the BPS field redefinition \cite{Bauer:2002nz} from the collinear modes. The soft scale associated with this measurement is $\ecfLa\sim \sje$, so the soft subjet is not resolved, and is simply described as part of the soft function. At this stage we have the usual SCET factorization formula for dijet production:
\begin{equation}
\frac{d\sigma(\outj;R)}{d\ecfLa d\ecfLb d\ecfres }\sim H(Q^2)J_n(\ecfLa,\ecfLb,\ecfres)\otimes J_{\bar n}(\outj) \otimes S_{n\bar n}(\ecfLa,\ecfLb,\ecfres;\outj) \,,
\end{equation}
where we have decoupled the collinear modes in the jet functions from the soft radiation at this scale using a BPS field redefinition. We have not yet performed the full multipole expansion required to separate all infrared scales \cite{Grinstein:1997gv,Beneke:2002ni}, since from the power-counting arguments given above we have not included enough modes to separate all scales.  Hence we let all modes resolved at this scale contribute to the jet measurements. 

As we lower the scale below $\ecfLa$, we resolve the soft subjet, and therefore we must further expand the soft function. However, since the collinear dynamics of the jets along the axes $n$ and $\bar n$ are at the scale $\ecfres$ and $\outj$, respectively, and they have been decoupled from the soft radiation, their scales can simply be lowered without any matching.\footnote{At this stage, the collinear modes cannot contribute to $\ecfLa,\ecfLb$ anymore, except via their overall direction.} This should be contrasted with the factorization theorem of \Ref{Bauer:2011uc} for the case of two collinear subjets, where additional matching must be performed in the jet function. In our case, an additional matching step must instead be performed in the soft function, corresponding to the known fact that the NGLs appear in the soft function.  At the scale $\ecfLa$, we expand the soft function and match onto a hard function $H_{n\bar{n}}^{sj}$ describing the production of a soft subjet, a jet function $\tilde J$ describing the dynamics of the soft subjet, and a global soft function involving three Wilson lines. This refactorization is shown schematically as the transition from the first to the second column in \Fig{fig:scales_factorization}. This is analogous to the  construction of the PDFs in SCET, see \Ref{Bauer:2002nz}.  The factorization theorem then becomes
\begin{align}
\frac{d\sigma(\outj;R)}{d\ecfLa d\ecfLb d\ecfres }&= H(Q^2)H^{sj}_{n\bar{n}}\Big(\ecfLa,\ecfLb\Big)J_n(\ecfres) \otimes J_{\bar n}(\outj)\nonumber \\
&\qquad\qquad \otimes S_{n\bar{n}\sja }\Big(\ecfres;\outj;R\Big) \otimes  \tilde J_{\sja}\Big(\ecfres\Big) \,.
\end{align}
In the final step of the matching, we lower to the scale set by the measurement of $\ecfres$. At this scale we probe the dynamics of the soft subjet, and must perform a final factorization of $\tilde J$ into a function describing the boundary soft modes and a function describing the soft subjet modes:
\begin{equation}\label{eq:ssubjetfact}
\tilde J_{\sja}\Big(\ecf{3}{\alpha}\Big)=  J_{\sja}\Big(\ecfres\Big)\otimes    S_{\sja\sjabar}(\ecfres;R)\,.
\end{equation}
The boundary soft modes can be BPS decoupled from the soft subjet modes, thus resulting in the final form of the factorization theorem in \Eq{fact_inclusive_form_1}. The virtualities of all the modes present in the final factorization theorem are listed in \Fig{fig:scales_factorization} for the specific case of $\alpha=2$.

The final factorization of the soft subjet function in \Eq{eq:ssubjetfact} is essential to resum all logarithms. In particular, the soft subjet modes, described by $J_{\sja}\Big(\ecfres\Big)$, are not sensitive to the jet boundary, as should be expected for a collinear mode, whereas the boundary soft modes are sensitive to the jet boundary.  Therefore, $S_{\sja\sjabar}(\ecfres;R)$ will have a logarithmic dependence on the angular distance of the soft subjet to the boundary $\Delta \sjtheta$, diverging as the soft subjet approaches the boundary.  The same logarithmic dependence of the boundary angle is found in the na\"ive global soft function $S_{n\bar{n}\sja }\Big(\ecfres;\outj;R\Big)$ both in the in-jet and out-of-jet regions of phase space.  To avoid double counting, one must subtract the contribution of the boundary soft region of phase space from the global soft function (this subtraction is implemented in the effective theory via a zero bin subtraction \cite{Manohar:2006nz}), which in turn removes the logarithmic dependence on $\Delta \sjtheta$ from the global soft contribution to the identified jet.  However, this dependence on $\Delta \sjtheta$ will exist in the boundary softs and the global soft radiation in the out-of-jet region. This will be important for the resummation of the NGLs associated with the soft subjet production by running from the boundary soft scale down to the out-of-jet scale.

In summary, we have presented a factorization theorem describing the region of phase space in which a soft subjet is identified within a jet. By performing multiple measurements to isolate a specific region of phase space, we were able to refactorize the multi-scale soft function. Effectively, the additional measurement converted soft scales to hard scales, so that logarithms associated with these ratios of scales can be resummed by standard renormalization group evolution, hence achieving the resummation of NGLs in this particular region of phase space. While this factorization theorem is interesting in its own right for studying the dynamics of the soft subjet, as relevant for example for the factorized description of jet substructure variables, we will not pursue this direction further in this paper, leaving it to a future publication \cite{usD2}. Instead, in this paper we will focus on using the factorization theorem in this region of phase space to understand properties of NGLs, and applying this understanding to the resummation of NGLs for more inclusive observables.

\section{Dressed Gluon Approximation}\label{sec:dressed_gluon}

In the previous section, we have shown how multiple measurements can be used to isolate a region of phase space involving a single soft subjet, and we presented a factorization theorem, \Eqs{fact_inclusive_form_1}{fact_inclusive_form_2}, in the framework of SCET describing this region of phase space.  By making multiple measurements, we are able to refactorize the soft function allowing for the resummation NGLs of $\ecfLa/\outj$ in this particular region of phase space. 

In this section, we discuss how the resummation of the NGLs in the soft subjet region of phase space allows us to understand more general properties of NGLs.  We propose a reorganization of the perturbative expansion for calculating NGLs of more inclusive observables. We call this the dressed gluon approximation, which we define in \Sec{sec:pants}.  We then explicitly demonstrate how the dressed gluon approximation can be used in a calculation, by calculating the one- and two-dressed gluon contribution to the NGLs for the hemisphere jet mass in \Sec{sec:dressed_calcs}.  In this section, we also develop an understanding for the emission of a soft subjet off of $N$ eikonal lines, and present a conjectured factorization theorem.  In \Sec{sec:nglcomp} we compare our dressed gluon approximations to both resummed and fixed-order solutions of the BMS equation, which captures the leading NGLs in the large $N_c$ limit.  Using the properties of the dressed gluon approximation, in \Sec{sec:insights} we discuss some insights into features of NGLs, and compare the expansion in the number of dressed gluons to other expansions of the BMS equation.

\subsection{
Putting the Pants on a Gluon One Leg at a Time 
}\label{sec:pants}

The dressed gluon approximation originates from the observation that the factorization theorem for the soft subjet resums a set of higher order corrections to the matrix element for the production of a single soft gluon from the $n\bar{n}$ dipole.  The matrix element for gluon emission from the dipole is given by the hard function, $H_{n\bar{n}}^{sj}(\sje,\sjtheta)$, in the factorization theorem of \Eq{fact_inclusive_form_2}.  These corrections include, however, more than just the threshold virtual corrections of the soft gluon current, as they also contain an arbitrary number of soft gluon emissions in the out-of-jet region.  Therefore, these are precisely the corrections associated with the NGLs.

However, our factorization theorem is multi-differential, and so to determine the NGLs for a more inclusive measurement requires marginalizing or integrating over observables that are not included in the measurement.  This integration can potentially be problematic from the point of view of our factorization theorem.  We have only resummed NGLs down to the scale set by $\ecf{3}{\alpha}$, but for an inclusive measurement we must integrate over all regions of phase space, including scales lower than $\ecf{3}{\alpha}$, beyond which no jet structure is resolved.  This would seem to indicate a loss of formal accuracy, as there may be large logarithms the factorization theorem is ignorant to that should be resummed.  Indeed, traditional strict logarithm counting is breaking down in this case, as there would be terms scaling like $\alpha_s \ln\sim 1$ that are not resummed by our factorization theorem.

The resolution of this is the realization that resummation of NGLs in our factorization theorem down to the scale $\ecf{3}{\alpha}$ exponentially suppresses the phase space for emissions at scales lower than $\ecf{3}{\alpha}$. This feature of the expansion in the number of dressed gluons will be discussed in detail in \Sec{sec:pert_vs_dressed}, after we have shown that the dressed gluon approximation gives rise to the buffer region \cite{Dasgupta:2002bw} in \Sec{sec:buffer}. Therefore, while a formal logarithmic accuracy of the resummation of NGLs may not necessarily be guaranteed, effects from emissions at unresolved scales are suppressed by their allowed phase space.  We refer to the resummation of a soft subjet according to the factorization theorem of \Eq{fact_inclusive_form_2} as the {\it dressed gluon approximation}.\footnote{We hope that this nomenclature explains the unusual title of this section.}  Resummation in the factorization theorem dresses the soft subjet at a scale defined by $\ecf{2}{\alpha}$ by an arbitrary number of soft emissions down to a scale set by $\ecf{3}{\alpha}$.  By the structure of the factorization theorem, the one-dressed gluon approximation is guaranteed to include the correct NGL at ${\cal O}(\alpha_s^2)$, and resums a tower of NGLs at higher orders in $\alpha_s$.  The accuracy of this approximation is controlled by the volume of allowed phase space for emissions at scales lower than $\ecf{3}{\alpha}$, so the one-dressed gluon approximation does not fully include, for example, the NGL at ${\cal O}(\alpha_s^3)$.  To fully describe this NGL, we must include the two-dressed gluon approximation, by resolving emissions down to a lower scale in the factorization theorem with further measurements.  This then correctly describes the NGL at ${\cal O}(\alpha_s^3)$, and resums a tower of NGLs to higher orders in $\alpha_s$.  One can continue the procedure, adding more and more dressed gluons to obtain an arbitrarily accurate description of the NGLs. However, as we will demonstrate, the phase space suppression accompanied by an increased number of dressed gluons causes the dressed gluon expansion to converge rapidly, so that for phenomenologically relevant values of the NGLs, only one or two dressed gluons are required for an accurate description. We will explicitly consider up to two dressed gluons in this paper.

We will first present a detailed explanation of the one-dressed gluon approximation and its construction from the factorization theorem of \Eq{fact_inclusive_form_2}, before discussing the extension to multiple dressed gluons.  Because we are interested in the NGLs for more inclusive jet measurements, we must integrate over the unresolved scale set by $\ecf{3}{\beta}$.  We will do this by Laplace transforming the multi-differential cross section for the soft subjet phase space region as:
\begin{align}\label{eq:margine3}
\frac{d\sigma\big(\ecfreslp,\outj;R\big)}{d\sje\, d\sjtheta}&=\int\limits_{0}^{\infty}d\ecfres\,e^{-\ecfreslp\ecfres}\frac{d\sigma(\outj;R)}{d\sje \,d\sjtheta \,d\ecfres} \,.
\end{align}
The cross section fully inclusive over $\ecfres$ is then found by the limit $\ecfreslp\to 0$.  This limit gives a prediction for a single soft gluon matrix element, with all possible low energy unresolved configurations produced by its subsequent splittings.  

The limit $\ecfreslp\to 0$ of the cross section defined in \Eq{eq:margine3} is formally singular at any fixed-order and is only regulated when resummed to all-orders.  In particular, the fixed-order anomalous dimensions present in the factorized form of \Eq{eq:margine3} are singular as $\ecfreslp\to 0$, which makes their interpretation challenging.  However, by reorganizing the functions present in the factorization theorem into in-jet and out-of-jet contributions, all dependence on the observable $\ecfreslp$ can be controlled and finite anomalous dimensions can be identified in the limit $\ecfreslp\to 0$.  Specifically, we rewrite the factorization theorem in the suggestive form:
\begin{align}\label{eq:factrewrite}
\frac{d\sigma\big(\ecfreslp,\outj;R\big)}{d\sje \,d\sjtheta}&=\left[H(Q^2)J_n\big(\ecfreslp\big)J_{\bar n}(\outj) S_{n\bar{n}}(\ecfreslp;\outj;R) \right]\\
&
\times\left[H^{sj}_{n\bar{n}}\left(\sje,\sjtheta\right)  \tilde J_{\sja}\big(\ecfreslp\big) \frac{S^{\text{(in)}}_{n\bar{n}\sja }\big(\ecfreslp;R\big)}{S^{\text{(in)}}_{n\bar{n} }\big(\ecfreslp;R\big)}\right]\left[\frac{S^\text{(out+NG)}_{n\bar{n}\sja }\big(\ecfreslp;\outj;R\big)}{S^\text{(out+NG)}_{n\bar{n}}(\outj;R)}\right] \,,\nonumber 
\end{align}
where we have refactorized the global soft function along the lines of \Refs{Hornig:2011iu,Hornig:2011tg}, separating out the global logarithms of $\ecfreslp$ and $\outj$ that are resummable within the factorization theorem.  All convolutions are implicit.  The (in) and (out) labels denote the in-jet and out-of-jet phase space regions, and (NG) denotes the non-global contributions.  For compactness, we have used the notation of \Eq{eq:ssubjetfact}, where $\tilde J_{\sja}$ is the unfactorized soft subjet function, which contains both the boundary soft and jet modes of the soft subjet.

Using the renormalization group equation that resums the global logarithms of $\ecfreslp$ and $\outj$, we define the refactorized global soft function via:
\begin{align}\label{eq:soft_fact_ng}
S_{n\bar{n}\sja }\big(\ecfreslp;\outj;R;\mu\big)=S^{\text{(in)}}_{n\bar{n}\sja }\big(\ecfreslp;R;\mu\big)S^\text{(NG)}_{n\bar{n}\sja }\big(\ecfreslp;\outj;R\big)S^\text{(out)}_{n\bar{n}\sja }\big(\outj;R;\mu\big) \,,
\end{align}
where we have included explicit dependence on the renormalization scale $\mu$.  Because $\mu$ corresponds to the scale for resummation of global logarithms, it does not appear in the non-global component of the soft function, $S^\text{(NG)}_{n\bar{n}\sja }$.  We also use the shorthand notation $S^\text{(out+NG)}_{n\bar{n}\sja }$ for the product of the out-of-jet and non-global soft functions.  In \Eq{eq:factrewrite}, we have removed any global contribution to $\outj$ from the initial $n\bar n$ dipole by the appropriate global part of the soft function $S_{n\bar{n}}(\ecfreslp;\outj;R)$.  This soft function has a similar factorization:
\begin{align}
S_{n\bar{n}}(\ecfreslp;\outj;R;\mu)=S^{\text{(in)}}_{n\bar{n}}\big(\ecfreslp;R;\mu\big)S^\text{(NG)}_{n\bar{n}}\big(\ecfreslp;\outj;R\big)S^\text{(out)}_{n\bar{n} }\big(\outj;R;\mu\big) \,,
\end{align}
where we have explicitly included dependence on the renormalization scale $\mu$.  The first factor in \Eq{eq:factrewrite}, $$H(Q^2)J_n\big(\ecfreslp\big)J_{\bar n}(\outj) S_{n\bar{n}}(\ecfreslp;\outj;R) $$ has the important property of itself being renormalization group invariant \cite{Ellis:2010rwa} for arbitrary jet radius $R$, assuming that $R$ is smaller than the angle between the $n$ and $\bar n$ directions.  

This fact has deep consequences.  We now introduce the two functions that define the dressed gluon's factorization theorem:
\begin{align}\label{eq:dressed_gluon_functions}
W_{n\bar{n}}(\sje,\sjtheta;R)&=\lim_{\ecfreslp\rightarrow 0}H^{sj}_{n\bar{n}}\left(\sje,\sjtheta\right)  \tilde J_{\sja}\big(\ecfreslp\big) \frac{S^{\text{(in)}}_{n\bar{n}\sja }\big(\ecfreslp;R\big)}{S^{\text{(in)}}_{n\bar{n} }\big(\ecfreslp;R\big)}\,,\nonumber\\
G_{n\bar{n}\sja}(\outj;R)&=\lim_{\ecfreslp\rightarrow 0}\frac{S^\text{(out+NG)}_{n\bar{n}\sja }\big(\ecfreslp;\outj;R\big)}{S^\text{(out+NG)}_{n\bar{n}}(\outj;R)} \,,
\end{align}
which are the second and third factors, respectively, in square brackets in \Eq{eq:factrewrite}.  By the renormalization group invariance of the total cross section and the first factor of \Eq{eq:factrewrite}, the product $W_{n\bar{n}} G_{n\bar{n}\sja}$ must also be renormalization group invariant.  That is, these functions have the renormalization group equations:
\begin{align}
\mu\frac{d}{d\mu}\ln W_{n\bar{n}}(\sje,\sjtheta;R)&=-\gamma_D\,,\\
\mu\frac{d}{d\mu}\ln G_{n\bar{n}\sja}(\outj;R)&=\gamma_D\,,\nonumber
\end{align}
where $\gamma_D$ is the anomalous dimension, which is given to one-loop in \App{app:Anom_Dim}.  The resummed dressed gluon with one-loop anomalous dimensions is
\begin{align}
W_{n\bar{n}}(\sje,\sjtheta;R;\mu)&G_{n\bar{n}\sja}(\outj;R;\mu)\\
&
\hspace{-1cm}
=W_{n\bar{n}}(\sje,\sjtheta;R;\mu)G_{n\bar{n}\sja}(\outj;R;\mu_i)\left(1-\frac{\tan ^2\frac{\sjtheta}{2}}{\tan ^2\frac{R}{2}}\right)^{\frac{\alpha_s C_A}{\pi}\ln \frac{\mu}{\mu_i}}\,,\nonumber
\end{align}
where the scale at which $G_{n\bar{n}\sja}$ is evaluated has been set to $\mu_i$.  Taking the tree-level matrix-elements then gives:
\begin{align}\label{eq:tree_dressed}
W_{n\bar{n}}(\sje,\sjtheta;R;\mu)G_{n\bar{n}\sja}(\outj;R;\mu)&=\frac{\alpha_s C_F}{4\pi^2\sje}\frac{2}{\sin^2\sjtheta}\left(1-\frac{\tan ^2\frac{\sjtheta}{2}}{\tan ^2\frac{R}{2}}\right)^{\frac{\alpha_s C_A}{\pi}\ln \frac{\mu}{\mu_i}}\,.
\end{align}
Note that the dressed gluon's matrix element vanishes as it approaches the jet boundary, where $\sjtheta\to R$.  Therefore, emissions are suppressed near the jet boundary corresponding to the buffer region identified in Monte Carlo simulations of NGLs \cite{Dasgupta:2002bw}.

\subsection{Calculating with a Dressed Gluon}\label{sec:dressed_calcs}

From the suggestive form of \Eq{eq:factrewrite}, we are able to define a generic procedure for incorporating non-global effects into the resummation of an arbitrary additive observable measured on a jet or other restricted phase space region.\footnote{Additivity of the energy correlation functions and the out-of-jet observable $B$ was necessary for the original form of the factorization theorem and its rewriting in \Eq{eq:factrewrite}.  However, we strongly suspect that the resummation of NGLs for non-additive observables, such as the fractional jet multiplicity \cite{Bertolini:frac} defined with the jets-without-jets algorithm \cite{Bertolini:2013iqa}, can be accomplished by extending the methods discussed here.  We thank Jesse Thaler for discussions on this point.}  As a concrete example, we will use the dressed gluon approximation, with one and two dressed gluons, to include non-global effects in the factorization theorem for the hemisphere mass observables in $e^+e^-$ collisions. This will be sufficient to clearly illustrate how the procedure can be extended to an arbitrary number of dressed gluons, and for an arbitrary additive observable.  We will denote the mass of the left (right) hemisphere as $m_L$ ($m_R$), and consider the cumulative cross section defined as
\begin{equation}
S(m_L,m_R)=\frac{1}{\sigma_0}\int\limits_0^{m_L} dm_L' \int\limits_0^{m_R}dm_R'\,\frac{d^2 \sigma}{dm_L'\, dm_R'}\,,
\end{equation}
where $\sigma_0$ is the Born cross section and $$\frac{d^2 \sigma}{dm_L'\, dm_R'}$$ is the double differential cross section of the hemisphere masses.
Non-global effects only appear in the soft function, so, in this section, we will demonstrate how to include the one- and two-dressed gluon approximations into the soft function for hemisphere mass measurements, with the generalization to other non-global measurements being straightforward.

We emphasize that the use of the dressed gluon will depend on the relation between $m_L$ and $m_R$. It is therefore important to clarify the notation used in describing the soft subjet factorization theorem compared with that used for the left-right hemisphere mass distribution. When we presented the soft subjet factorization theorem, we referred to an in-jet region, in which we were multi-differential, and an out-of-jet region, in which a single observable $B$ was measured. Importantly, as discussed in \Sec{sec:modes}, the factorization theorem presented is valid in the case that the out-of-jet scale is lower than the in-jet scale.\footnote{A factorization theorem for the opposite case will be presented in a future publication \cite{us:inclusive}.} To make the correspondence with the left-right hemisphere mass distribution, we must make an assumption for the relation between $m_L$ and $m_R$.  For the case of the hemisphere masses considered here, this is somewhat of a trivial point since the distinction is just a matter of relabeling; however, for general geometries it is important. In this section, we will therefore write explicit expressions for the case $m_L>m_R$, and simply indicate with $R\leftrightarrow L$ the opposite case. For $m_L> m_R$, the measurement $m_R$ corresponds to the out-of-jet measurement $B$ in the soft subjet factorization theorem, as it sets the lower scale. The multi-differential measurement is then made in the left hemisphere, which in the dressed gluon approximation, will give rise to a dressed gluon which will be integrated over the phase space in the left hemisphere. We will use somewhat interchangeably the notation left (right) and in (out) depending on the context, with the hope that this does not cause confusion.

\subsubsection{A Single Dressed Gluon}\label{sec:onedress}

We begin with the simplest case of one-dressed gluon. It is important to note that one cannot just simply replace the standard fixed-order calculation for the dressed gluon, since the presence of the resummation of the dressed gluon would change the renormalization of the global divergences.  These we do not want to change, since in the hemisphere mass factorization theorem these divergences are tied to the resummation of the large global logarithms. The domain of consistency of our soft subjet factorization automatically imposes this constraint: we must only dress the gluon when it is energetic enough. Therefore, we only dress the gluon in the left hemisphere if $m_L > m_R$; otherwise, we dress the gluon in the right hemisphere.

At one-loop, the cumulative soft function is therefore dressed as:
\begin{align}
S^{(1)}_{n\bar{n}}(m_L,m_R)\big|_\text{dressed}&=2C_Fg^2\mu^{2\epsilon}\int[d^dp]_+\frac{n\cdot\bar{n}}{(n\cdot p)(p\cdot\bar{n})}\Theta(m_L-n\cdot p)\Theta(\bar{n}\cdot p-n\cdot p)\nonumber\\
&\qquad\quad\times\left\{\Theta(m_R-n\cdot p)+\Theta(n\cdot p-m_R)\left(1-\frac{n\cdot p}{\bar{n}\cdot p}\right)^{\frac{\alpha_s C_A}{\pi}\ln \frac{n\cdot p}{m_R}}\right\}\nonumber\\
&\quad+\{R\leftrightarrow L, n\leftrightarrow \bar{n}\}\,,
\end{align}
where $d=4-2\epsilon$, $\mu$ is the scale in dimensional regularization, and $[d^dp]_+$ is the Lorentz-invariant phase space for an on-shell, positive energy gluon.  To connect with the expression as written in \Eq{eq:tree_dressed}, note that 
\begin{equation}
\frac{n\cdot p}{\bar n \cdot p} = \tan^2\frac{\sjtheta}{2} \,.
\end{equation}
  The $\Theta$-functions enforce the phase space constraints and, in particular, only turn the dressing factor on if the mass in a hemisphere is not set by the identified gluon emission in that hemisphere.  The purely global contributions to the mass can be separated out by simple rearrangement:
\begin{align}\label{eq:reorgdressed}
\Theta(m_R-n\cdot p)&+\Theta(n\cdot p-m_R)\left(1-\frac{n\cdot p}{\bar{n}\cdot p}\right)^{\frac{\alpha_s C_A}{\pi}\ln \frac{n\cdot p}{m_R}}\nonumber \\
&\qquad\qquad=1+\Theta(n\cdot p-m_R)\left[\left(1-\frac{n\cdot p}{\bar{n}\cdot p}\right)^{\frac{\alpha_s C_A}{\pi}\ln \frac{n\cdot p}{m_R}}-1\right]\,.
\end{align}
The global contribution, corresponding to the ``1'' in \Eq{eq:reorgdressed}, has been studied in great detail in the literature, so for our purposes here, we will ignore it.  Further, because there are no divergences associated with the dressed gluon, we can work in strict $d=4$.  Then, the one-dressed gluon contribution is
\begin{align}\label{eq:one_dress_integrand}
S^{(1,\text{NG})}_{n\bar{n}}(m_L,m_R)&=\Theta(m_L-m_R)\frac{\alpha_s C_F}{\pi}\int\limits_{m_R}^{m_L}\frac{d (n\cdot p)}{ n\cdot p}\int\limits_{ n\cdot p}^{\infty}\frac{d(\bar{n}\cdot p)}{\bar{n}\cdot p}\left[\left(1-\frac{n\cdot p}{\bar{n}\cdot p}\right)^{\frac{\alpha_s C_A}{\pi}\ln \frac{n\cdot p}{m_R}}-1\right]\nonumber\\
&
\quad+\{R\leftrightarrow L, n\leftrightarrow\bar{n}\}\,.
\end{align}
The ``NG'' notation in the superscript denotes that we are only considering the non-global contribution to the soft function as captured by the one-dressed gluon.

The integrals can be evaluated and one finds
\begin{align}\label{eq:bms_1glue}
S^{(1,\text{NG})}_{n\bar{n}}(m_L,m_R) &=\Theta(m_L-m_R)\left\{
-\frac{\alpha_s C_F}{\pi} \gamma_E\ln \left(\frac{m_L}{m_R}\right)-\frac{C_F}{C_A}\ln \Gamma\left[1+\frac{\alpha_s C_A}{\pi}\ln \left(\frac{m_L}{m_R}\right)\right]
\right\}
\nonumber\\
  &
\qquad  +\{R\leftrightarrow L\} \,,
\end{align}
where $\gamma_E$ is the Euler-Mascheroni constant and $\Gamma[x]$ is the Euler Gamma function.
Expanding to the first few orders, we find
\begin{align}
S^{(1,\text{NG})}_{n\bar{n}}(m_L,m_R) &=\Theta(m_L-m_R)\left\{-\frac{\pi^2}{12}\frac{C_F}{C_A}L^2+\frac{\zeta(3)}{3}\frac{C_F}{C_A}L^3 + {\cal O}(\alpha_s^4)\right\}\nonumber \\
&\quad \  \ +\{R\leftrightarrow L\} \,,
\end{align}
where $\zeta(3)=1.202...$ is the Riemann $\zeta$-function and we have used
\begin{equation}\label{eq:def_L}
L=\frac{\alpha_s}{\pi} C_A  \ln\left(\frac{m_L}{m_R}\right)\,.
\end{equation}
The ${\alpha_s^2}$ term is correct, while higher order terms are in general not, but this is expected for reasons discussed earlier as the dressed gluon approximation is not an expansion with a fixed logarithmic counting. In particular, the term at ${\alpha_s^3}$ differs by a factor of 2 from the true result in the large-$N_c$ limit, where $C_F \to N_c/2$ \cite{Schwartz:2014wha}. Nevertheless, the two-dressed gluon approximation will fully capture this term, and produce a more accurate approximation for terms at even higher orders. We will shortly discuss in more detail the organization of the perturbative expansion in terms of dressed gluons.  

In this section we have focused on extracting only the NGLs for the hemisphere mass distribution; however, it should be clear from the presentation that the dressed gluon approximation can be used to perform a complete calculation also including global logarithms. In particular, the expansion in dressed gluons reorganizes the perturbative expansion in terms of ordinary gluons in the low scale matrix element.  The global resummation factor $U(\mu_f,\mu_i)$ multiplies the perturbative expansion in dressed gluons. Explicitly, if we have calculated to the $\ell$th loop order, then formally we have
\begin{align}
S_{n\bar{n}}(m_L,m_R;\mu_f) &=U(\mu_f,\mu_i)\sum_{i=0}^{\ell}\left[S_{n\bar{n}}^{(i)}(m_L,m_R;\mu_i)-c_iL^i+\text{DG}_i\right] \,.
\end{align}
We have subtracted all the fixed order NGLs $c_iL^i$ that are included in the $i$ dressed gluons, denoted by DG$_i$. This is apparent from the factor $``1"$ appearing in \Eq{eq:reorgdressed}, which was ignored in this section, and will be further seen in subtractions necessary to extract the non-global contribution from the two-dressed gluon approximation in \Sec{sec:twodress}.

\subsubsection{Generalization to $N$ Eikonal Lines}\label{sec:ndressed}

In this section we generalize the construction of the dressed gluon approximation from the emission of a soft subjet from two eikonal lines in the $n,\bar n$ directions, to the case of the emission of a soft subjet from $N$ eikonal lines. This construction is necessary to go beyond the one-dressed gluon approximation. Since the leading NGLs arise in the strongly-ordered limit, we are interested in studying multiple strongly-ordered soft subjets, which will become the multiple dressed gluons. In the strongly-ordered limit, the factorization theorem can be obtained by performing a sequence of matchings, where at each stage, all more energetic subjets can be treated as eikonal lines, and all less energetic subjets are unresolved. This is shown schematically in \Fig{fig:doublesoft} for the case of two strongly-ordered subjets, which will be discussed in detail in \Sec{sec:twodress} where we consider the calculation of the two-dressed gluon contribution to the NGLs for the hemisphere jet mass. Because we can perform this sequence of matchings, to generalize the dressed gluon approximation it suffices to understand how to add a soft subjet to $N$ eikonal lines.

When we considered the factorization for a single soft subjet in \Sec{sec:Fact}, we emphasized that the soft subjet region of phase space can be isolated in an IRC safe manner by a multi-differential measurement of the energy correlation functions. This construction can be generalized to isolating $m$ soft subjets by measuring the energy correlation functions $\ecf{2}{\alpha},\cdots, \ecf{m+2}{\alpha}$. The explicit condition on the phase space in terms of the energy correlation functions is not of particular interest, but has been discussed for isolating three prong structure in \cite{Larkoski:2014zma}. We therefore simply assume that sufficiently many IRC safe measurements have been made to isolate the desired region of phase space.

We now consider the addition of a soft subjet to $N$ eikonal lines. We first discuss the organization of color before describing the anomalous dimension of the generalized dressed gluon. All color matrices encoding the color entanglement of the eikonal lines resides in the hard matching coefficient describing the soft subjet production from the initial $N$ eikonal lines.  To understand this organization of color, we start with an $N$ (sub)jet factorization theorem in the region of phase space that has been isolated by a sufficiently differential measurement. This defines the initial hard matching coefficient $\mathbf{H}_N$ and soft function $\mathbf{S}_N$. 

\begin{figure}
\begin{center}
\includegraphics[scale=.35]{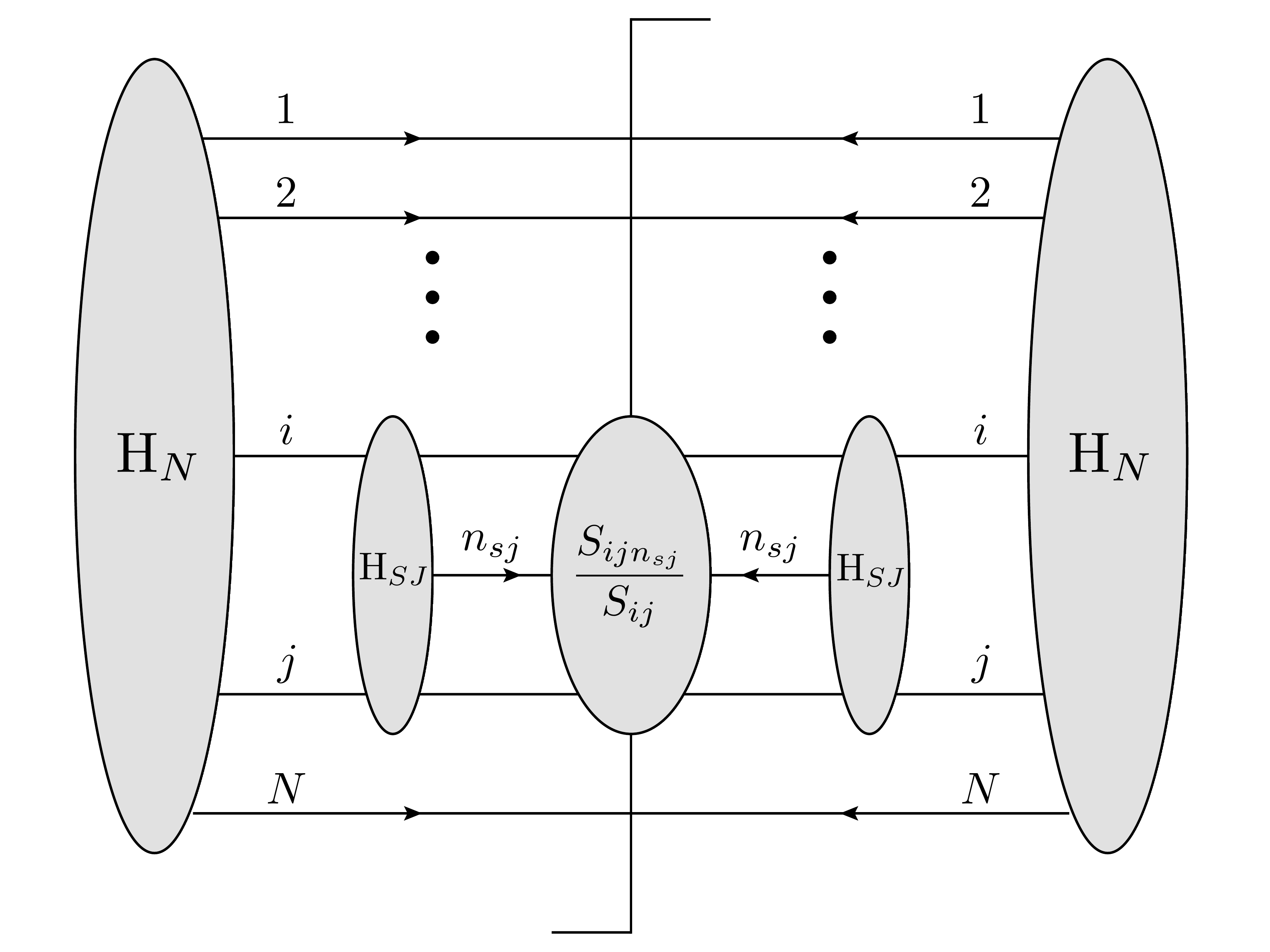}
\caption{The diagrammatic structure of the cut diagrams of the soft jet production in the factorization equation \eqref{eq:N-eikonal_lines_soft}. For concreteness, we have taken the soft jet to be created off of the $i, j$ eikonal lines of the parent $N$-jet factorization. The color matrices of each function are to be inserted at the indicated regions along the eikonal lines. Thus a $\mathbf{T}$ matrix of the soft jet production is inserted between the hard function $H_N$ and any global soft radiation. Note that the new soft jet eikonal line enters only into the color multipole function of the $i, j$ lines.}
\label{fig:schmatic_sj_production}
\end{center}
\end{figure}

To understand how to add the soft subjet to this factorization theorem, we consider the diagrams that can contribute to the matching coefficient describing the soft subjet production off of the $N$ eikonal lines. Only diagrams which are fully color connected\footnote{Fully connected soft diagrams generalize the notion of webs in the two-eikonal line soft function. Webs can always be given a topological condition on the diagram of two-eikonal particle irreducible. These are the diagrams that appear naturally in the logarithm of the soft function.} that contain the real soft gluon that will form the soft subjet will contribute \cite{Gardi:2010rn, Gardi:2011wa, Gardi:2013ita}. The disconnected diagrams will not be color entangled with the soft subjet evolution, and are reproduced by a soft function containing only the original $N$ eikonal lines with the appropriate measurement constraints. Thus we are lead to conjecture the following factorization theorem that describes the addition of a soft subjet:
\begin{center}
{\bf Soft Subjet Factorization Conjecture}
\end{center}
\begin{equation}\label{eq:N-eikonal_lines_soft}
\hspace{-.26cm}
\boxed{
\hspace{-0.1cm}
\begin{aligned}
\frac{d\sigma(\outj_N)}{d\sje \,d\sjOmega \,d\ecfresgen}&=\hspace{-0.5cm} \sum_{\{i,...,k\}\subset \{1,...,N\}}   \hspace{-0.5cm}  \text{tr}\Big[\mathbf{H}_N\cdot \mathbf{H}_{sj}^{(i...k)AB}
 \tr_{sj}\big[\mathbf{S}_{i...k\sja}^{AB}(\ecfresgen;\outj_N )
 \otimes\mathbf{S}_{i...k}^{-1}(\ecfresgen;\outj_N )\big] \\
 &\hspace{5cm}
\otimes \mathbf{S}_{N}(\ecfresgen;\outj_N)\Big]\otimes \tilde J_{sj}\otimes_{\ell=1}^N J_\ell\,.
\end{aligned}
\hspace{-0.1cm}
}
\end{equation}
Here $\ecfresgen$ is the observable that sets the unresolved infrared scale and the $\outj_N$ are all the possible out-of-jet measurements.  Explicit dependence on the jet radius and the arguments of the jet functions have been suppressed. The $\mathbf{S}_{i...k}$ denotes a soft function with eikonal lines $i...k$. Note that for compactness, we have written the soft subjet jet function following the notation of \Eq{eq:ssubjetfact}, and have not explicitly written it as factorized into jet and boundary soft modes, although this factorization must be performed for a completely factorized description of the soft subjet dynamics. The notation $\tr_{sj}$ denotes that we are to trace over all the color indicies of the eikonal lines $i,...,k$. The adjoint indices $AB$ are tied to the soft subjet's Wilson line, and contracts with the $\mathbf{T}$ color matrices of the matching coefficient $\mathbf{H}_{sj\,AB}^{(i...k)}$.  The form of the factorization at the level of cut diagrams is given in \Fig{fig:schmatic_sj_production}.

The key feature of \Eq{eq:N-eikonal_lines_soft} is the appearance of a new soft function, which we refer to as the {\it color multipole function}:
\begin{align}
\text{tr}_{sj}\Big[\mathbf{S}_{i...k\sja}^{AB}(\ecfresgen;\outj_N )\otimes\mathbf{S}_{i...k}^{-1}(\ecfresgen;\outj_N)\Big]\,,
\end{align}
which encodes that the soft subjet modifies the color structure of only the eikonal lines that participated in its production. The subtraction removes overlap of the soft subjet soft function with the original $N$ eikonal line soft function. Effectively, the eikonal line introduced by the soft subjet is only sensitive to the color multipole involved in its creation. The soft subjet production matching coefficients are determined by the finite part of:
\begin{align}\label{eq:soft_jet_hard_matching}
\sum_{\{i,...,k\}\subset \{1,...,N\}}\mathbf{H}_{sj}^{(i...k)AB}&=\langle 0|T\{\mathbf{S}_{1}...\mathbf{S}_{N}\} |sj^{A}\rangle\langle sj^{B}|\bar{T}\{\mathbf{S}_{1}...\mathbf{S}_{N}\}|0\rangle\Big|_\text{finite}\,,
\end{align}
This matrix element selects all connected diagrams with a single soft jet state crossing the cut, while all disconnected diagrams cancel between the time-ordered and anti-time ordered products.  Thus it is directly related to the logarithm of the diagrammatic expansion of a soft function.  In \App{app:twosubjet}, we will use this conjecture for the factorization theorem to explicitly write the factorization theorem for two soft subjet production.

There is substantial evidence for this factorization theorem.  First, it preserves the renormalization group structure of the parent factorization theorem: there is a precise cancellation in the anomalous dimensions of the $N$-jet hard function and the $N$-jet soft function. Were we to introduce a soft eikonal line into the global soft function, this would violate the cancellation. In effect, we would not be able to factorize the soft jet production matching coefficient from the underlying hard process. The underlying $N$-hard jets have been fixed by the measurements imposed, and cannot be modified by subsequent infrared evolution. Second, the factorization formula is consistent with the known expression for the matrix elements for two soft partons, as given in \Ref{Catani:1999ss}, which will be discussed in detail below. Third, the generation of the hard matching coefficient for soft jet production can be considered as a differential operator acting on the soft function of parent eikonal lines that generate the soft subjet.  Since the soft function is known to exponentiate only a strict subset of the diagrams involved in its calculation, this naturally leads to the hard matching given in \Eq{eq:soft_jet_hard_matching}. 

From this factorization theorem, we can then define generalized dressed gluon factors ${\bf W}$ and $G$ as
\begin{align}\label{eq:genWG}
{\bf W}^{(i...k)}(\sje,\sjOmega)&=\lim_{\ecfresgenLP\rightarrow 0}\mathbf{H}_{sj}^{(i...k)AB}(\sje,\sjOmega)\,\tr_{sj}\big[\mathbf{S}_{i...k\sja}^{AB}(\ecfresgenLP)\otimes\mathbf{S}_{i...k}^{-1}(\ecfresgenLP)\big]^{\text{(in)}}
 \tilde J_{sj}(\ecfresgenLP,R)\,,\nonumber\\
G_{i...k}(B_N)&=\lim_{\ecfresgenLP\rightarrow 0}
\tr_{sj}\big[\mathbf{S}_{i...k\sja}(\ecfresgenLP;\outj_N )\cdot\mathbf{S}_{i...k}^{-1}(\ecfresgenLP;\outj_N )\big]^{\text{(out+NG)}} \,.
\end{align}
As with the one-dressed gluon, we Laplace transform $\ecfresgen$ so that the limit $\ecfresgenLP\rightarrow 0$ of its Laplace conjugate results in being fully inclusive over $\ecfresgen$.  These ${\bf W}$ functions depend on the global color structure of the $N$ eikonal lines in the system, but importantly is independent of the out-of-jet scales $B_N$.  Similar to the factorization of the global soft function in \Eq{eq:soft_fact_ng}, in-jet and out-of-jet scales are split into different soft functions in the refactorization of the color multipole function.  The $G$ function depends on the directions of the lines participating in the soft subjet product.  As was the case with the factorization theorem for one soft subjet, for hemisphere jets, the product of each ${\bf W}$ and $G$ are renormalization group invariant.  The structure of ${\bf W}$ can be expressed in a color multipole expansion as
\begin{equation}
{\bf W}^{(i...k)}(\sje,\sjOmega) = {\bf T}^i\cdots {\bf T}^k\, W_{i...k}(\sje,\sjOmega) \,,
\end{equation}
where $W_{ij}(\sje,\sjOmega)$ describes the emission of the dressed gluon from a color dipole, $W_{ijk}(\sje,\sjOmega)$ describes the connected emission of the dressed gluon from three Wilson lines, etc.

A possible modification to the conjecture is that soft subjet soft functions do not themselves form a color singlet as postulated above in the formulation of the color multipole function, so that
\begin{align}
 \mathbf{H}_{sj}^{(i...k)AB}\cdot\mathbf{S}_{i...k\sja}^{AB}(\ecfresgen;\outj_N )\otimes\mathbf{S}_{i...k}^{-1}(\ecfresgen;\outj_N )\,
\end{align}
is a generic color tensor over the color indices of the eikonal lines $i...k$. This color tensor must then be inserted between the contractions of the soft and hard functions $\mathbf{H}_N$ and $\mathbf{S}_N$, following the $\mathbf{T}$ matrix conventions of \Ref{Catani:1999ss} (since effectively it can be expressed as a sum over such color matrices). For soft functions with 2 or 3 Wilson lines, as considered in this paper, the function can always be written as a color singlet, consistent with the conjectured form discussed above.

\subsubsection{Anomalous Dimension for Dressed Gluons with $N$ Eikonal Lines}\label{sec:anom_dressed}

To dress the resolved gluon requires renormalization of the generalized ${\bf W}$ and $G$ functions, as defined in \Eq{eq:genWG}. At one-loop, the anomalous dimension of these objects can be determined by the one and two soft gluon emission matrix elements from an arbitrary $N$-point squared amplitude. This can be determined directly from the factorization equation \eqref{eq:N-eikonal_lines_soft}, or from \Ref{Catani:1999ss}.\footnote{These give equivalent results: the first term of \Eq{eq:two_real_soft_gluons} arises from the global soft function, while the second is generated by the color multipole function.} In the notation of \Ref{Catani:1999ss},  for soft gluons with momenta $q_1,q_2$ these squared amplitudes are  
\begin{align}
\label{eq:one_real_soft_gluons}\left|{\cal A}(q_1,p_1^{a_1},...,p_N^{a_N})\right|^2&=\left(4\pi\alpha_s\mu^{2\epsilon}\right)\sum_{i,j=1}^N S_{ij}(q_1)|{\cal A}^{(ij)}(p_1^{a_1},...,p_N^{a_N})|^2 \,,\\
\label{eq:two_real_soft_gluons}
\hspace{-0.3cm}
\left|{\cal A}(q_1,q_2,p_1^{a_1},...,p_N^{a_N})\right|^2&=\left(4\pi\alpha_s\mu^{2\epsilon}\right)^2\left\{\frac{1}{2}\sum_{i,j=1}^N\sum_{k,l=1}^NS_{ij}(q_1)S_{kl}(q_2)\left|{\cal A}^{(ij)(kl)}(p_1^{a_1},...,p_N^{a_N})\right|^2\right.\nonumber\\
&
\hspace{2.8cm}
\left.-C_A\sum_{i,j=1}^NS_{ij}(q_1,q_2)\left|{\cal A}^{(ij)}(p_1^{a_1},...,p_N^{a_N})\right|^2\right\}\,,
\end{align}
respectively, where the indices in parentheses denote the dipole from which the soft gluon has been emitted and $p_i$ and $a_i$ are the momenta and color of particle $i$.  While the factorization above is valid for arbitrary soft gluon emissions, in the strongly-ordered limit, the soft gluon emission factors are
\begin{align}
S_{ij}(q)&=\frac{p_i\cdot p_j}{(p_i\cdot q)(q\cdot p_j)}\,,\\
S_{ij}^{\text{(s-o)}}(q_1,q_2)&=\frac{p_i\cdot p_j}{(p_i\cdot q_1)(q_1\cdot q_2)(q_2\cdot p_j)}+\frac{p_i\cdot p_j}{(p_i\cdot q_2)(q_2\cdot q_1)(q_1\cdot p_j)}\nonumber \\
&
\hspace{1.5cm}
-\frac{p_i\cdot p_j}{(p_i\cdot q_1)(q_1\cdot p_j)}\frac{p_i\cdot p_j}{(p_i\cdot q_2)(q_2\cdot p_j)}\,,\nonumber
\end{align}
where (s-o) denotes strongly-ordered.

To continue, we assume that at least one of the eikonal lines lives in the jet whose substructure we wish to probe.  Then, \Eq{eq:one_real_soft_gluons} has the interpretation as the tree-level hard matching coefficient for the soft subjet while \Eq{eq:two_real_soft_gluons} describes the real emission contribution from the soft subjet.  This emission can be either in the jet, where it contributes at the resolution scale of the jet $\ecfresgen$, or it is outside the jet, where it contributes to the out-of-jet measurements $B_N$.  The soft real emission in the jet at the resolution scale could be either global soft, boundary soft, or radiation from other soft modes formed by having multiple eikonal lines in the jet.  To determine the anomalous dimension of the ${\bf W}$ and $G$ functions, and therefore to dress the gluon, we are only interested in emissions from the soft subjet that leave the jet.

The anomalous dimensions of ${\bf W}$ and $G$ are determined solely by the term in \Eq{eq:two_real_soft_gluons} that is explicitly proportional to $C_A$.  Contributions from this term can entangle in- and out-of-jet scales, corresponding to sensitivity to non-global structure.  By contrast, the emissions from first term of \Eq{eq:two_real_soft_gluons} are color disconnected and therefore can only contribute to the global structure of either the in- or out-of-jet regions.  These observations allow us to rewrite the $C_A$ term for emission of a strongly-ordered soft gluon with momentum $q$ from the soft subjet in the direction $n_{sj}$ as
\begin{align}
&-C_A\sum_{i,j=1}^NS^{\text{(s-o)}}_{ij}(\sja,q)\left|{\cal A}^{(ij)}(p_1^{a_1},...,p_N^{a_N})\right|^2=\\
&\sum_{i,j=1}^N\left[S_{ij}(\sja)\left|{\cal A}^{(ij)}(p_1^{a_1},...,p_N^{a_N})\right|^2\right] 
  C_A\left(\frac{p_i\cdot p_j}{(p_i\cdot q)(q\cdot p_j)}-\frac{\sja\cdot p_j}{(\sja\cdot q)(q\cdot p_j)}  -\frac{p_i\cdot \sja}{(p_i\cdot q)(q\cdot \sja)}\right) \nonumber
\end{align}
We immediately recognize the factor in square brackets as the hard function for the soft gluon created from the eikonal line $i,j$.  Therefore, to one-loop, the anomalous dimension of the generalized dressed gluon is
\begin{align}\label{eq:gen_dress}
\gamma_{ij\sja}^D&=-\frac{\alpha_s C_A}{\pi}\int\limits_\text{out}\frac{d\Omega_q}{4\pi}\frac{S_{ij}^{\text{(s-o)}}(\sja,q)}{S_{ij}(\sja)} \nonumber \\
&=\frac{\alpha_s C_A}{\pi}\int\limits_\text{out}\frac{d\Omega_q}{4\pi}\left[\frac{p_i\cdot p_j}{(p_i\cdot q)(q\cdot p_j)}-\frac{\sja\cdot p_j}{(\sja\cdot q)(q\cdot p_j)}-\frac{p_i\cdot \sja}{(p_i\cdot q)(q\cdot \sja)}\right]\,,
\end{align}
where $q=(1,\hat{q})$ and $\hat{q}$ is a unit vector in the direction of the emission from the soft subjet and ``out'' means that the angular integral is only evaluated in the out-of-jet region.

The coefficient of the dressed gluon anomalous dimension is given the by the gluon color Casimir, $C_A$, multiplied by the $\mathcal{O}(\alpha_s)$ cusp anomalous dimension \cite{Polyakov:1980ca,Brandt:1981kf,Korchemsky:1987wg}, $\Gamma^0_{\text{cusp}}(\alpha_s)=\alpha_s/\pi$. We conjecture that to all orders in perturbation theory the coefficient of the dressed gluon anomalous dimension is given by $C_A\, \Gamma_{\text{cusp}}$. While this may seem unmotivated from the above calculation, it appears naturally from the factorization theorem for the soft subjet in \Eq{fact_inclusive_form_1}. Recall that the dynamics of the soft subjet which are sensitive to the non-global structure is described by the boundary soft mode.  Because it is soft, the boundary soft mode sees the soft subjet as eikonalized and because it is collinear-soft, only sees the eikonal likes $n_{sj}$ and $\bar n_{sj}$. Thus it is natural that the anomalous dimension associated with the dressed gluon should be given by the cusp anomalous dimension.\footnote{See also \Refs{Ivanov:1985np,Korchemsky:1988si,Korchemsky:1991zp,Korchemsky:1992xv} for similar arguments for the appearance of the cusp anomalous dimension in other contexts.}

For hemisphere jets, the renormalization group evolution of the $\mathbf{W}$ and $G$ functions is straightforward to determine.  Recall that we can expressed the $\mathbf{W}$ function in a basis of color multipoles as  
\begin{equation}
\mathbf{W}^{(i...k)}(\sje,\sjOmega)=\mathbf{T}^i\cdots \mathbf{T}^k \,W_{i...k}(\sje,\sjOmega)+\dotsc\,\\
\end{equation}
where the $\dotsc$ denotes higher multipoles.  The tree-level matrix element for the dipole factor is
\begin{equation}
W_{ij}^\text{(tree)}(\sje,\sjOmega) = \frac{\alpha_s}{4\pi^2\sje}\frac{p_i\cdot p_j}{(p_i\cdot \sja)(\sja\cdot p_j)} \,,
\end{equation}
where $\sje$ is the energy fraction of the soft subjet and $\sja$ is a lightlike vector along the direction of the soft subjet.  The renormalization group evolution of $\mathbf{W}^{(ij)}$ is  
\begin{equation}
\mu\frac{d}{d\mu}\mathbf{W}^{(ij)}(\sje,\sjOmega)= -\gamma_{ij\sja}^D\,\mathbf{W}^{(ij)}(\sje,\sjOmega)\,,
\end{equation}
which, to one-loop, the anomalous dimension is given by \Eq{eq:gen_dress}.  At one- and two-loop order, only dipoles contribute to the evolution of the soft subjet \cite{Catani:1999ss,Becher:2009qa,Gardi:2009qi,Dixon:2009ur}, and so only the evolution of $\mathbf{W}^{(ij)}(\sje,\sjOmega)$ is necessary to resum logarithms associated with dressing a gluon through next-to-leading logarithmic accuracy.  

For reasons which will become clear when we discuss the relation of the dressed gluon to the BMS equation, we will use the following notation for the exponentiated anomalous dimension with fixed coupling
\begin{equation}\label{eq:exp_dressed}
U_{ij n_{sj}}(L)=\exp \left[ L \frac{\gamma_{ij\sja}^D \pi}{\alpha_s C_A}  \right]\,,
\end{equation}
where the normalization is chosen to correspond to the definition of the logarithm appearing in \Eq{eq:def_L}.

\subsubsection{Two Dressed Gluons}\label{sec:twodress}

We now continue and present the two-dressed gluon approximation as applied to the hemisphere jet mass.  This discussion will be limited to leading logarithmic accuracy for the two-dressed gluon, but the extension to higher logarithmic orders can be accomplished straightforwardly by calculating the objects in the appropriate factorization theorem to higher perturbative orders.  Unlike the one-dressed gluon, for which we presented explicit calculations, in this section we will only present the form of the integrand for the non-global component of the soft function for hemisphere jet masses which comes from the two-dressed gluon approximation.  Numerical comparison of the two-dressed gluon to the BMS equation will be presented in \Sec{sec:nglcomp}.

\begin{figure}
\begin{center}
\subfloat[]{\label{fig:twosoft}
\includegraphics[scale = .2225]{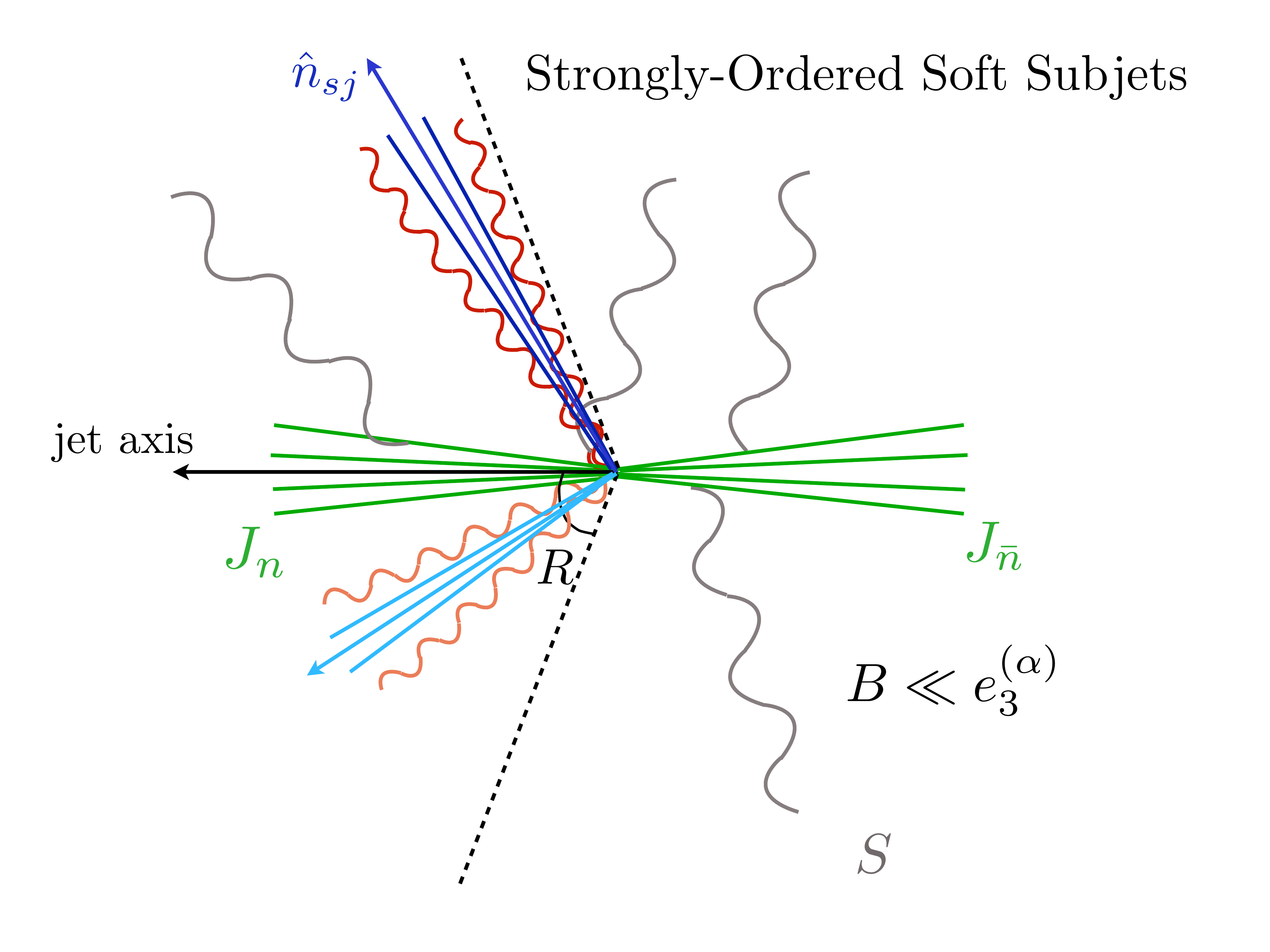}
}
$\qquad$
\subfloat[]{\label{fig:twosoft_scales}
\includegraphics[scale = 0.225]{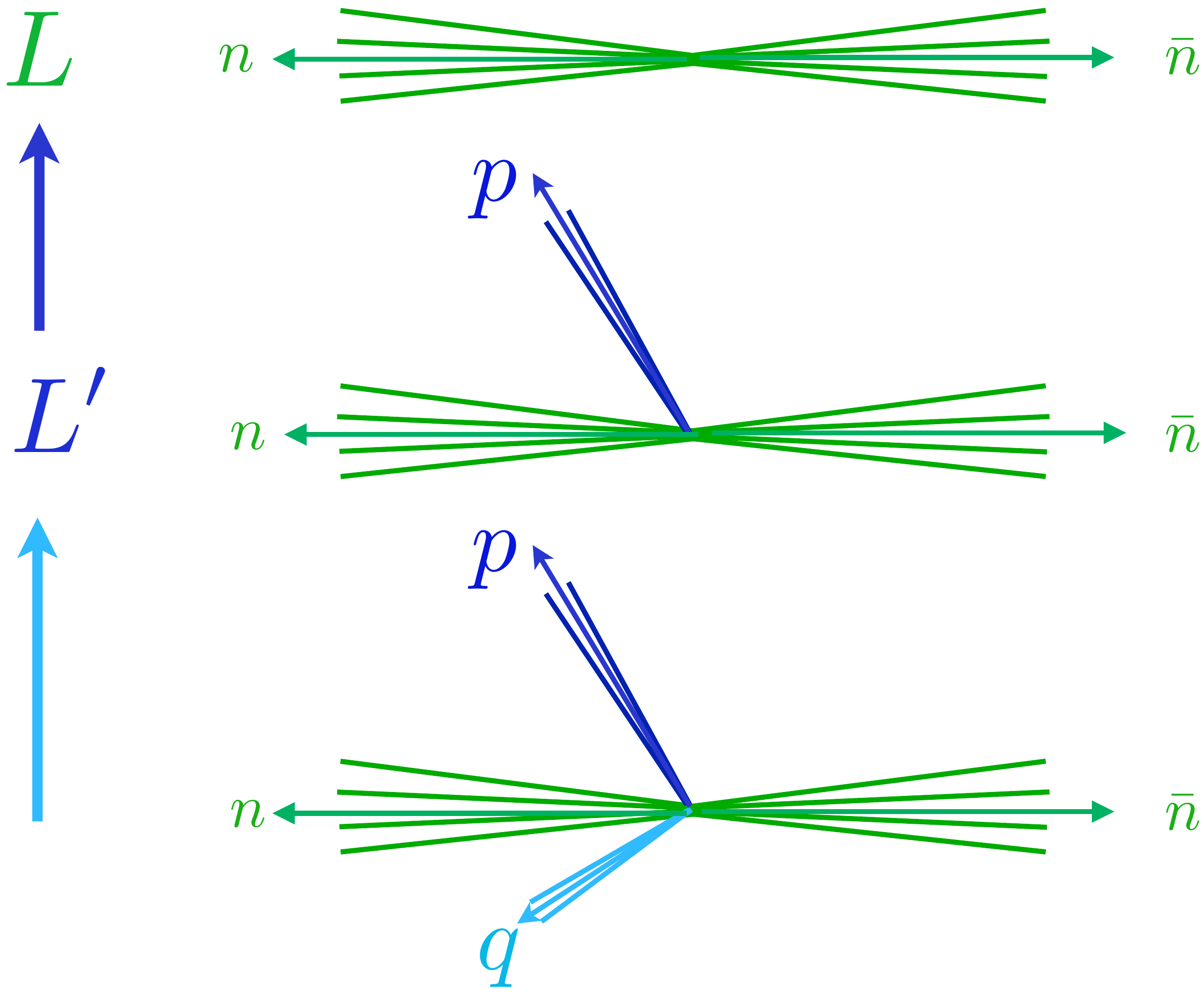}
}
\end{center}
\caption{ (a) Schematic depiction of the region of phase space defined by two strongly-ordered soft subjets, which gives rise to the two-dressed gluon approximation. (b) Illustration of the resolved subjets as a function of the resolution scale, as implemented by the matching procedure in this region of phase space.
}
\label{fig:doublesoft}
\end{figure}

As with the identification of the one soft subjet region of phase space, we perform a series of measurements on the jet to identify two soft subjets and a hard core of radiation.  As discussed in \Ref{Larkoski:2014zma}, this three-prong region of phase space can be isolated by measuring $\ecf{2}{\alpha}$, $\ecf{3}{\alpha}$, and $\ecf{4}{\alpha}$ on the jet and demanding parametric relations between them.  As the precise relationships are not vital to the two-dressed gluon calculation, we do not present them.  To ensure that the soft subjets are well-separated, we can measure an additional three-point energy correlation function, $e_3^{(\beta)}$, and demand that $\ecf{3}{\alpha}\sim \ecf{3}{\beta}$. These measurements fully isolate the two soft subjet region, and further measurements can be performed to determine the subjet energy fractions and relative angles.  The two soft subjet region of the cross section can then be factorized into appropriate hard, jet and soft functions describing the various modes that contribute to the various observables. As in the case of a single soft subjet,  we must assume that we are in a region of phase space where $B\ll \ecf{3}{\alpha}$, so that the NGLs are parametrically large. The precise structure of this factorization theorem, while potentially interesting for particular applications, is not relevant for the dressed gluon approximation and will not be presented here.  As with the one-dressed gluon, to obtain the description of the two-dressed gluon from the factorization theorem, we associate functions in the factorization theorem and integrate over unresolved scales.  \Fig{fig:twosoft} illustrates the two soft subjet region of phase space, where we assume that the energy of the subjets is strongly-ordered.  

Once we have isolated two strongly-ordered soft subjets, we can then connect to the two-dressed gluon approximation by a sequence of matchings between different effective theories.  This is illustrated in \Fig{fig:twosoft_scales}, where we assume the subjets have momenta $p$ and $q$ with $p\gg q$, and the procedure is similar in spirit to that proposed for describing the parton shower in SCET \cite{Bauer:2006mk,Bauer:2006qp,Baumgart:2010qf}.  At the lowest scale, the softest subjet with momenta $q$ is produced from the $np$ or $\bar n p$ dipoles.  This emission is dressed by the anomalous dimension of the dressed gluon, \Eq{eq:gen_dress}, integrated up to the scale of the harder subjet, $L'$ (where the strongly-ordered limit breaks down).  Once integrated to the scale of the harder subjet with momentum $p$, the procedure is repeated, with this subjet dressed by the anomalous dimension of the dressed gluon, and integrated up to the scale $L$, corresponding to the highest soft subjet scale.

To illustrate this procedure, we will first apply it to the one-dressed gluon, and show that it reproduces the calculation from \Sec{sec:onedress}.  Using the exponentiated anomalous dimension from \Eq{eq:exp_dressed} for dressing the gluon, the soft function for a  dressed gluon in the left hemisphere is
\begin{align}
S^{(1)}_{n\bar{n}}(L)\big|_\text{dressed}=\int_{-\infty}^L dL' \int\limits_{\text{left}} \frac{d\Omega_p}{4\pi} \left\{
\Theta(-L')\, H_{n\bar n}(p) + \Theta(L')\, H_{n\bar n}(p) U_{n\bar n p}(L')
\right\}\,,
\end{align}
where, for example, $\Theta(L')$ is satisfied only if the energy of the gluon is sufficiently large.  Here, $H_{n\bar n}(p)$ is the eikonal matrix element for the emission of a soft gluon from Wilson lines in the $n$ and $\bar n$ directions.  The gluon with momentum $p$ is only dressed, that is, multiplied by the exponentiated dressing anomalous dimension $U_{n\bar n p}(L')$, if its energy is sufficiently large.  Re-associating the $\Theta$-functions, we find
\begin{align}\label{eq:catani_onedress}
S^{(1)}_{n\bar{n}}(L)\big|_\text{dressed}=\int_{-\infty}^L dL' \int\limits_{\text{left}} \frac{d\Omega_p}{4\pi} \left\{
 H_{n\bar n}(p) + \Theta(L')\, H_{n\bar n}(p) \left(U_{n\bar n p}(L')-1\right)
\right\}\,.
\end{align}
We now immediately recognize the first term in \Eq{eq:catani_onedress} as the global hemisphere soft function, and therefore, the second term is the non-global soft function for the one-dressed gluon:
\begin{equation}\label{eq:catani_1ngl}
S^{(1,\text{NG})}_{n\bar{n}}(L)=\int_0^L dL' \int\limits_{\text{left}} \frac{d\Omega_p}{4\pi}\,  H_{n\bar n}(p) \left(U_{n\bar n p}(L')-1\right)\,.
\end{equation}
One can verify that \Eq{eq:catani_1ngl} agrees with the analytic expression for the non-global soft function for the one-dressed gluon in \Eq{eq:bms_1glue}.

Now, we apply the same procedure to determine the two-dressed gluon contribution, in the strongly-ordered and large $N_c$ limit.  The two-dressed gluon soft function is
\begin{align}
\hspace{-.1cm}
S^{(2)}_{n\bar{n}}(L)&= \int\limits_{-\infty}^{L} dL'  \int\limits_{-\infty}^{L'} dL''\int\limits_{\text{left}} \frac{d\Omega_q}{4\pi}\int\limits_{\text{left}} \frac{d\Omega_p}{4\pi}\, \left\{ 
\Theta(-L')\Theta(-L'')H_{n\bar n}(p)\left[
H_{np}(q)+H_{\bar n p}(q)
\right]
\right.\nonumber
\\
&
\hspace{-1.2cm}
+ \Theta(L')\Theta(-L'')\, H_{n\bar n}(p)U_{n\bar n p}(L')\left[
H_{np}(q)+H_{\bar n p}(q)
\right]\nonumber \\
&
\hspace{-1.2cm}
\left.+\,\Theta(L')\Theta(L'') \, H_{n\bar n}(p)U_{n\bar n p}(L')\left[
H_{np}(q)U_{n p q}(L'')+H_{\bar n p}(q)U_{\bar n pq}(L'')
\right]
\right\}  \,,
\end{align}
where we assume that $p\gg q$.  Again, to isolate the non-global contribution, we re-associate the $\Theta$-functions which yields
\begin{align}\label{eq:catani_2dress}
\hspace{-.1cm}
S^{(2)}_{n\bar{n}}(L)&= \int\limits_{-\infty}^{L} dL'  \int\limits_{-\infty}^{L'} dL''\int\limits_{\text{left}} \frac{d\Omega_q}{4\pi}\int\limits_{\text{left}} \frac{d\Omega_p}{4\pi}\, \left\{ 
H_{n\bar n}(p)\left[
H_{np}(q)+H_{\bar n p}(q)
\right]
\right.
\\
&
\hspace{-1.5cm}
+ \Theta(L')\left[\, H_{n\bar n}(p)\left(U_{n\bar n p}(L')-1\right)\left[
H_{np}(q)+H_{\bar n p}(q)
\right]
+\Theta(L'')H_{n\bar n}(p)H_{n\bar n}(q)\left(
U_{n\bar n q}(L'')-1
\right)
\right]
\nonumber \\
&
\hspace{-1.5cm}
+\Theta(L')\Theta(L'')\left[  H_{n\bar n}(p)U_{n\bar n p}(L')\left[
H_{np}(q)\left(U_{n p q}(L'')-1\right)+H_{\bar n p}(q)\left(U_{\bar n pq}(L'')-1\right)
\right]\right. \nonumber \\
&
\hspace{7.3cm}
\left.\left.
-H_{n\bar n}(p)H_{n\bar n}(q)\left(
U_{n\bar n q}(L'')-1
\right)
\right]
\right\} \nonumber \,.
\end{align}
The strongly-ordered global hemisphere soft function for two gluons is immediately identified as the first line of \Eq{eq:catani_2dress}.  The second line is the strongly-ordered non-global soft function for the one-dressed gluon, with either gluon $p$ or gluon $q$ dressed.  Because of the subtractions in this term, when dressed gluon $p$ is collinear with $n$, its contribution vanishes, and similar for the contribution when dressed gluon $q$ is collinear to $n$.  The last line in \Eq{eq:catani_2dress} is the strongly-ordered non-global soft function for the two-dressed gluon:
\begin{align}
\label{eq:catani_2dress_ngl}
&S^{(2,\text{NG})}_{n\bar{n}}(L)= \int\limits_0^{L} dL'  \int\limits_0^{L'} dL''\int\limits_{\text{left}} \frac{d\Omega_q}{4\pi}\int\limits_{\text{left}} \frac{d\Omega_p}{4\pi}\\
&
\qquad
\times \left\{  H_{n\bar n}(p)U_{n\bar n p}(L')\left[
H_{np}(q)\left(U_{n p q}(L'')-1\right)+H_{\bar n p}(q)\left(U_{\bar n pq}(L'')-1\right)
\right]\right. \nonumber \\
&
\hspace{9cm}
\left.
-H_{n\bar n}(p)H_{n\bar n}(q)\left(
U_{n\bar n q}(L'')-1
\right)
\right\}  \,.\nonumber
\end{align}
The subtractions that appear in the expression above from re-associating the phase space constraints from dressing are necessary from the effective theory perspective to force the non-global soft function to be restricted to the two-dressed gluon phase space region.  These remove appropriate collinear limits of the two resolved gluons in the soft function, acting as zero bin subtractions in the effective theory \cite{Manohar:2006nz}.

The form of the one- and two-dressed gluon calculations suggest that the all-orders non-global soft function calculated in terms of dressed gluons takes a form motivated by non-abelian exponentiation of the soft function \cite{Gatheral:1983cz,Frenkel:1984pz,Gardi:2010rn}.  We conjecture that the full non-global soft function can be written schematically as 
\begin{equation}
S^{(\text{NG})}_{n\bar{n}}(L) = 
1+\sum_{i=1}^\infty  \tilde C_i \,S^{(i,\text{NG})}_{n\bar{n}}(L)\,,
\end{equation}
where $\tilde C_i$ represents an appropriate color factor. A detailed study of this conjecture is, however, beyond the scope of this paper.

\subsection{Numerical Comparison to the BMS Equation}\label{sec:nglcomp}

In this section we compare our dressed gluon approximations with different calculations of the leading, large-$N_c$ NGLs for the hemisphere jet mass.  We will compare to both fixed-order and resummed distributions for the NGLs.  In the large-$N_c$ limit, the leading fixed-order NGLs were calculated in \Ref{Schwartz:2014wha} to 5 loops by explicit iteration of the BMS equation.  The expansion is
\begin{equation}\label{eq:5loop}
S^{(\text{NG})}_{n\bar n}=1-\frac{\pi^2}{24}L^2+\frac{\zeta(3)}{12}L^3+\frac{\pi^4}{34560}L^4+\left( -\frac{\pi^2\zeta(3)}{360}+\frac{17\zeta(5)}{480} \right )L^5 + {\cal O}(L^6)\,,
\end{equation}
where we recall that $$L=\frac{\alpha_s}{\pi}N_c \ln\left(\frac{m_L}{m_R}\right)\,.$$ Leading logarithmic resummation of NGLs in the large-$N_c$ limit can be achieved by a numerical or Monte Carlo solution to the BMS equation.  Dasgupta and Salam (DS) \cite{Dasgupta:2001sh} proposed the following fit to their Monte Carlo result:
\begin{equation}
S^{(\text{NG})}_{n\bar n}=\exp \left [ -\frac{\pi^2}{24} L^2\frac{1+0.180625L^2}{1+0.325472L^{1.33}}  \right]\,.
\end{equation}
We have written our own implementation of the DS Monte Carlo algorithm and will include it in the comparisons.

 \begin{figure}
\begin{center}
\subfloat[]{\includegraphics[width=6.9cm]{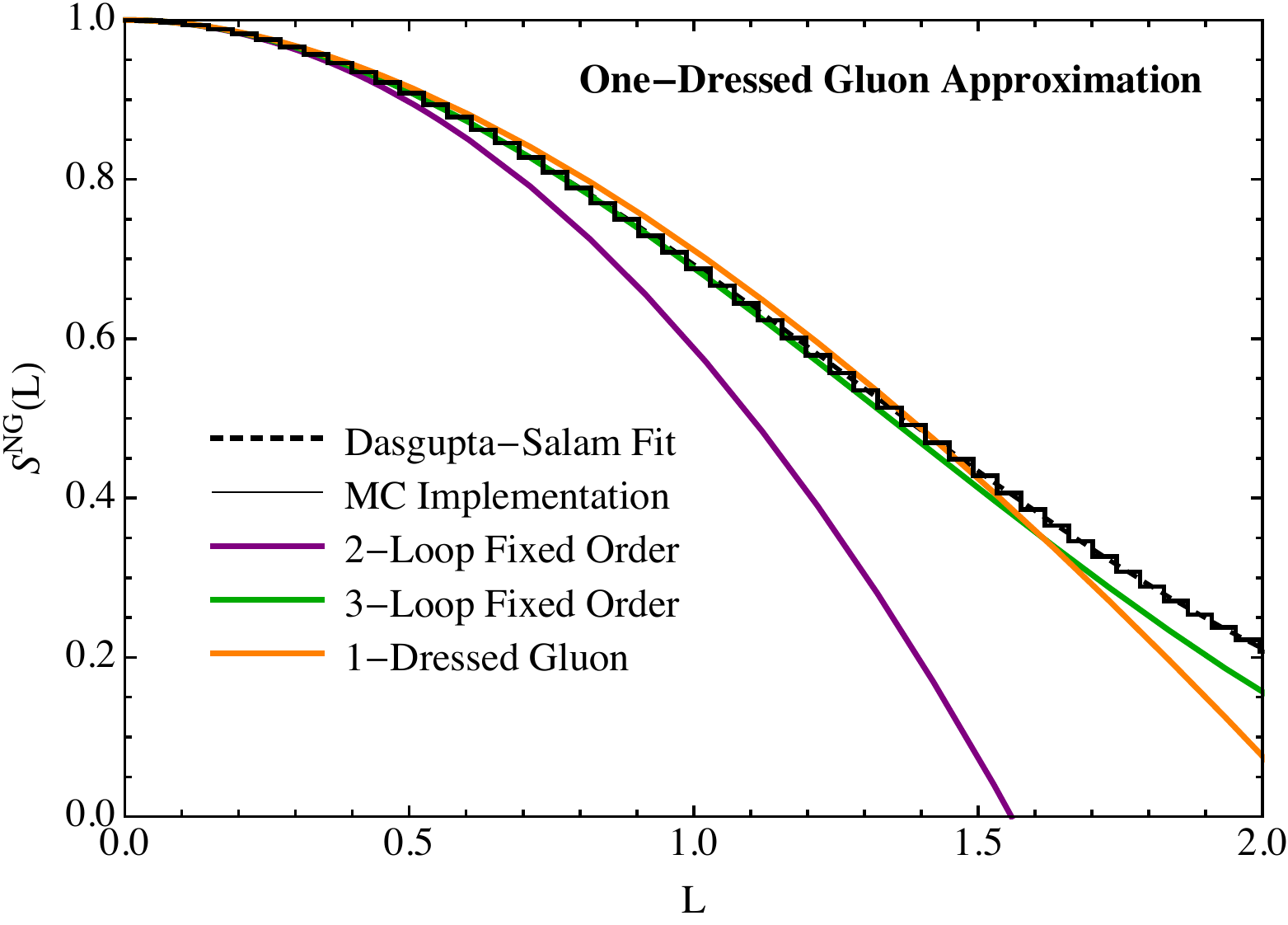} 
}\qquad
\subfloat[]{\includegraphics[width=7cm]{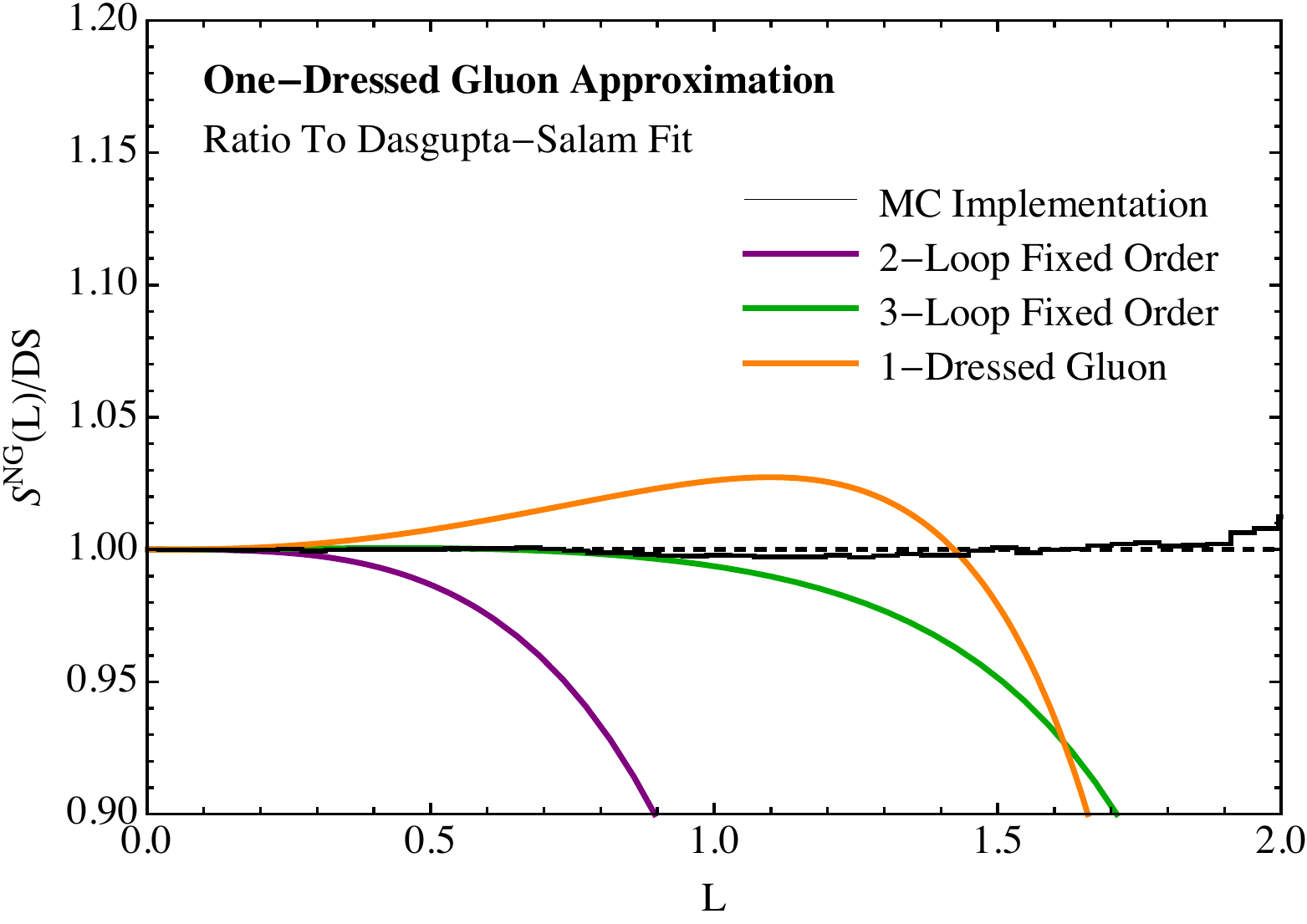} 
}
\end{center}
\caption{Comparison of the one-dressed gluon approximation to the 2- and 3-loop fixed order results for the non-global piece of the hemisphere mass soft function, the Dasgupta-Salam fit for the leading logarithmic resummation of the NGLs from a Monte Carlo, and our implementation of the DS Monte Carlo.
}
\label{fig:single_dressed}
\end{figure}

We begin in \Fig{fig:single_dressed} where we compare the one-dressed gluon approximation to the fixed-order and resummed results listed above.  In the comparison, we include the 2- and 3-loop fixed-order expansions, the DS fit, and our Monte Carlo output.  As expected, because it reproduces the full 2-loop result, the one-dressed gluon agrees with the 2-loop expansion at small values of $L$.  As discussed earlier, the one-dressed gluon does not include the full 3-loop result and so the 3-loop expansion is a better approximation of the full resummed result out to about $L=1$.  Nevertheless, the one-dressed gluon is accurate to better than 5\% of the resummed distribution out to $L=1.5$, corresponding to a ratio of several hundred between the hemisphere jet masses.  A distinction between the resummed distribution and the one-dressed gluon is that the one-dressed gluon diverges at sufficiently large $L$.  This behavior will be present at any fixed order in the dressed gluon expansion, and therefore does not produce the physical large $L$ distribution.  However, such large $L$ values where the dressed gluon diverges are well beyond the range of phenomenological applications.
\Fig{fig:single_dressed} also shows the output of our Monte Carlo, which agrees to within 1\% with the DS fit out to $L=2$.

\begin{figure}
\begin{center}
\subfloat[]{\includegraphics[width=7cm]{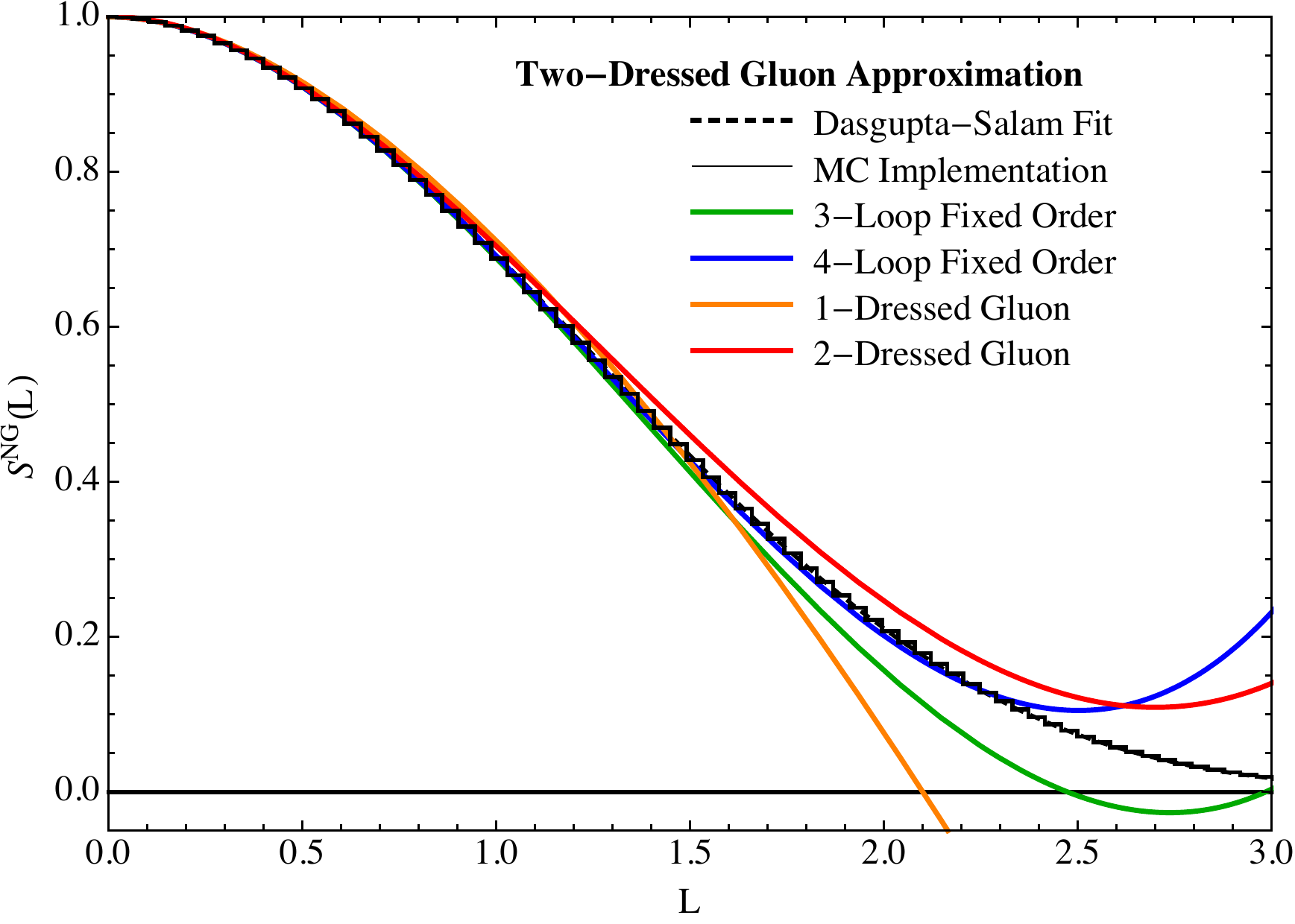} 
}\qquad
\subfloat[]{\includegraphics[width=7cm]{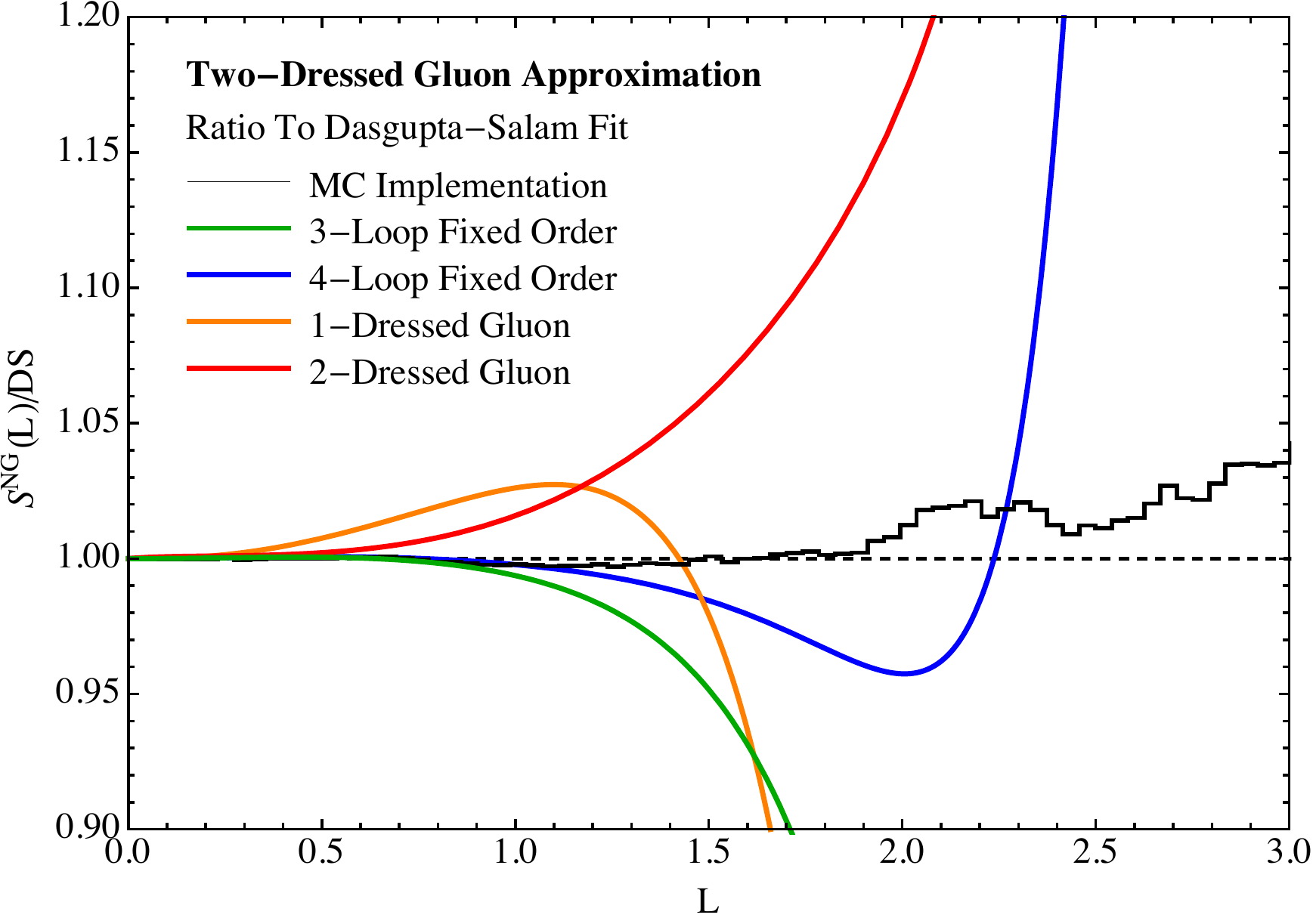} 
}
\end{center}
\caption{Comparison of the two-dressed gluon approximation to the 3- and 4-loop fixed order results for the non-global piece of the hemisphere mass soft function, the Dasgupta-Salam fit for the leading logarithmic resummation of the NGLs, and our implementation of the DS Monte Carlo.  
}
\label{fig:double_dressed}
\end{figure}

Improved accuracy can be obtained by including the two-dressed gluon approximation which we show in \Fig{fig:double_dressed}.  As with the one-dressed gluon, we compare with the DS fit and our implementation of the Monte Carlo, but now we compare with the 3- and 4-loop fixed-order expansions.   The inclusion of the two-dressed gluon approximation fully accounts for the 3-loop fixed order result, and so nicely agrees out to about $L=1$.  Perhaps more interesting is that the two-dressed gluon approximation agrees with the Monte Carlo to better than 5\% out to $L=1.5$, and slowly diverges from the DS fit beyond there.  This slow divergence at large $L$ values, especially as compared to fixed-order expansions, may suggest that the dressed gluon approximation is a convergent expansion.  We also compare to the output of our Monte Carlo, which begins to diverge from the DS fit near $L=2.5$, where finite cutoff effects or finite statistics are important.

In addition to the numerical comparisons presented in \Figs{fig:single_dressed}{fig:double_dressed}, we could directly compare the 5-loop result in \Eq{eq:5loop} to the fixed-order expansion of the dressed gluons.  However, such a comparison would potentially be misleading and obscure many important features for the following reasons.\footnote{Nevertheless, for completeness we will do this comparison later.}  As mentioned in \Ref{Schwartz:2014wha}, the fixed order expansion of the leading non-global logarithms appears to be an asymptotic series.  This is in contrast to leading global logarithms, which, for observables like the jet mass, can be resummed into an exponential.  In the case of global logarithms, because the series expansion of the exponential function has infinite radius of convergence, comparing to its fixed-order expansion is meaningful.  As one calculates to higher and higher orders, one exactly builds up the exponentiated form for the leading global logarithms.  However, if the fixed-order expansion of non-global logarithms is indeed asymptotic, then this is precisely the wrong way to organize it.  The behavior of the one- and two-dressed gluons, on the other hand, suggests that the dressed gluon expansion is convergent.  If this is the case, then there is no sense in which the fixed-order expansion builds up the dressed gluon approximation.  Additionally, as we will show in \Sec{sec:buffer}, the dressed gluons manifest emergent phenomena of non-global logarithms that are not present at any fixed order.

\begin{figure}
\begin{center}
\subfloat[]{\includegraphics[width=7cm]{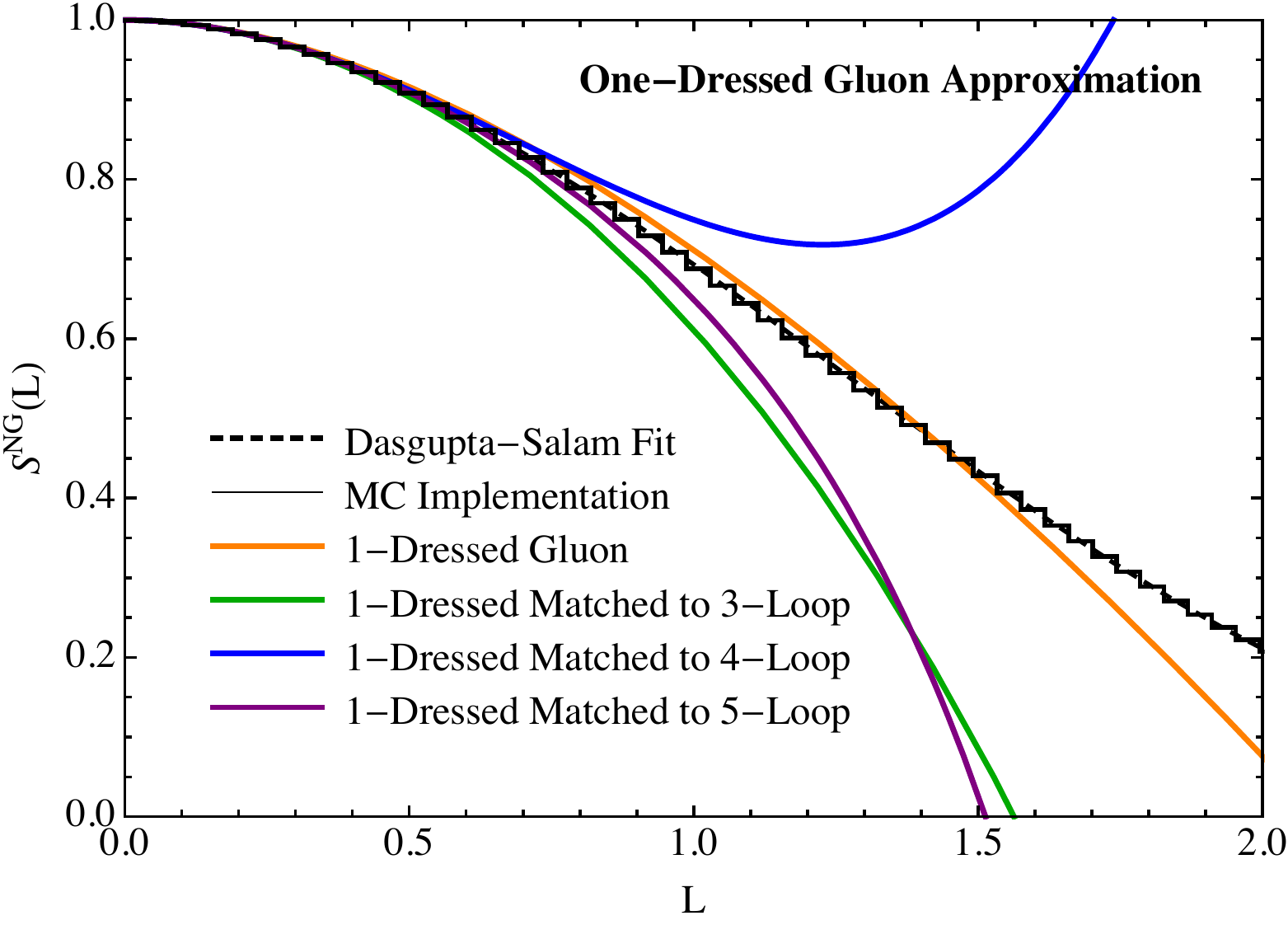} 
}\qquad
\subfloat[]{\includegraphics[width=7cm]{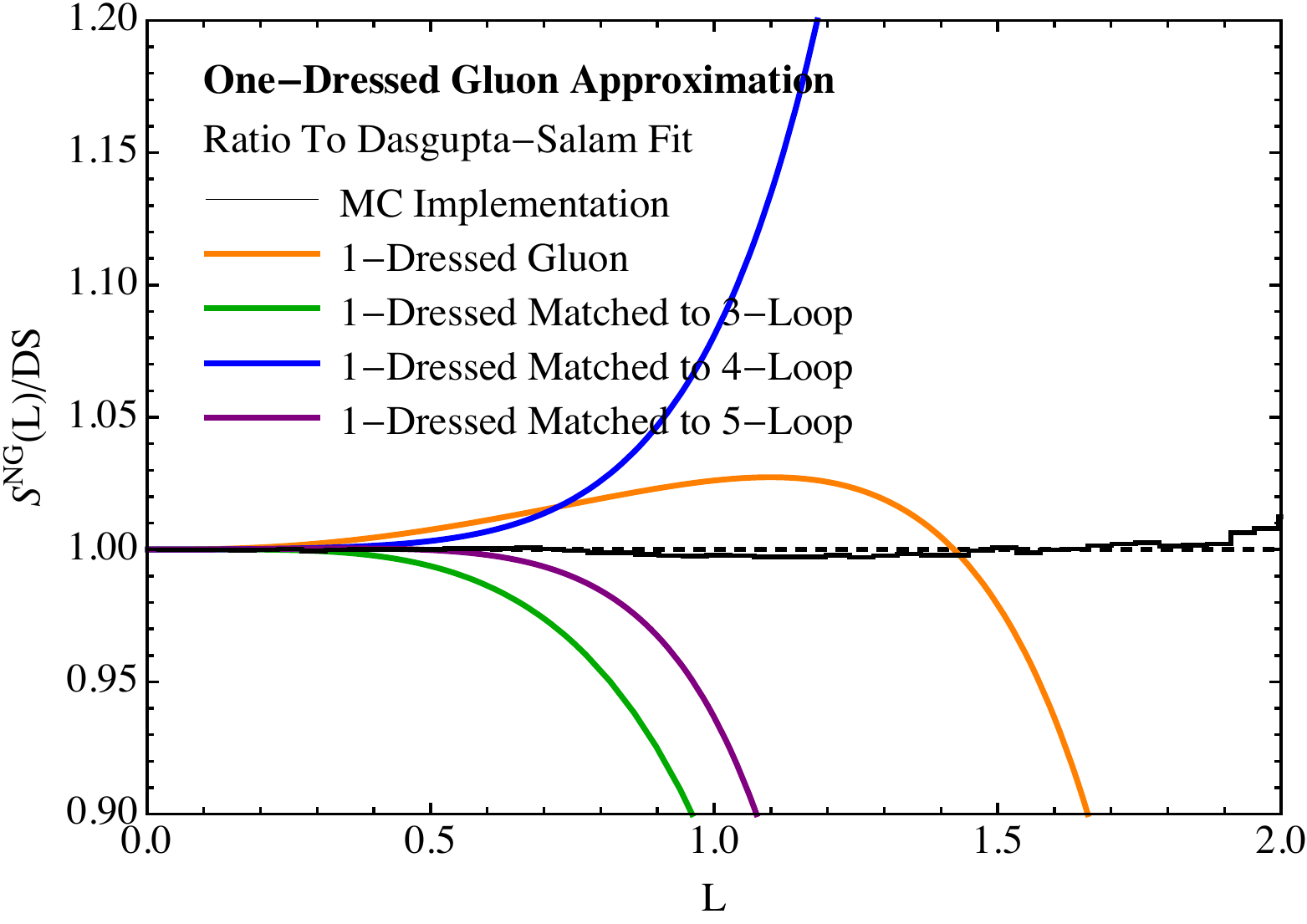} 
}
\end{center}
\caption{Comparison of the one-dressed gluon approximation, matched to the 3-, 4- or 5-loop fixed-order non-global piece of the hemisphere mass soft function, the Dasgupta-Salam fit for the leading logarithmic resummation of the NGLs, and our implementation of the DS Monte Carlo.
}
\label{fig:double_dressed_matched}
\end{figure}

To emphasize this point, in \Fig{fig:double_dressed_matched} we compare the one-dressed gluon approximation to matching the one-dressed gluon to the 3-, 4-, or 5-loop fixed-order NGLs from \Eq{eq:5loop}.  While matching to the fixed-order NGLs does improve the accuracy of the one-dressed gluon at small $L$, it does so at the cost of greatly decreasing the accuracy at higher $L$ values.  Even when matched to the 5-loop fixed-order NGLs, the one-dressed gluon with no matching is more accurate over a wider range of $L$ values.  This is concrete evidence that the fixed-order expansion of the NGLs is not the correct way to organize their expansion.  The dressed gluon resums a highly non-trivial subset of the NGLs to produce an approximation to the full result that is accurate over a wide dynamic range.

\begin{table}
\begin{center}
\begin{tabular}{c| c c c c}
 & $L^1$ & $L^2$ & $L^3$ & $L^4$ \\[0.1cm]
 \hline
 One-Dressed& 0 & $-\frac{\pi^2}{24}$ & $\frac{\zeta(3)}{6}$ & $-\frac{\pi^4}{720}$  \\
 Two-Dressed & 0 & 0 & $-\frac{\zeta(3)}{12}$ & $\frac{\pi^4}{480}(1\pm0.05)$  \\
\hline
 Sum & 0 & $-\frac{\pi^2}{24}$ & $\frac{\zeta(3)}{12}$ & $\frac{\pi^4}{1440}(1\pm0.2)$ \\
Exact & 0 &  $-\frac{\pi^2}{24}$ & $\frac{\zeta(3)}{12}$ & $\frac{\pi^4}{34560}$
\end{tabular}
\end{center}
\caption{
\label{tab:fotodg}
Coefficients of the NGLs as calculated from the one- and two-dressed gluon approximations through $L^4$.  The sum of the one- and two-dressed gluon approximation is compared to the exact fixed order result from \Eq{eq:5loop}.
}
\end{table}

For completeness, in \Tab{tab:fotodg}, we compare the numerical coefficients of the NGLs as found from the dressed gluon approximation to the exact fixed-order results, through $L^4$.  For both the one-dressed gluon and the fixed-order results, the coefficients are known analytically, while for the two-dressed gluon, we have determined the coefficients numerically by Monte Carlo integration of \Eq{eq:catani_2dress_ngl}.  Correspondingly, the uncertainty in the exact value of the $L^4$ coefficient for the two-dressed gluon is included in the table.  As discussed earlier, the one-dressed gluon gets the $L^2$ term correct exactly, but not higher order terms in the expansion.  When the two-dressed gluon contribution is included, the dressed gluon approximation exactly reproduces the correct coefficient at order $L^3$.  The coefficient at order $L^4$ is numerically small, which is manifest as a large cancellation between the one- and two-dressed gluons.  However, to fully reproduce this term requires the three-dressed gluon, which we do not compute here.  Nevertheless, the absolute value of the three-dressed gluon contribution must be significantly smaller than that from the one- and two-dressed gluons at this order.  Again, the fixed-order expansion may be asymptotic and so this comparison should be taken cautiously, but because corrections due to higher-order dressed gluons to the fixed-order NGLs seem to decrease in magnitude, this is further evidence that the dressed gluon expansion converges.

\subsection{Insights into features of NGLs and the BMS Equation}\label{sec:insights}

In addition to the accurate description of NGLs, the dressed gluon approximation also provides a physical picture for the effects of a phase space boundary.  In this section we discuss some features of NGLs and methods of expansion of the BMS equation, for which the dressed gluon approximation provides insight.  From these examples, we are able to more precisely define the formal expansion of the dressed gluon approximation as an expansion in the unresolved phase space volume, which we discuss in \Sec{sec:pert_vs_dressed}.

\subsubsection{Buffer Region}\label{sec:buffer}

In \Ref{Dasgupta:2002bw} a scenario for the possible underlying dynamics for NGLs was proposed, which identified a ``buffer region'' near the phase space boundaries where emissions are suppressed.  Consider some phase space region $\Omega$ in which an energy veto is applied.  \Ref{Dasgupta:2002bw} proposed that the mechanism for suppressing radiation emitted into $\Omega$ was due to a buffer region around the boundary of $\Omega$ which itself contained little radiation.  A particularly interesting consequence of this proposal is the approximate geometry independence of the NGLs, as the buffer region smoothes the detailed shape of the boundary of $\Omega$. 

In \Ref{Dasgupta:2002bw} an evolution equation for the width of the buffer region as a function of the in-$\Omega$ and out-of-$\Omega$ scales was proposed based on some simple assumptions.  The solution they found was
\begin{align}\label{eq:buffer}
\eta_{\text{buffer}}\simeq (L-L')  \left\langle \frac{\delta \eta}{\delta L}\right\rangle\,,
\end{align}
where $\eta_\text{buffer}$ is the width of the buffer region in pseudorapidity, $L$ and $L'$ are logarithms of two scales in $\Omega$, and $\big\langle \frac{\delta \eta}{\delta L}\big\rangle$ acts as an average speed of the evolution of the border of the buffer region in $\eta$, assumed to be independent of $L$ and postulated to be proportional to $C_A$. A Monte Carlo study was performed, which provided some qualitative support for the buffer mechanism; nevertheless, the exact linear relation of \Eq{eq:buffer} was not observed.  However, it was not clear if this was due to $L$ dependence of $\big\langle \frac{\delta \eta}{\delta L}\big\rangle$ for some unknown dynamics or simply because the system had not reached its asymptotic behavior.

\begin{figure}
\begin{center}
\subfloat[]{\label{fig:buffer_pic}\includegraphics[width=7cm]{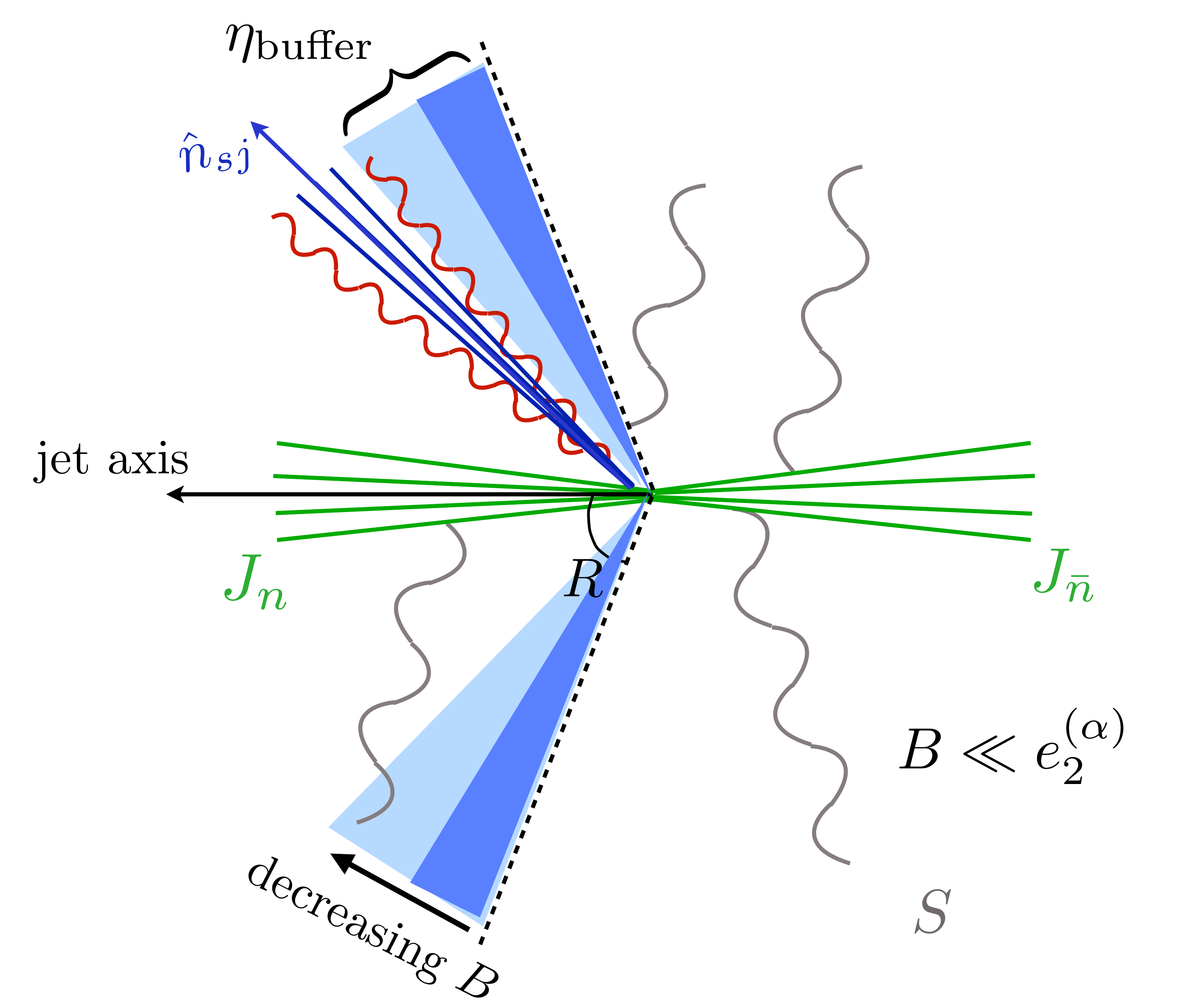} 
}
\qquad
\subfloat[]{\label{fig:buffer_plot}\includegraphics[width=7cm]{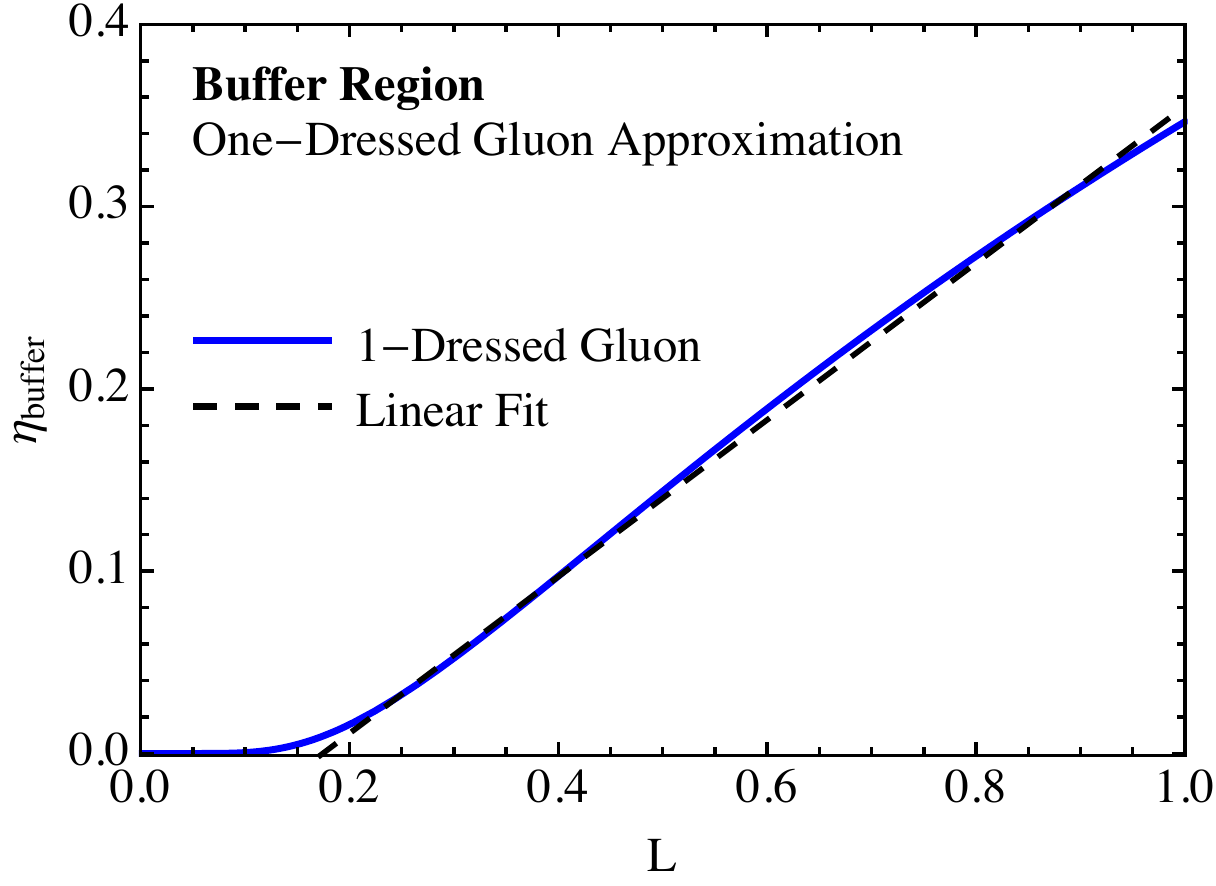} 
}
\end{center}
\caption{(a) A schematic depiction of the buffer region of \Ref{Dasgupta:2002bw} which arises analytically in the dressed gluon approximation. As the out-of-jet scale is lowered (or correspondingly the in-jet scale is raised), the buffer region grows.  (b) The width of the buffer region $\eta_\text{buffer}$ in pseudorapidity as a function of $L$, which exhibits linear growth. The width of the buffer region is only plotted for values of $L$ for which the one-dressed gluon approximation provides a valid approximation to the NGL dynamics.  
}
\label{fig:buffer}
\end{figure}

The existence and properties of such a buffer region can be addressed by studying the soft jet region of phase space, as it describes the dynamics of a single gluon which itself radiates more gluons. The boundary soft function, its anomalous dimension, and its manifestation in the dressed gluon approximation provide strong support for the buffer mechanism. The buffer region is manifest in the one-dressed gluon approximation through the tree-level matrix-element
\begin{align}\label{eq:buff}
W_{n\bar{n}}(\sje,\sjtheta;R;\mu)G_{n\bar{n}\sja}(\outj;R;\mu)&=\frac{\alpha_s C_F}{4\pi^2\sje}\frac{2}{\sin^2\sjtheta}\left(1-\frac{\tan ^2\frac{\sjtheta}{2}}{\tan ^2\frac{R}{2}}\right)^{\frac{\alpha_s C_A}{\pi}\ln \frac{\mu}{\mu_i}}\,,
\end{align}
which vanishes as the dressed gluon approaches the boundary of the jet.  A schematic depiction of the buffer region for our geometrical setup is shown in \Fig{fig:buffer_pic}. 

As a proxy for the width of the buffer region as defined by the one-dressed gluon approximation, we consider the half maximum of the suppression factor in \Eq{eq:buff}.  That is, we define the width of the buffer region $\eta_\text{buffer}$ for hemisphere jet masses via
\begin{equation}
\frac{1}{2}=\left(
1-e^{-2\eta_\text{buffer}}
\right)^L\,.
\end{equation}
The width of the buffer region $\eta_\text{buffer}$ is plotted in \Fig{fig:buffer_plot} for the range of $L$ over which the one-dressed gluon approximation is valid. It appears to rapidly asymptote to linear growth as a function of $L$, which continues throughout the range of validity of the one-dressed gluon approximation. While these values of $L$ are smaller than those considered in \Ref{Dasgupta:2002bw}, and so it is possible that different dynamics are involved, we believe that the dressed gluon approximation provides support for the buffer region.  At higher values of $L$, additional dressed gluons are required to accurately describe NGLs, which is beyond the scope of this paper.

\subsubsection{Expansions of the BMS Equation} \label{sec:expand_BMS}

In this section we relate our dressed gluon approximation with the BMS equation \cite{Banfi:2002hw}, and comment on various expansions proposed in the literature. The BMS equation for the purely non-global piece of the hemisphere mass distribution, $g_{ab}(L)$, is given by 
\begin{equation}\label{eq:BMS_eq}
\partial_L g_{ab}(L)=\int\limits_\text{left} \frac{d\Omega_j}{4\pi}  \mathcal{W}_{ab}^j \left[  U_{abj}(L)g_{aj}(L) g_{jb}(L)-g_{ab}(L)  \right]\,,
\end{equation}
with the boundary condition $g_{ab}(0)=1$. For simplicity, in this section we will follow closely the notation in the BMS literature, in particular \Ref{Schwartz:2014wha}. We will use the left/right instead of in/out labeling, and we define 
\begin{equation}
\mathcal{W}_{ab}^j=\frac{1-\cos(\theta_{ab})}{[1-\cos(\theta_{aj})][1-\cos(\theta_{jb})]}=\frac{n_a\cdot n_b}{(n_a\cdot n_j)(n_j\cdot n_b)}\,,
\end{equation}
and
\begin{equation}\label{eq:exp_BMS_anom}
U_{abj}(L)=\exp \left[ L \int\limits_\text{right} \frac{d\Omega_1}{4\pi}(\mathcal{W}_{ab}^1-\mathcal{W}_{aj}^1-\mathcal{W}_{jb}^1)   \right]\,.
\end{equation}
Note that \Eq{eq:exp_BMS_anom} is the exponentiated dressed gluon anomalous dimension from \Eq{eq:exp_dressed}. The evolution equation resums all leading NGLs in the large $N_c$ limit.  As it is a non-linear equation, solving it exactly analytically is challenging, and so several expansions have been proposed to systematically approximate it.

One expansion that we presented in \Eq{eq:5loop} is the fixed-order expansion to 5-loops from \Ref{Schwartz:2014wha}.  While this expansion provides insight at small values of NGLs, it exhibits poor convergence and may only be an asymptotic series for $L>1$.  Another approximation used in the literature \cite{Banfi:2002hw,Schwartz:2014wha} is to rewrite the BMS equation in the form
\begin{equation}\label{eq:trunc_BMS}
\partial_L g_{ab}(L)=\int\limits_\text{left} \frac{d\Omega_j}{4\pi} g_{ab}(L)\mathcal{W}_{ab}^j  \left[  U_{abj}(L)-1  \right]  +\int\limits_\text{left} \frac{d\Omega_j}{4\pi}  \mathcal{W}_{ab}^j  U_{abj}(L)  \left[  g_{aj}(L) g_{jb}(L)-g_{ab}(L)  \right]\,.
\end{equation}
The second term does not contribute to three loops, and the first term gives a linear evolution equation which is straightforward to solve. Specializing to the case where $a,b=n,\bar n$ and performing the integrals, this linear equation becomes
\begin{equation}\label{eq:lin_bms}
\partial_L g_{n\bar n}=-\frac{1}{2} \left ( \gamma_E +\frac{\Gamma'(1+L)}{\Gamma(1+L)}   \right ) g_{n \bar n}\,.
\end{equation}
While the linear equation is easily solved in closed form, it has no particular region of validity and a numerical comparison to the solution to the BMS equation demonstrates that it is a poor approximation above $L\sim 0.5$. 

The dressed gluon approximation, by contrast to either of these other expansions, is not an expansion in the coupling, but rather in the number of resolved soft subjets.  It is therefore interesting to ask how the dressed gluon approximation manifests as an expansion of the BMS equation.  We expand the non-global function $g_{ab}(L)$ as
\begin{equation}\label{eq:dressed_exp}
g_{ab}(L)=
1+\tilde g^{(1)}_{ab}(L)+\tilde g^{(2)}_{ab}(L)+\cdots \,.
\end{equation}
 We will relate $\tilde g^{(1)}_{ab}(L)$ to the one-dressed gluon, $\tilde g^{(2)}_{ab}(L)$ to the two-dressed gluon, and so forth.  

Inserting this expansion into the BMS equation, the differential equation for the one-dressed gluon function $\tilde g^{(1)}_{ab}(L)$ is then
\begin{equation}\label{eq:onedress_bms}
\partial_L \tilde g^{(1)}_{ab}(L)=\int\limits_\text{left} \frac{d\Omega_j}{4\pi} \mathcal{W}_{ab}^j  \left[  U_{abj}(L)-1  \right] =-\frac{1}{2} \left ( \gamma_E +\frac{\Gamma'(1+L)}{\Gamma(1+L)}   \right ) \,,
\end{equation}
which is equivalent to the one-dressed gluon result calculated in \Eq{eq:bms_1glue}.  This truncation is also very similar to the linearized equation, \Eq{eq:lin_bms}.  However, unlike \Eq{eq:lin_bms}, the one-dressed gluon is not itself exponentiated.  It is interesting that this simple approximation to the BMS equation, which is a valid expansion independent of its interpretation in terms of dressed gluons, to our knowledge has not been well-studied in the literature.  The approximation of \Eq{eq:onedress_bms} is also significantly more accurate than the linearized BMS approximation, \Eq{eq:lin_bms}.

While we have shown that this expansion reproduces the one-dressed gluon, it is interesting to compare the interpretation of the subtraction term in the BMS equation and the dressed gluon approximation. Note that the $-1$ subtraction appears explicitly in both the dressed gluon integrand of \Eq{eq:one_dress_integrand}, and in the above expansion of the BMS equation \Eq{eq:onedress_bms}, where it arises from the subtraction term $-g_{ab}(L)$ in the full BMS equation, \Eq{eq:BMS_eq}. In the derivation of the BMS equation presented in \Ref{Banfi:2002hw}, this term was included as a virtual subtraction by unitarity. While this is indeed the unique subtraction which renders the BMS evolution equation infrared finite, it does not necessarily correspond to virtual corrections from our point of view since jets can genuinely be collinear. On the other hand, in the effective field theory approach the origin of this subtraction is clear: it arises from restricting the effective field theory description of the soft subjet region to its regime of validity, similar to a zero bin subtraction \cite{Manohar:2006nz}.\footnote{Indeed, this subtraction can be implemented as a subtraction that removes overlap with the SCET$_+$ factorization of \Ref{Bauer:2011uc} of two collinear subjets (see also forthcoming \cite{usD2}). This is consistent with the fact that to have the correct sum of factorized bare matrix elements, a zero-bin is required \cite{Chiu:2009yx}.} Indeed, in the effective field theory approach, because we have completely factorized the dynamics describing the soft jet, the real and virtual infrared divergences cancel separately within each function in the factorization theorem of \Eq{fact_inclusive_form_1}.

The two-dressed gluon is found by inserting \Eq{eq:dressed_exp} in to the BMS equation and expanding to higher order.  The differential equation for $\tilde g^{(2)}_{ab}(L)$ is
\begin{align}
&\partial_{L'} \tilde g^{(2)}_{ab}(L')=  \int\limits_0^{L'} dL''\int\limits_{\text{left}} \frac{d\Omega_q}{4\pi}\int\limits_{\text{left}} \frac{d\Omega_p}{4\pi}\\
&
\times \left\{  \mathcal{W}_{ab}^p U_{ab p}(L')\left[
\mathcal{W}_{ap}^q\left(U_{a p q}(L'')-1\right)+\mathcal{W}_{b p}^q\left(U_{b pq}(L'')-1\right)
\right]
-\mathcal{W}_{ab}^p\mathcal{W}_{ab}^q\left(
U_{ab q}(L'')-1
\right)
\right\} \,, \nonumber
\end{align}
which upon integrating, is exactly the equation for the two-dressed gluon in \Eq{eq:catani_2dress_ngl} with $a,b=n,\bar n$.  One can work to higher orders in the dressed gluon expansion and build up the full solution to the BMS equation.

\subsubsection{Perturbative Expansion vs.~Dressed Gluon Expansion}\label{sec:pert_vs_dressed}

We conclude this section by discussing the distinction between a fixed-order expansion of the BMS equation and the dressed gluon expansion, defined in \Eq{eq:dressed_exp}. We argue that the dressed gluon approximation is the correct way to organize the perturbative expansion of NGLs, as supported by the convergence and accuracy of the dressed gluon approximation demonstrated in \Sec{sec:nglcomp}.

The dressed gluon approximation is not an expansion in the coupling or traditional logarithmic counting, but in the number of resolved soft subjets in a jet as defined by the soft subjet factorization theorem.  An arbitrary number of unresolved gluons are emitted from the soft subjets so that the dressed gluon approximation includes terms to all orders in the coupling, yet for a fixed number of dressed gluons, it does not fully capture the complete logarithmic series to any formal accuracy using a traditional logarithmic counting. There are always contributions at the same logarithmic accuracy which arise from a higher number of dressed gluons that are ignored.  As discussed in the study of the buffer region, the anomalous dimension of the soft subjet/dressed gluon suppresses emissions near the boundary of the jet, as well as suppressing the region of phase space when multiple resolved soft subjets approach one another.  The interpretation of this is that higher-order dressed gluons have a significantly reduced phase space volume in which they can live.  This has important consequences for how the formal expansion of the dressed gluon should be addressed.

In a fixed logarithmic counting, one implicitly counts the phase space for soft gluons as $\mathcal{O}(1)$, and so it does not affect the parametric scaling of the logarithms or the manner in which they should be counted.\footnote{Historically, the vast majority of global observables that were studied, like thrust \cite{Farhi:1977sg}, $C$-parameter \cite{Parisi:1978eg,Donoghue:1979vi,Ellis:1980wv},  heavy jet mass \cite{Catani:1991bd}, broadening \cite{Rakow:1981qn,Ellis:1986ig,Catani:1992jc}, etc., define scales for which soft emissions are at lower, or at least the same, virtuality as collinear emissions.  Thus the traditional logarithmic counting $\alpha_s \ln \sim 1$ accurately captures the dominant singular structure.  However, there are phenomenologically-relevant examples of observables for which collinear emissions have lower virtuality than soft emissions; for example, recoil-free angularities with angular exponent $\beta < 1$ \cite{Larkoski:2014uqa}.  For these observables the na\"ive logarithmic counting must also be modified, instead taking $(\alpha_s/\beta) \ln \sim 1$, otherwise perturbative predictions will be extended well beyond their range of applicability. } This is appropriate when resolved emissions do not affect the allowed phase space volume for further emissions in the system; however, this is not the case with NGLs, as is nicely illustrated with the one-dressed gluon and its buffer region shown in \Fig{fig:buffer}. To take the phase space suppression into account in the logarithmic counting, we write the expansion in the schematic form
\begin{align} \label{eq:pert_expand}
S^{(\text{NG})}\sim&\ L^2(1-\Delta \eta^{(1)})+
\alpha_s L(1-\Delta \eta^{(1)})
 \nonumber \\
+&\ L^3(1-\Delta \eta^{(2)})+
\alpha_s L^2(1-\Delta \eta^{(2)})+
\alpha_s^2 L(1-\Delta \eta^{(2)})
\nonumber \\
+&\ L^4(1-\Delta \eta^{(3)})+
\alpha_s L^3(1-\Delta \eta^{(3)})+
\alpha_s^2 L^2(1-\Delta \eta^{(3)})+
\alpha_s^3 L(1-\Delta \eta^{(3)})
\nonumber \\
+&\ \cdots\,,
\end{align}
where the $\Delta \eta^{(i)}$ correspond to the phase space suppression from the buffer region with $i$ dressed gluons, with $\Delta \eta^{(i+1)}>\Delta \eta^{(i)}$. The $\Delta \eta^{(i)}$ are treated as negligible in a traditional logarithmic counting, but will be important for NGLs.

For small $L$, the buffer region is small, and therefore the $\Delta \eta^{(i)}$ can be formally neglected.  In this case, higher powers of $L$ are suppressed and we expect that the  resummation of subleading terms, like $\alpha_s L^n$, will be as important as the leading logarithms. For $L\sim1$, the buffer region is also ${\cal O}(1)$, and so the available phase space for resolved gluons is suppressed. With traditional logarithmic counting, $L\sim 1$, and the entire first column of \Eq{eq:pert_expand} must be resummed, as done by the BMS equation.  However, including the power counting of the volume of phase space, higher order logarithmic terms are increasingly suppressed by the allowed phase space, a fact which should be taken into account in the organization of the perturbative expansion.  Consistent incorporation of logarithmic and phase space counting is accomplished by the dressed gluon approximation through the resummation of the unresolved emissions associated with the dressed gluon.  This explains why even the one-dressed gluon approximation exhibits such good convergence even to $L\sim 1$.  As $L$ increases beyond $1$, there is a competition between the higher powers of $L$ and the phase space suppression, and so convergence of the expansion at large $L$ requires including greater numbers of dressed gluons. However, since the total phase space volume is finite and the buffer region becomes large at large $L$, we expect that the expansion in the number of dressed gluons will rapidly converge.

These dynamics are not incorporated in a fixed-order expansion in which there is no distinction between the counting of resolved and unresolved gluons. Indeed, the fixed-order expansion does not seem to uniformly converge to the leading logarithmic resummation, suggesting it is an asymptotic series. A possible cause of the asymptotic nature of the series could be the perturbative expansion of the factor
\begin{equation}
U_{abj}(L)=\exp \left[ L \int\limits_\text{right} \frac{d\Omega_1}{4\pi}(\mathcal{W}_{ab}^1-\mathcal{W}_{aj}^1-\mathcal{W}_{jb}^1)   \right ]\,.
\end{equation}
In the dressed gluon approximation, this arises from renormalization group evolution of the generalized dressed gluon anomalous dimension of \Eq{eq:gen_dress}. Traditional logarithmic counting assumes that the integrand appearing in the exponential is $\mathcal{O}(1)$ in all regions of phase space, which is not true.   Maintaining this factor, as the dressed gluon does, appears to be vital for convergence of the expansion to arbitrary $L$ values.

\section{Resummation of NGLs to Higher Accuracy}\label{sec:nglimp}

The resummation of NGLs has thus far been restricted to leading logarithmic accuracy, where it is described by the BMS equation. We have described how the leading NGLs can be captured, and systematically calculated, by a sequence of factorization theorems producing the dressed gluon approximation. In this section we discuss how our approach can be extended beyond leading logarithmic accuracy. 

\subsection{Subleading Soft Corrections}

We begin by briefly discussing the NGLs that can be calculated and understood using the soft subjet factorization theorem and the corresponding dressed gluon approximation. Working within the soft approximation, there are three ways in which the accuracy of the dressed gluon approximation can be improved: by including a greater number of dressed gluons, by calculating the anomalous dimensions and matching for the dressing to higher orders, or including soft dijets, whose energies are not strongly ordered. Because we have established a factorization theorem for soft subjets, and a set of observables that factorize in all regions of phase space, all improvements are well defined and in principle are straightforward to implement. However, the various types of improvements will affect the distributions in different ways, correspond to different expansions, and be relevant in different regions of phase space. As discussed in \Sec{sec:pert_vs_dressed} the expansion to higher orders in the dressed gluon approximation is an expansion in the allowed phase space volume, while the expansion to higher orders in the dressed gluon anomalous dimensions and matching corresponds to a more familiar logarithmic expansion of the factorization theorem.

At large values of $L$, the most important corrections will be those from including an increasing number of dressed gluons at leading-logarithmic accuracy.  However, calculating higher dressed gluons is of limited phenomenological interest, as this simply resums a more complete set of leading NGLs, and the two-dressed gluon approximation is already accurate at the percent-level for a wide dynamic range. Further, we argued that the dynamics of the buffer region and its phase space suppression modifies the na\"ive logarithmic counting, and therefore we expect convergence and high accuracy of the dressed gluon approximation to large values of $L$ even for a limited number of dressed gluons.

On the other hand, for $L \lesssim 1$, we expect the most important corrections from the resummation of subleading NGLs are captured by calculating the dressed gluon anomalous dimensions to higher order, and capturing effects that are simply beyond the strongly-ordered soft approximation. This is evident from the numerical comparison of the one-dressed gluon approximation to the leading logarithmic resummation of the NGLs, which agree at the percent level for $L \lesssim 1$.

\subsection{Going Beyond the Soft Approximation}

\begin{figure}
\begin{center}
\includegraphics[width=7.5cm]{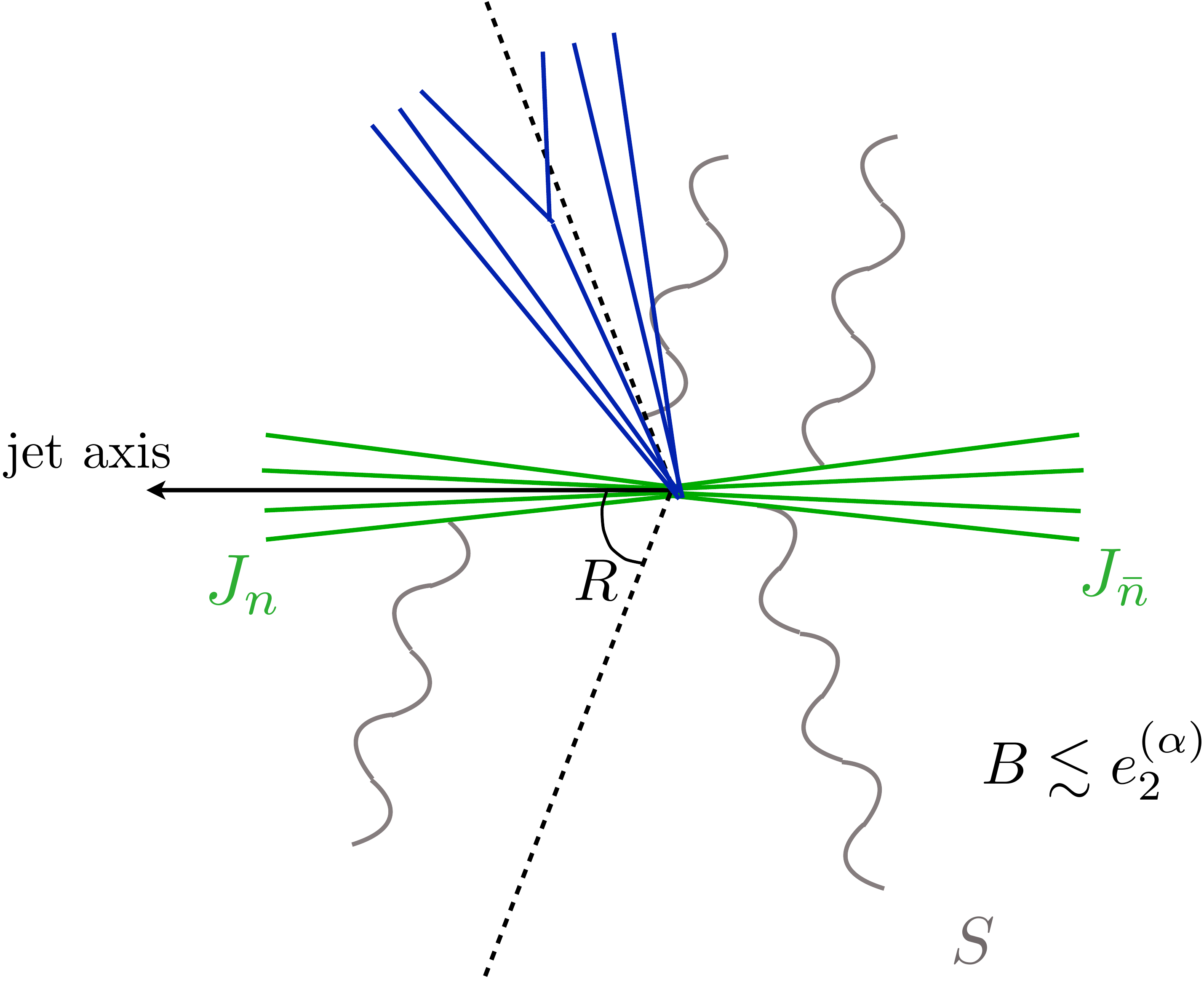}
\end{center}
\caption{Schematic depiction of the region of phase space relevant when $B \lesssim \ecf{2}{\alpha}$, but the relation is not parametric. In this region of phase space a soft subjet is pressed up against the boundary of the jet, and subleading NGLs arise from collinear splittings which cross the jet boundary. 
}
\label{fig:collinear_split}
\end{figure}

Subleading corrections to the leading logarithmic resummation are most important at small $L$. As $L$ becomes increasingly small, the out-of-jet scale $B$ approaches the soft jet scale, $\ecf{2}{\alpha}$. Recall from \Sec{sec:modes} that an important part of the factorization theorem in the soft subjet region of phase space was the inclusion of the boundary soft mode, which appeared due to the fact that $B \ll \ecf{2}{\alpha}$ effectively implemented a measurement on the soft jet emissions.  This required a refactorization of the dynamics of the soft jet into soft jet modes, which do not resolve the boundary of the jet, and boundary soft modes, which are lower energy but do resolve the boundary. In the case that $B\lesssim \ecf{2}{\alpha}$ this argument is no longer valid, the dynamics of the soft jet should no longer be refactorized in this manner, and instead one should have a soft jet function which itself is sensitive to the jet boundary. Therefore, in this region of phase space, collinear splittings from the soft subjet are sensitive to the boundary of the jet. The contributions to subleading NGLs from such splittings is apparent from a detailed study of the fixed order calculation at two loops \cite{Hornig:2011iu, Kelley:2011aa}. This region of phase space is schematically illustrated in \Fig{fig:collinear_split}.

While the NGLs arising in this region of phase space are formally subleading by traditional logarithmic counting, they may nevertheless be as important or even more important than the multiple dressed gluon effects of the leading NGLs.  Furthermore, in the soft subjet factorization theorem, logarithms of the ratios of the in-jet to out-of-jet scale are resummed. In the collinear region of phase space, however, this ratio is order $1$, and instead one must resum logarithms of the angle between the subjet axis and the jet boundary. Contributions from this region of phase space could therefore be important phenomenologically, and are also important for achieving a greater understanding of the dynamics of the buffer region. The approach to the resummation of NGLs presented in this paper, namely isolating a region of phase space, providing a factorized description of the dynamics, and resuming the NGLs by renormalization group evolution, can again be applied to understand this region of phase space. A factorization theorem describing this resummation, and allowing for the resummation of subleading logarithms arising from collinear splittings, along with a numerical study of their relevance will be presented in a forthcoming paper \cite{us_subNGL}.

\subsection{Ingredients for Subleading NGLs}
Given these considerations, we summarize the needed ingredients for resumming the subleading NGLs:
\begin{itemize}
\item One real emission with a virtual correction, and two real emissions in the color multipole function of a single dressed gluon.
\item The matching of a single dressed gluon to two-loops. This requires the two loop soft gluon current, calculated in \Refs{Duhr:2013msa,Li:2013lsa}.
\item Soft dijet matching in the not strongly-ordered regime, with a single real emission in their color multipole function.
\item Incorporation of the subleading collinear effects at the jet boundary matched to the wide angle soft emissions.
\end{itemize}
Each of these contributions form indepedent physical processes, and due to the dynamics of the buffer region modifying the logarithmic counting, may not all contribute equally to subleading NGLs. Since the resummation of subleading NGLs has not yet been accomplished, it would be extremely interesting to determine if such a resummation has phenomenological implications.

\section{Conclusions}\label{sec:conc}

In this paper we have presented a novel approach to the resummation of NGLs. By performing a multi-differential measurement on a jet, we are able to identify a phase space region involving a soft subjet, in which the NGLs can be factorized and resummed down to an unresolved infrared scale. Resummation is accomplished by renormalization group evolution in the factorization theorem and anomalous dimensions can be calculated to any perturbative accuracy.

An understanding of the dynamics of the NGLs from the factorization theorem in the soft subjet region of phase space led us to introduce the dressed gluon approximation. We demonstrated how the dressed gluon approximation can be used to calculate NGLs by explicitly calculating the NGLs for the hemisphere jet mass distribution in the one- and two-dressed gluon approximations. These computations were compared with numerical solutions to the BMS equation, and expansions of it. The dressed gluon approximation exhibited excellent convergence over phenomenological values of $L$, and provided considerably better convergence than other expansions, including the fixed-order perturbative result. Indeed, the dressed gluon approximation does not correspond to a fixed $L$ counting, but rather a novel expansion in the number of resolved gluons, which can be thought of as a perturbative expansion in distinct factorization theorems. We showed how this expansion at leading logarithmic accuracy can be obtained from the BMS equation. The dressed gluon approximation also gives an analytic realization of the buffer region, a proposed underlying description of the dynamics involved in the physics of NGLs. Futhermore, we have also discussed how our approach of isolating regions of phase space with multi-differential measurements to resum NGLs can be extended systematically beyond leading logarithmic accuracy.

We have also realized the BMS equation through a sequence of effective theories that produce the dressed gluon approximation. Part of this was due to the organization of the infrared degrees of freedom into the modes of soft-collinear effective theory. Repeating the soft subjet factorization decorates the original factorization theorem with new terms whose renormalization group invariance is independent of the parent factorization theorem's renormalization group structure.\footnote{This is not suprising from a measurement point of view. One is implicitly imposing additional measurements on the same events whose distribution the parent factorization theorem describes.} Thus each term corresponding to a different number of dressed gluons in the dressed gluon approximation is realized as a renormalization group invariant object, using the SCET modes. This helps to illuminate why the derivation of the BMS equation has resisted a renormalization group derivation within the SCET framework. It would be interesting to see if one could formulate a different effective theory that gave the BMS equation directly as a renormalization group equation.  A potential starting point may be reorganizing the infrared degrees of freedom using the so-called group space variables of \cite{Blaizot:2002np} used to derive and solve the B-JIMWLK hierarchy \cite{Mueller:1994jq, Balitsky:1995ub, Kovchegov:1999yj, JalilianMarian:1996xn,JalilianMarian:1997gr,Iancu:2001ad} (see also \cite{Caron-Huot:2013fea}).  The relationship between the the B-JIMWLK hierarchy and the BMS equation has been widely studied \cite{Marchesini:2003nh, Weigert:2003mm, Avsar:2009yb, Hatta:2008st, Hatta:2013iba,Caron-Huot:2015bja} and has been used to calculate leading NGLs with full color dependence. Nevertheless, it is difficult to see how one would incorporate the collinear degrees of freedom in such an approach, as well as the observable dependence of the global renormalization group that is the hallmark of soft-collinear factorization.

Our analysis in this paper has been limited to leading logarithmic accuracy to illustrate our approach to the resummation of non-global logarithms.  In addition to addressing the contributions to subleading non-global logarithms and calculation of the necessary objects as discussed in \Sec{sec:nglimp}, there are several open questions that one would want to understand to further validate the picture that we have constructed here.  We have provided a qualitative understanding of the importance of phase space suppression and the buffer region to the convergence of the dressed gluon expansion.  An all-orders in the dressed gluon expansion understanding of the buffer region, extending our discussion in \Sec{sec:buffer}, could lead to an explicit proof that the dressed gluons is a convergent expansion.  Even without an explicit proof of convergence, an understanding of the large $L$ properties of dressed gluons would be desirable.   If it can be explicitly shown that at large $L$, dressed gluons are produced in a stochastic process, for example, this would suggest a particular form for the exponentiated non-global soft function.

Understanding the factorization and all orders resummation properties of non-global observables is essential for connecting with many phenomenologically relevant jet observables. In this paper we have presented a first step towards this goal by presenting an effective field theory understanding of NGLs, and their relation to the soft substructure of jets. Since this paper presents the first step in understanding the soft substructure of jets, we conclude by discussing several important applications where we believe that our factorization theorem and understanding of the soft subjet region can be fruitfully applied.

\subsubsection*{$0\to1$ Jet Bin Transition for Electroweak Boson Production}

Jet binning plays an important role in many LHC analyses, for example $H\to WW$.  In this example, the experimental sensitivity is highest in the exclusive zero- and one-jet bins due to the large $t\bar t$ background. There has been considerable study of the resummation for the exclusive zero-jet bin \cite{Berger:2010xi,Banfi:2012jm,Banfi:2012yh,Becher:2012qa,Stewart:2013faa,Becher:2013xia,Moult:2014pja}, as well as for the exclusive one- \cite{Jouttenus:2013hs,Liu:2012sz} , and even two-jet bin \cite{Gangal:2013nxa}. However, in these cases a factorization theorem only exists in the case that the jets are at the hard scale. An important open problem is how to describe the transition from the zero-jet to one-jet region, where the jet has small $p_T$. Attempts at understanding this region by combining information from different jet bins has been discussed in \Ref{Boughezal:2013oha}.

To understand the zero-jet to one-jet transition requires understanding the factorization theorem in the regime that a soft (sub)jet is formed. Our soft subjet factorization theorem provides a description of the dynamics in this region of phase space, and therefore can be used to study the transition.

\subsubsection*{Soft Subjet Region for Jet Substructure Observables}

Another important application of our factorization theorem for the soft subjet region of phase space is towards the analytic understanding of jet substructure observables which resolve a two-prong structure, as required for boosted $W/Z/H$ tagging. A complete description of the relevant phase space requires factorization theorems for one-prong jets, jets with hard, collinear subjets (described by the SCET$_+$ effective theory \cite{Bauer:2011uc}), and jets with a hard core and a soft subjet, as presented here.  While the focus of this paper has been of those aspects of the soft subjet factorization theorem as relevant for understanding NGLs, our factorization theorem provides a complete description of the dynamics in the soft subjet region of phase space. It can therefore be incorporated into a complete study of the phase space for two-prong jets.  An analytic calculation for the substructure observable $D_2$ \cite{Larkoski:2014gra} combining the factorization theorems in each relevant region of phase space will be presented in a forthcoming publication \cite{usD2}.

\subsubsection*{Improving Monte Carlo Generators}

Monte Carlo generators play a vital role in the accurate and realistic description of QCD processes at colliders.  The soft subjet factorization theorem may have consequences for developing Monte Carlos that are accurate to beyond leading logarithmic accuracy.  With the one-prong and collinear subjet regions, the soft subjet region completes the description of the $\ecf{2}{\alpha}, \ecf{3}{\alpha}$ phase space \cite{us_subNGL}, which completely characterizes a $1\to2$ splitting.  A possible implementation of a Monte Carlo parton shower would be to first randomly choose a point in the $\ecf{2}{\alpha}, \ecf{3}{\alpha}$ phase space plane.  At this phase space point, the emission is weighted with a probability determined by a generalized Sudakov factor, which in the soft region of phase space is a dressing at the scale set by $\ecf{2}{\alpha}$ by emissions at the scale $\ecf{3}{\alpha}$. Such a Monte Carlo would then accurately describe the complete phase space for a $1\to 2$ splitting. These techniques and way of thinking could be extended to a description of the multi-differential phase space of the set of energy correlation functions $\{e_2^{(\alpha)},e_3^{(\alpha)},\dotsc,e_{n+1}^{(\alpha)}\}$ which completely characterize a $1\to n$ splitting. By randomly choosing a point in the multi-dimensional phase space and implementing the appropriate resummation for that region of phase space, one could envision a fully differential parton shower, accurate to the logarithmic accuracy of the factorization theorems in all regions of phase space.

\begin{acknowledgments}
We thank Jesse Thaler, Iain Stewart, Ira Rothstein, Bob Jaffe, Simone Marzani, Chris Lee, Dan Kolodrubetz, Piotr Pietrulewicz, Frank Tackmann, and Jon Walsh for helpful discussions.  We also thank Jan Balewski for the use of, and assistance with, the Ganglia cluster at MIT for generation of the NGL Monte Carlo.  This work is supported by the U.S. Department of Energy (DOE) under grant Contract Numbers DE-SC00012567 and DE-SC0011090. D.N. is also supported by an MIT Pappalardo Fellowship.  I.M. is also supported by NSERC of Canada.  D.N.~thanks the hospitality of the Los Alamos theory group.  We thank the Erwin Schr\"odinger Institute and the organizers of the ``Jets and Quantum Fields for LHC and Future Colliders'' workshop for hospitality and support where this seeds of this work were sown.
\end{acknowledgments}

\appendix

\section{Definitions of Factorized Functions for Soft Subjet Production}\label{app:Factorized_Functions}

In this appendix we give operator definitions in the formalism of SCET for all functions appearing in the soft subjet factorization theorem presented in  \Sec{sec:Fact}, 
\begin{align}
\frac{d\sigma(\outj;R)}{d\ecfLa d\ecfLb d\ecfres }&=H(Q^2) H^{sj}_{n\bar{n}}\Big(\ecfLa,\ecfLb\Big) J_{n}\Big(\ecfres\Big)\otimes J_{\bar{n}}(\outj) \nonumber\\
&\qquad\otimes S_{n\bar{n}\sja }\Big(\ecfres;\outj;R\Big)\otimes J_{\sja}\Big(\ecfres\Big)\otimes S_{\sja\sjabar}(\ecfres;R)\,,
\end{align}
whose structure we have recalled for convenience.
The one-loop calculation of these functions will be given in \App{app:oneloopcalcs} along with their anomalous dimensions. We will only give results for the case that the soft subjet is produced by a gluon, off of the initial $q\bar q$ pair in $e^+e^-$ annihilation. Other partonic configurations are straightforward, and obey the same type of factorization, but their hard production coefficient is not enhanced by the soft singularity $1/\sje$.

The operator definitions in this section are given in terms of the collinear gauge invariant quark and gluon SCET fields \cite{Bauer:2000yr,Bauer:2001ct}, which we denote $\mathcal{B}_{\perp_{\sja}}^{\mu},\chi_{n}$, as well as (lightlike) Wilson lines, $S_q$. The Wilson lines extend from the origin to infinity along the direction of their specifying vector, $q$. Explicitly
\begin{align}
S_q={\bf P} \exp \left( ig \int\limits_0^\infty ds\, q\cdot A(x+sq)    \right)
\end{align}
where $\bf P$ denotes path ordering, and $A$ is the appropriate gauge field, and the color representation has been suppressed. Since we only consider the case of $e^+e^-$, all Wilson lines are outgoing. The soft Wilson lines carry the color representation of their parent collinear sectors, that is, adjoint representation for gluons and fundamental representation for quarks. Since we have no more than three Wilson lines in a soft function, the soft functions can always be written as color-singlet traces. In the more general case, the soft function is a color matrix, which must be traced against the hard functions, $H(Q^2)$ and $H^{sj}$ appearing in the factorization theorem ( see e.g. \Refs{Ellis:2010rwa,Jouttenus:2011wh} for more details). We will also use the large label momentum operator $\mathcal{P}^\mu$ \cite{Bauer:2000yr} in the function definitions, which extracts the large component of the momentum for a particle in a given sector. We denote by $Q$ the center of mass energy of the $e^+e^-$ collisions, so that $Q/2$ is the energy deposited in a hemisphere, and $Q_{SJ}\ll Q$ is the large component of the soft jet momentum.

The functions appearing in the soft subjet factorization theorem of \Eq{fact_inclusive_form_1} have the following SCET operator definitions:
\begin{itemize}
\item Soft Subjet Jet Function:
{\small\begin{align}
&J_{\sja }\Big(\ecf{3}{\beta}\Big)=\\
& \hspace{.25cm}
\frac{(2\pi)^3}{C_A}\text{tr}\langle 0|\mathcal{B}_{\perp_{\sja}}^{\mu}(0)\Theta_{O}(B)\delta(Q_{SJ}-\sjabar \cdot{\mathcal P})\delta^{(2)}(\vec{{\mathcal P}}_{\perp_{SJ}})\delta\Big(\ecf{3}{\beta}-\Theta_{FJ}\ecfop{3}{\beta}\big|_{SJ}\Big)\,\mathcal{B}_{\perp_{\sja}\mu}(0)|0\rangle \nonumber
\end{align}}
\item Jet Function:
{\small\begin{align}
\hspace{-1cm}
J_{n}\Big(\ecf{3}{\beta}\Big)&=\frac{(2\pi)^3}{C_F}\text{tr}\langle 0|\frac{\bar{n}\!\!\!\slash}{2}\chi_{n}(0) \Theta_{O}(B)\delta(Q-\bar{n}\cdot{\mathcal P})\delta^{(2)}(\vec{{\mathcal P}}_{\perp})\delta\Big(\ecf{3}{\beta}-\Theta_{FJ}\ecfop{3}{\beta}\big|_{HJ}\Big)\bar{\chi}_n(0)|0\rangle
\end{align}}
\item Boundary Soft Function:
{\small\begin{align}
S_{\sja \,\sjabar }\Big(\ecf{3}{\beta};R\Big)&=\frac{1}{C_{A}}\text{tr}\langle 0|T\{S_{\sja } S_{\sjabar }\} \Theta_{O}(B) \delta\Big(\ecf{3}{\beta}-\Theta_{FJ}\ecfop{3}{\beta}\big|_{BS}\Big)\bar{T}\{S_{\sja } S_{\sjabar }\} |0\rangle
\end{align}}
\item Soft Subjet Soft Function:
{\small\begin{align}
\hspace{-1cm}S_{\sja \,n\,\bar{n}}\Big(\ecf{3}{\beta},B;R\Big)&=\text{tr}\langle 0|T\{S_{\sja } S_{n} S_{\bar{n}}\}\Theta_{O}(B)\delta\Big(\ecf{3}{\beta}-\Theta_{FJ}\ecfop{3}{\beta}\big|_{S}\Big)\bar{T}\{S_{\sja } S_{n} S_{\bar{n}}\} |0\rangle
\end{align}}
\end{itemize}

The definitions of these functions include measurement operators, which when acting on the final state, return the value of a given observable. The operator $\ecfop{3}{\beta}$ measures the contribution to $\ecf{3}{\beta}$ from final states, and must be appropriately expanded following the power counting of the sector on which it acts. The operators $\Theta_{FJ}$, and $\Theta_{O}$ constrain the measured radiation to be in the jet or out of the jet, respectively, and will be defined shortly. 

The action of the measurement function $\ecfop{3}{\beta}$ on a arbitrary state for each of the factorized sectors contributing to the three-point energy correlation function measurement is given by
{\small\begin{align}\label{eq:action_measurements}
\ecfop{3}{\beta}\big|_{SJ}\Big|X_{sj}\Big\rangle&=\sum_{k_i,k_j\in X_{sj}} N_{SJ} \frac{\sjabar\cdot k_i}{Q}\frac{\sjabar\cdot k_j}{Q}\left(\frac{k_i\cdot k_j}{\sjabar\cdot k_i\sjabar\cdot k_j}\right)^{\frac{\beta}{2}}\Big|X_{sj}\Big\rangle\,,\\
\ecfop{3}{\beta}\big|_{HJ}\Big|X_{hj}\Big\rangle&=\sum_{k_i,k_j\in X_{hj}} N_{HJ}\frac{\nbar\cdot k_i}{Q}\frac{\nbar\cdot k_j}{Q}\left(\frac{k_i\cdot k_j}{\nbar\cdot k_i\nbar\cdot k_j}\right)^{\frac{\beta}{2}}\Big|X_{hj}\Big\rangle\,,\\
\ecfop{3}{\beta}\big|_{BS}\Big|X_{bs}\Big\rangle&=\sum_{k\in X_{bs}}N_{BS} \frac{\sjabar\cdot k}{Q}\left(\frac{\sja\cdot k}{\sjabar\cdot k}\right)^{\frac{\beta}{2}}\Big|X_{bs}\Big\rangle\,,\\
\ecfop{3}{\beta}\big|_{S}\Big|X_{s}\Big\rangle&=\sum_{k\in X_{s}}N_S\frac{k^0}{Q}\left(\frac{\sja\cdot k}{k^0}\frac{n\cdot k}{k^0}\right)^{\frac{\beta}{2}}\Big|X_{s}\Big\rangle\,,
\end{align}}
where, for simplicity, we have extracted the normalization factors
\begin{alignat}{2}\label{eq:norm_factors}
N_{SJ}&=2^{-3+\beta}\frac{Q_{hj}}{Q}(n\cdot\sja)^{\beta}\,, & \qquad
N_{HJ}&=2^{-3+\beta}\frac{Q_{sj}}{Q}(n\cdot\sja)^{\beta}\,,\\
N_{BS}&=2^{-1+\frac{\beta}{2}}N_S(n\cdot\sja)^{\beta/2}\,,& \qquad
N_{S}&=\frac{Q_{hj}Q_{sj}}{4Q^2}(n\cdot\sja)^{\beta/2}\,,\\
Q_{hj}&=\bar n \cdot p_{hj}\,, &\qquad Q_{sj}&=\sjabar \cdot p_{sj}.
\end{alignat}%
$Q_{hj}$ and $Q_{sj}$ are the large light-cone momentum components for the hard jet and the soft subjet, respectively.
These expressions follow from properly expanding the definition of the energy correlation function measurements in the power counting of each of the sectors. Note that on the jet sectors, the three-point correlation measurement becomes an effective two-point correlation measurement, since the two-point energy correlation function is set by the initial splitting of the subjet.

The in-jet restriction, $\Theta_{FJ}$, is given by
\begin{align}
\Theta_{FJ}(k)&=\Theta\left(\tan ^2\frac{R}{2}-\frac{n\cdot k}{\bar{n}\cdot k}\right)\,.
\end{align}
The jet restriction must also be expanded following the power counting of the given sector. We will see that this is actually quite subtle for the soft subjet modes, since the angle between the soft subjet axis and the boundary of the jet has a non-trivial power counting. In particular, the expansion of $\Theta_{FJ}(k)$ is different for the soft subjet jet and boundary soft modes, and will demonstrate the necessity of performing the complete factorization of the soft subjet dynamics into jet and boundary soft modes. The explicit expansions in each sector's power counting will be given in \App{app:oneloopcalcs}, when we consider the one-loop calculation of the functions appearing in the factorization theorem. Finally, since we are considering the case where the out-of-jet scale $B$ is much less than the in-jet scale, the operator
$$
\Theta_{O}(B)
$$
must also be included in the definition of the soft subjet functions. This operators vetoes out-of-jet radiation above the scale $B$. The explicit expression for $\Theta_{O}(B)$ expanded in the power counting of each of the factorized sectors will be given in the one-loop calculations of \App{app:oneloopcalcs}.

\section{One-Loop Calculations of Soft Subjet Functions}\label{app:oneloopcalcs}

In this appendix we present the one-loop calculation of all the functions appearing in the soft subjet factorization theorem of \Sec{sec:fact_theorem}, whose operator definitions are given in \App{app:Factorized_Functions}.  As discussed when defining the soft subjet functions \App{app:Factorized_Functions}, we will only give results for the case that the soft subjet is produced by a gluon, although it is straightforward to extend the calculation to other partonic configurations. Throughout this section, we will make use of the convenient shorthand notation
\begin{align}
[d^dk]_+&=\frac{d^dk}{(2\pi)^d}2\pi\Theta(k^0)\delta(k^2),
\end{align}
for the integration measure of an on-shell, massless, final-state parton. For the jet and soft functions, we only give the final expressions in the Laplace space of $\ecfres$, where they satisfy a multiplicative renormalization group evolution. This allows for a straightforward comparison of the anomalous dimensions.

\subsection{Hard Matching for Dijet Production}

The hard matching coefficient, $H(Q^2)$, is the well known hard function for the production of a $q \bar q$ pair in $e^+e^-$ annihilation. It is defined by
\begin{equation}
H(Q^2, \mu)= |C(Q^2,\mu)|^2\,,
\end{equation}
where $C(Q^2,\mu)$ is the Wilson coefficient obtained from matching the full theory QCD current $\bar \psi \gamma^\mu \psi$ onto the SCET dijet operator $\bar \chi_n \gamma^\mu_\perp \chi_{\bar n}$. This Wilson coefficient is well known (see e.g. \cite{Bauer:2003di,Manohar:2003vb,Ellis:2010rwa,Bauer:2011uc} ), and is given at one-loop by
\begin{equation}
C(Q^2,\mu)=1+\frac{\alpha_s(\mu)\, C_F}{4\pi}\left ( -\log^2\left[ \frac{-Q^2}{\mu^2}  \right]+3\log \left[ \frac{-Q^2}{\mu^2} \right]-8+\frac{\pi^2}{6}   \right )\,.
\end{equation}
The branch cut in the logarithms must be taken as $-Q^2 \to -Q^2-i\epsilon$.

\subsection{Hard Matching for Soft Jet Production}
The hard matching coefficient $H^{sj}(\sje,\sjtheta)$ is determined by the finite parts of the soft matrix element for a single soft state
\begin{align}
H^{sj}(\sje,\sja)&=\,\text{tr}\langle 0|T\{S_nS_{\bar{n}}\}|sj \rangle\langle sj|\bar{T}\{S_nS_{\bar{n}}\}|0\rangle_{\text{fin}}\,.
\end{align}
The virtual corrections of the effective theory cancel the IR divergences of this matrix element, giving a finite matching coefficient. This matrix element can be calculated from the square of the soft gluon current \cite{Berends:1988zn,Catani:2000pi}, which is known to two loop order \cite{Duhr:2013msa,Li:2013lsa}. Here, for simplicity, we restrict ourselves to one-loop accuracy.  The tree level and one-loop hard matching coefficients for the soft subjet production are given by
{\small\begin{align}
H^{sj(\text{tree})}_{n\bar{n}}(\sje,\sja)&=\frac{\alpha_s C_F}{4\pi^2\sje}\frac{n\cdot \bar{n}}{n\cdot\sja\,\sja\cdot\bar{n}}\,,\\
H^{sj(1)}_{n\bar{n}}(\sje,\sja)&=H^{sj(\text{tree})}_{n\bar{n}}(\sje,\sja)\left(\frac{\alpha_sC_A}{\pi}\right)   \left[ -\frac{1}{4}\ln ^2\left(\frac{2\mu^2\bar{n}\cdot n}{Q_{sj}^2n\cdot\sja\,\sja\cdot\bar{n}}\right) +\frac{5\pi^2}{24}\right]\,.
\end{align}}
The results of \cite{Catani:2000pi} can be used to determine the soft-jet production matching from an arbitrary number of hard jets at one loop.

\subsection{Jet Function}\label{sec:hard_jet}

In this section we calculate the jet function for the energetic subjet along the $n$ direction. The one-loop expression for the na\"ive (before zero bin subtraction) jet function is
{\small\begin{align}
J_{n}^{(1)}(Q_{J},\eec{3}{\beta})&=\mu^{2\epsilon}C_i\, g^2\int[d^dk_1]_+\int[d^dk_2]_+(2\pi)^{d-1}\delta^{d-2}(\vec{k}_{1\perp_{sj}}+\vec{k}_{2\perp_{sj}})\nonumber\\
&\hspace{3cm}\times\Theta_J\Big(\eec{3}{\beta},B,R,Q_{J},k_1,k_2\Big)\frac{Q_{J}\, P_{qg}\left(\frac{\bar n \cdot k_1}{Q_{J}},\frac{\bar n \cdot k_2}{Q_{J}}\right)}{2k_1\cdot k_2}\,.
\end{align}}%
Here we have chosen to calculate the jet function by integrating against the splitting function \cite{Ritzmann:2014mka}. Since we have assumed the partonic configuration in which the soft subjet is a gluon jet, the jet in the $n$ direction is assumed to be described by a collinear quark field.  For the splitting functions we use the (slightly unconventional) notation
\begin{align}\label{eq:split}
\langle P_{qg}(z_1, z_2) \rangle&=\left[ \frac{1+z_1^2}{z_2}-\epsilon z_2 \right]\,,\\
\langle P_{gg}(z_1, z_2) \rangle&=2\left [\frac{z_1}{z_2}+\frac{z_2}{z_1}+z_1z_2 \right]\,, \\
\langle P_{q \bar q}(z_1, z_2) \rangle &=\left [1-\frac{2z_1 z_2}{1-\epsilon} \right]\,,
\end{align}
where the $\langle \rangle$ denote that the splitting functions are spin averaged.

The jet algorithm and measurement constraint are given by
{\small\begin{align}\label{eq:total_sj_constraints}
\Theta_J\Big(\eec{3}{\beta},B,R,k_1,k_2\Big)&=\Theta\left(\tan ^2\frac{R}{2}-\frac{n\cdot k_1}{\bar{n}\cdot k_1}\right)    \Theta\left( \tan ^2 \frac{R}{2}-\frac{n\cdot k_2}{\bar{n}\cdot k_2}\right)     \delta(Q_{J}-\bar n \cdot k_1-\bar n \cdot k_2)\nonumber\\
&\hspace{1cm}\delta\left(     \eec{3}{\beta}-N_{HJ}\frac{\nbar\cdot k_1}{Q}\frac{\nbar\cdot k_2}{Q}\left(\frac{k_1\cdot k_2}{\nbar\cdot k_1\nbar\cdot k_2}\right)^{\frac{\beta}{2}}  \right)\nonumber\\
&\hspace{-3cm}+\delta(\eec{3}{\beta})    \Theta\left(\frac{n\cdot k_1}{\bar{n}\cdot k_1}-\tan ^2\frac{R}{2}\right)     \Theta\left(\tan ^2\frac{R}{2}-\frac{n\cdot k_2}{\bar{n}\cdot k_2}\right)    \Theta\left(B-\frac{1}{2}\bar n \cdot k_1\right)\delta\left(Q_{J}-\bar n \cdot k_2\right)\nonumber\\
&\hspace{-3cm}+\delta(\eec{3}{\beta})    \Theta\left (\frac{n\cdot k_2}{\bar{n}\cdot k_2}-\tan ^2\frac{R}{2}\right)    \Theta\left(\tan ^2\frac{R}{2}-\frac{n\cdot k_1}{\bar{n}\cdot k_1}\right)  \Theta\left(B-\frac{1}{2}\bar n \cdot k_2\right)\delta\left(Q_{J}-\bar n \cdot k_1\right)\,,
\end{align}}%
where we have used the expression for the action of the $\ecf{3}{\beta}$ measurement on a hard jet state from \Eq{eq:action_measurements}.

In the power counting of the $n$ collinear sector, the second two terms vanish upon performing the multipole expansion on the jet function constraint. The first term simplifies since
\begin{align}
\Theta\left(\tan ^2\frac{R}{2}-\frac{n\cdot k_i}{\bar{n}\cdot k_i}\right)\to 1,
\end{align}
for any particle $i$ in the $n$ collinear sector.  The phase space for the two partons in the jet with these constraints imposed is then given by
{\small\begin{align}\label{eq:jet_pspace}
(2\pi)^{d-1}\int[d^dk_1]_+\int[d^dk_2]_+\delta^{d-2}(\vec{k}_{1\perp}+\vec{k}_{2\perp})\delta(Q_{J}-\bar n \cdot k_1-\bar n \cdot k_2)\nonumber\\
=\frac{2\pi^{\frac{1}{2}-\epsilon}Q_{J}}{(2\pi)^{3-2\epsilon}\,\Gamma(\frac{1}{2}-\epsilon)}\int_0^1 \frac{dZ}{Z(1-Z)}\int_{0}^{\pi}d\phi\,\text{sin}^{-2\epsilon}\phi\,\,,
\end{align}}%
where $Z$ defines the large momentum fractions of the partons as
\begin{align}
\bar n \cdot k_1=Q_{J}Z\,,\qquad
\bar n \cdot k_2=Q_{J}(1-Z)\,,
\end{align}
and the angle $\phi$ is defined by
\begin{equation}
k_\perp \cdot n_\perp=\cos\phi |k_\perp| |n_\perp|\,.
\end{equation}
Substituting this into the expression for the jet function gives
{\small\begin{align}
J_{n}^{(1)}(Q_{J},\eec{3}{\beta})&=\mu^{2\epsilon}C_F\, g^2 \,
 \frac{2\pi^{\frac{1}{2}-\epsilon}Q_{J}}{(2\pi)^{3-2\epsilon}\,\Gamma(\frac{1}{2}-\epsilon)}\int_0^1 \frac{dZ}{Z(1-Z)}\int_{0}^{\pi}d\phi\,\text{sin}^{-2\epsilon}\phi\,   \frac{Q_{J}\, P_{qg}\left(\frac{\bar n \cdot k_1}{Q_{J}},\frac{\bar n \cdot k_2}{Q_{J}}\right)}{2k_1\cdot k_2}\nonumber \\
 &\hspace{1.5cm}\times\delta\left(     \eec{3}{\beta}-  N_{HJ}\frac{\nbar\cdot k_1}{Q}\frac{\nbar\cdot k_2}{Q}\left(\frac{k_1\cdot k_2}{\nbar\cdot k_1\nbar\cdot k_2}\right)^{\frac{\beta}{2}}  \right)\,.
\end{align}}%
Performing the integrals as an expansion in $\epsilon$ and transforming to Laplace space, we find 
{\small\begin{align}
J_{n}^{(1,\,\text{div})}(Q_{J},\eec{3}{\beta})&= \frac{\alpha_s C_F}{2\pi} \left[ \frac{-\beta}{(1-\beta)\epsilon^2} +\frac{3}{2\epsilon} -\frac{2}{(1-\beta)\epsilon}\log\left( H\left(\eeclp{3}{\beta}\right)   \right) \right] \,,  \\
J_{n}^{(1,\,\text{fin})}(Q_{J},\eec{3}{\beta})&=\frac{\alpha_s C_F}{2\pi}\left[ \frac{-9 \pi ^2 \beta ^2+78 \beta ^2+16 \pi ^2 \beta -150 \beta -4 \pi ^2+72}{12 (\beta -1) \beta }  \right. \\
&\hspace{1.0cm}\left.   
+\frac{3}{\beta}\log\left( H \left(\eeclp{3}{\beta}\right) \right) +\frac{2}{\beta(\beta-1)}\log^2\left( H\left(\eeclp{3}{\beta}\right)\right)
 \right] +\mathcal{O}(\epsilon)\,, \nonumber
\end{align}}
where $\eeclp{3}{\beta}$ is the Laplace conjugate of $\eec{3}{\beta}$, and we have explicitly separated the finite and divergent pieces. The argument of the logarithms is given by
{\small\begin{align}
H\left(\eeclp{3}{\beta}\right)&=2^{-\frac{\beta}{2}}e^{ \gamma_E} \left(\frac{\mu}{Q}\right)^\beta  N_{HJ}\,  \eeclp{3}{\beta} \,.
\end{align}}%
Note that all zero bins for the jet function vanish.

\subsection{Soft Subjet Jet Function}\label{sec:ssj}

In this section we calculate the jet function for the soft subjet itself. Since the soft subjet is near the boundary of the jet, we will see that we must carefully treat the jet boundary constraint, emphasizing the role of the boundary soft mode.  The one-loop expression for the na\"ive (before zero bin subtraction) jet function is
{\small\begin{align}\label{eq:subjet_jetfunc_integrand}
J_{\sja}^{(1)}(Q_{SJ},\eec{3}{\beta})&=\mu^{2\epsilon}\, g^2\int[d^dk_1]_+\int[d^dk_2]_+(2\pi)^{d-1}\delta^{d-2}(\vec{k}_{1\perp_{sj}}+\vec{k}_{2\perp_{sj}})\\
&\hspace{0cm}\times\Theta_J\Big(\eec{3}{\beta},B,R,Q_{SJ},k_1,k_2\Big)\frac{Q_{SJ}\, \left[C_A P_{gg}\left(\frac{\bar n_{sj} \cdot k_1}{Q_{SJ}},\frac{\bar n_{sj} \cdot k_2}{Q_{SJ}}\right) +n_f T_F\, P_{q \bar q}\left(\frac{\bar n_{sj} \cdot k_1}{Q_{SJ}},\frac{\bar n_{sj} \cdot k_2}{Q_{SJ}}\right)    \right]}{2k_1\cdot k_2}\,, \nonumber
\end{align}}%
where $n_f$ denotes the number of light flavors, and $T_F=1/2$ specifies our normalization convention for the SU$(3)$ algebra. Here we have taken the soft subjet to be a gluon jet, and have again chosen to calculate the jet function by integrating over the splitting functions, where the arguments of the splitting function denote the energy fraction of the two partons, as defined in \Eq{eq:split}. The jet algorithm and measurement constraint are given by
{\small\begin{align}
\Theta_J\Big(\eec{3}{\beta},B,R,k_1,k_2\Big)&=\Theta\left(\tan ^2\frac{R}{2}-\frac{n\cdot k_1}{\bar{n}\cdot k_1}\right)    \Theta\left( \tan ^2 \frac{R}{2}-\frac{n\cdot k_2}{\bar{n}\cdot k_2}\right)     \delta(Q_{SJ}-\sjabar \cdot k_1-\sjabar \cdot k_2)\nonumber\\
&\hspace{1cm}\delta\left(     \eec{3}{\beta}-N_{SJ} \frac{\sjabar\cdot k_1}{Q}\frac{\sjabar\cdot k_2}{Q}\left(\frac{k_1\cdot k_2}{\sjabar\cdot k_1\sjabar\cdot k_2}\right)^{\frac{\beta}{2}}  \right)\nonumber\\
&\hspace{-3cm}+\delta(\eec{3}{\beta})    \Theta\left(\frac{n\cdot k_1}{\bar{n}\cdot k_1}-\tan ^2\frac{R}{2}\right)     \Theta\left(\tan ^2\frac{R}{2}-\frac{n\cdot k_2}{\bar{n}\cdot k_2}\right)    \Theta\left(B-\frac{1}{2}\sjabar \cdot k_1\right)\delta\left(Q_{SJ}-\sjabar \cdot k_2\right)\nonumber\\
&\hspace{-3cm}+\delta(\eec{3}{\beta})    \Theta\left (\frac{n\cdot k_2}{\bar{n}\cdot k_2}-\tan ^2\frac{R}{2}\right)    \Theta\left(\tan ^2\frac{R}{2}-\frac{n\cdot k_1}{\bar{n}\cdot k_1}\right)  \Theta\left(B-\frac{1}{2}\sjabar \cdot k_2\right)\delta\left(Q_{SJ}-\sjabar \cdot k_1\right)\,,
\end{align}}%
where we have used the expression for the action of the $\ecf{3}{\beta}$ measurement on a soft jet state from \Eq{eq:action_measurements}.  Since we are considering the case where the out-of-jet scale $B$ is lower than the in-jet scale, we can multipole expand the constraint in the out-of-jet region as
{\small\begin{align}
\Theta\Big(B-\frac{1}{2}\sjabar \cdot k_i\Big)\rightarrow \Theta\Big(-\frac{1}{2}\sjabar \cdot k_i\Big)=0\,,
\end{align}}%
which eliminates the second two terms in \Eq{eq:total_sj_constraints}. This implies that the jet boundary effectively acts as a hard wall for radiation in the soft subjet jet function.  For the jet modes of the soft subjet, we can also multipole expand the jet function constraints
\begin{equation}
\Theta\left(\tan ^2\frac{R}{2}-\frac{n\cdot k_1}{\bar{n}\cdot k_1}\right)    \Theta\left( \tan ^2 \frac{R}{2}-\frac{n\cdot k_2}{\bar{n}\cdot k_2}\right) \to 1\,.
\end{equation}
The fact that this constraint can be multipole expanded follows from the power counting in \Sec{sec:modes}, where we found that the angle between the soft subjet modes and the soft subjet axis scales like $\theta_{cs}^\alpha\sim \frac{\ecf{3}{\alpha}}{\left(\ecf{2}{\alpha}\right)^2}$, while the angle between the soft subjet axis and the jet boundary satisfies $\Delta \theta_{sj}\gg \frac{\ecf{3}{\alpha}}{\left(\ecf{2}{\alpha}\right)^2}$. This can also be seen from expanding the jet constraints in the local soft subjet coordinates, where we find
{\small\begin{align}\label{eq:fjet_constraint_expanded}
\tan ^2\frac{R}{2}-\frac{n\cdot k}{\bar{n}\cdot k}&=\tan ^2\frac{R}{2}-\frac{n\cdot \sja }{\bar{n}\cdot \sja }+4\frac{k_{\perp_{sj}}\cdot n_{\perp_{sj}}}{(\bar{n}\cdot \sja )^2\sjabar \cdot k}+...\,,\nonumber\\
&=\tan ^2\frac{R}{2}-\tan ^2\frac{\theta_{sj}}{2}+4\frac{k_{\perp_{sj}}\cdot n_{\perp_{sj}}}{(\bar{n}\cdot \sja )^2\sjabar \cdot k}+... >0 \,,
\end{align}}%
where we have used that $n_{\perp_{sj}}=-\bar{n}_{\perp_{sj}}$.

The ability to perform this multipole expansion relies crucially on the fact that we have fully factorized the dynamics of the soft subjet into jet modes and boundary soft modes. For the boundary soft modes, we \emph{cannot} perform the above multipole expansion. This implies that the soft subjet jet function will not depend on the factor $\anglediff$, as should be the case for a collinear function, while the boundary soft function carries the entire dependence of the soft subjet dynamics on the difference  $\anglediff$. Since both particles are constrained to lie within the jet, and the jet boundary constraint is multipole expanded, the phase space constraints for the soft subjet jet function are identical as for the standard jet function for the variable $\ecf{3}{\beta}$, but for a gluon jet. The explicit expression will be given shortly. Alternatively, it is possible to calculate the soft subjet jet function without performing the multipole expansion on the jet constraint. In this case one finds that the phase space for the jet function is corrected by a term which depends on $\anglediff$, leading to a correction to the jet function depending on $\anglediff$. However, we have explicitly checked that performing the appropriate boundary soft zero bin subtraction entirely removes this correction, again emphasizing the importance of this mode. We therefore stress the importance of a proper power counting analysis when analyzing the effective theories for more complicated jet configurations. A similar feature was also noted in \Ref{Ellis:2010rwa} for the calculation of different individual jet functions with a jet algorithm constraint. 

We now give explicit expressions for the gluon jet function. Since the phase space is identical to that given in \Eq{eq:jet_pspace}, but with the integration performed against the splitting functions as indicated in \Eq{eq:subjet_jetfunc_integrand}, we simply give the final result. Performing the integrals as an expansion in $\epsilon$ and transforming to Laplace space, we find
{\small \begin{align}
J_{n_{sj}}^{(1,\,\text{div})}(Q_{J},\eec{3}{\beta})&= \frac{\alpha_s }{2\pi} \left[\frac{\beta  C_A}{ (\beta -1) \epsilon ^2} +\frac{\beta_0}{  2\epsilon }+\frac{2C_A   \log H\left(\eeclp{3}{\beta}\right)}{(\beta -1) \epsilon }\right] \,,  \\
J_{n_{sj}}^{(1,\,\text{fin})}(Q_{J},\eec{3}{\beta})&=\frac{\alpha_s }{2\pi}\left[    \frac{2 C_A \log ^2 H\left(\eeclp{3}{\beta}\right)}{(\beta -1) \beta }+\frac{11 C_A \log H\left(\eeclp{3}{\beta}\right)}{3 \beta }-\frac{4 n_f T_F \log H\left(\eeclp{3}{\beta}\right)}{3 \beta }  \right.    \\  
&\hspace{-1.5cm}   \left . -\frac{\pi ^2
   \beta  C_A}{12 (\beta -1)}-\frac{67 C_A}{9 \beta }+\frac{\pi ^2 C_A}{3 (\beta -1) \beta }+\frac{2 \pi ^2 C_A}{3
   \beta }+\frac{67 C_A}{9}-\frac{2 \pi ^2 C_A}{3}+\frac{26 n_f T_F}{9 \beta }-\frac{23 n_f T_F}{9}
\right] \,, \nonumber
\end{align}}
where $\beta_0$ is defined with the normalization
\begin{equation}
\beta_0=\frac{11 C_A}{3}-\frac{4n_f T_F}{3 } \,,
\end{equation}
and where $\eeclp{3}{\beta}$ is the Laplace conjugate of $\eec{3}{\beta}$. We have explicitly separated the finite and divergent pieces. The argument of the logarithms is given by
{\small\begin{align}
H\left(\eeclp{3}{\beta}\right)&=2^{-\beta/2}e^{ \gamma_E} \frac{Q_{sj}^2}{Q^2}\left(\frac{\mu}{Q_{sj}}\right)^\beta  N_{SJ}\,  \eeclp{3}{\beta}\,.
\end{align}}%
Note that all zero bins for the soft subjet jet function vanish.

\subsection{Global Soft Function}\label{sec:global_soft}

In this section we calculate the one-loop global soft function. The soft function involves three eikonal lines in the $n, \bar n$, and $n_{sj}$ directions, since the angle between the soft subjet axis and the $n$ and $\bar n$ axes is $\mathcal{O}(1)$, and is therefore resolved by the soft radiation. This is distinct from the situation in the SCET$_+$ factorization theorem of \Ref{Bauer:2011uc}. We will see the importance of the performing the appropriate zero bin subtractions, and the role of the boundary soft mode. Indeed, the fact that the soft function has a non-trivial zero bin is itself unusual.

The general form  of the one-loop soft function is (see e.g. \Ref{Ellis:2010rwa})
\begin{align}\label{eq:oneloop_soft_general}
S^{(1)}(\eec{3}{\beta})&=\frac{1}{2}\sum_{i\neq j}\mathbf{T}_i\cdot\mathbf{T}_j S_{ij}^{(1)}(\eec{3}{\beta})\,,
\end{align}
where $\mathbf{T}_i$ is the color generator of leg $i$, and the sum runs over all pairs of legs. The global soft radiation is at a scale such that it can contribute to both the in-jet and out-of-jet observables. Since we work only to one-loop in this appendix, the integral in the soft function is over the phase space for a single parton. We can therefore straightforwardly separate the in and out-of-jet contributions through the measurement functions 
\begin{equation}
\text{in:}\qquad \Theta\left(\tan ^2\frac{R}{2}-\frac{n\cdot k}{\bar{n}\cdot k}\right),\qquad \text{out:}\qquad \Theta\left(\frac{n\cdot k}{\bar{n}\cdot k}-\tan ^2\frac{R}{2}\right)\,,
\end{equation}
where $k$ denotes the momentum of the soft parton. In this section we will split the calculation into two pieces, considering first the in-jet contribution, and then the out-of-jet contribution. This is important to emphasize that contributions to the soft function which depend on large logarithms of $\anglediff$ arise only from the out-of-jet region of integration. Although such logs naively appear in the in-jet contribution to the soft function, they are removed by the boundary soft zero-bin subtraction.

To one-loop, the soft function for the exchange between the eikonal lines $n_a$ and $n_b$ is given by
{\small\begin{align}\label{eq:soft_in}
\hspace{-0.4cm}S_{\,n_a n_b}^{(1)}(\eec{3}{\beta})&=\int[d^dk]_+\frac{2n_a\cdot n_b}{n_{a}\cdot k\,k\cdot n_b}\Theta\left(\tan ^2\frac{R}{2}-\frac{n\cdot k}{\bar{n}\cdot k}\right)\delta\left(\eec{3}{\beta}-N_S\frac{k^0}{Q}\left[\frac{\sja\cdot k}{k^0}\frac{n\cdot k}{k^0}\right]^{\frac{\beta}{2}}\right)\,,
\end{align}}
for the in-jet region, and 
{\small\begin{align}\label{eq:soft_out}
\hspace{-0.4cm}S_{\,n_a n_b}^{(1)}(B)&=\int[d^dk]_+\frac{2n_a\cdot n_b}{n_{a}\cdot k\,k\cdot n_b}\Theta\left(\frac{n\cdot k}{\bar{n}\cdot k}-\tan ^2\frac{R}{2}\right)  \delta\left(B-  n \cdot k \right)\,,
\end{align}}
for the out-of-jet region. Following the decomposition in \Eq{eq:oneloop_soft_general}, we have explicitly extracted the color factor, so that it does not appear in these expressions. The dressed gluon approximation holds for an arbitrary additive observable, $B$, for example, in \Sec{sec:dressed_calcs} we used the mass as an example. Here, for simplicity we have chosen to measure the energy in the out-of-jet region. Since the soft subjet soft function contains in its definition the three eikonal lines $n, \bar n, n_{sj}$, we must sum over contributions from exchanges between all possible pairs.

\subsubsection*{Na\"ive In-Jet Soft Function}

We begin by calculating the na\"ive (i.e. without zero-bin subtraction) contributions to the in-jet soft function. For simplicity, we give only the finite pieces, dropping $\epsilon$-divergences. The anomalous dimensions will be given in \App{app:Anom_Dim}. The contributions from the exchange between the three possible pairs of eikonal lines are given by
{\small\begin{align}
\tilde{S}_{n \,\bar n}^{(1,\,\text{fin})}(\eeclp{3}{\beta})&=\frac{\alpha_s}{\pi(1-\beta)}\text{ln}[T] \Bigg(\text{ln}[T]-2 (1-\beta ) \text{ln}\Bigg[2 \frac{\tan\frac{R}{2}}{\tan\frac{\sjtheta}{2}}\Bigg]\Bigg)+R_{n \,\bar n}^{(1)}(\sjtheta,R)+C_{n\bar{n}}^{(1)}\,,\\
\tilde{S}_{n \,\sja}^{(1,\,\text{fin})}(\eeclp{3}{\beta})&=\frac{\alpha_s}{\pi(1-\beta)}\text{ln}[T] \left(2 \text{ln}[T]-(1-\beta ) \text{ln}\left[4(\bar{n}\cdot \sja)^2 \Big(1-\frac{\tan^2\frac{\sjtheta}{2}}{\tan^2\frac{R}{2}}\Big)\right]\right)\nonumber\\
&\qquad+R_{n \,\sja}^{(1)}(\sjtheta,R)+B_{n\,\sja}^{(1)}(\sjtheta,R)+C_{n\,\sja}^{(1)}\,,\\
\tilde{S}_{\bar n \,\sja}^{(1,\,\text{fin})}(\eeclp{3}{\beta})&=\frac{\alpha_s}{\pi(1-\beta)}\text{ln}[T] \left(\text{ln}[T]-(1-\beta ) \text{ln}\left[(\bar{n}\cdot \sja)^2\Big(1-\frac{\tan^2\frac{\sjtheta}{2}}{\tan^2\frac{R}{2}}\Big) \Big(\frac{\tan\frac{R}{2}}{\tan\frac{\sjtheta}{2}}\Big)^2\right]\right)\nonumber\\
&\qquad+R_{\bar n \,\sja}^{(1)}(\sjtheta,R)+B_{\bar n\,\sja}^{(1)}(\sjtheta,R)+C_{\bar{n}\sja}^{(1)}\,.
\end{align}}
Here we have extracted the common factor
{\small\begin{align}
T&= e^{\gamma_E} N_S\frac{ \eeclp{3}{\beta}\, \mu }{Q \,\text{tan}^{1-\beta }\frac{\sjtheta}{2}}\Big(\frac{n\cdot \sja}{2}\Big)^{\beta/2}
\end{align}}%
as well as the functions $R_{~}^{(1)}, B_{~}^{(1)},$ and constants $C_{~}^{(1)}$. The functions $R_{~}^{(1)}$ depend only on $\theta_{sj}$ and $R$, and are given by
{\small\begin{align}
R_{n \,\bar n}^{(1)}(\sjtheta,R)&=\frac{\alpha_s}{\pi}(1-\beta )\text{ln}\Bigg[\frac{\tan\frac{R}{2}}{\tan\frac{\sjtheta}{2}}\Bigg] \text{ln}\Bigg[4 \frac{\tan\frac{R}{2}}{\tan\frac{\sjtheta}{2}}\Bigg]+I_{n\,\bar{n}}(\sjtheta,R)\,,\\
R_{n \,\sja}^{(1)}(\sjtheta,R)&=-\frac{\alpha_s}{2\pi}(1-\beta)\Bigg(\text{ln}\Bigg[\frac{\tan\frac{R}{2}}{\tan\frac{\sjtheta}{2}}\Bigg] \text{ln}\left[\frac{\Big(\frac{\tan\frac{R}{2}}{\tan\frac{\sjtheta}{2}}\Big)^3}{\Big(1+\frac{\tan\frac{R}{2}}{\tan\frac{\sjtheta}{2}}\Big)^6}\right]\nonumber\\
&\qquad-2 \text{ln}\Big[\frac{\bar{n}\cdot \sja}{2}\Big] \text{ln}\left[4\bar{n}\cdot \sja\frac{ \Big(\frac{\tan\frac{R}{2}}{\tan\frac{\sjtheta}{2}}\Big)^2}{\Big(1+\frac{\tan\frac{R}{2}}{\tan\frac{\sjtheta}{2}}\Big)}\right]+3\text{ln}^2\Bigg[1+\frac{\tan\frac{R}{2}}{\tan\frac{\sjtheta}{2}}\Bigg]\nonumber\\
 &\qquad+6 \text{Li}_2\left[-\frac{1}{2}\right]+3 \text{Li}_2\left[\frac{3}{4}\right] -6 \text{Li}_2\left[1-\frac{\tan\frac{\sjtheta}{2}}{\tan\frac{R}{2}}\right]+6 \text{Li}_2\left[\frac{\tan\frac{R}{2}}{\tan\frac{\sjtheta}{2}+\tan\frac{R}{2}}\right]\Bigg)\nonumber\\
&\qquad+I_{n\,\sja}(\sjtheta,R)\,,\\
R_{\bar n \,\sja}^{(1)}(\sjtheta,R)&=-\frac{\alpha_s}{2\pi}(1-\beta)\Bigg(2 \text{ln}\Big[\frac{\bar{n}\cdot \sja}{2}\Big] \text{ln}\left[4\bar{n}\cdot \sja\frac{ \Big(\frac{\tan\frac{R}{2}}{\tan\frac{\sjtheta}{2}}\Big)^2}{\Big(1+\frac{\tan\frac{R}{2}}{\tan\frac{\sjtheta}{2}}\Big)}\right]+\text{ln}^2\Bigg[1+\frac{\tan\frac{R}{2}}{\tan\frac{\sjtheta}{2}}\Bigg]\nonumber\\
&\qquad-\text{ln}\Bigg[\frac{\tan\frac{R}{2}}{\tan\frac{\sjtheta}{2}}\Bigg] \text{ln}\left[16 \Big(\frac{\tan\frac{R}{2}}{\tan\frac{\sjtheta}{2}}\Big)^5 \Big(1+\frac{\tan\frac{R}{2}}{\tan\frac{\sjtheta}{2}}\Big)^2\right]\nonumber\\
&\qquad-2 \text{Li}_2\left[1-\frac{\tan\frac{\sjtheta}{2}}{\tan\frac{R}{2}}\right]+2 \text{Li}_2\left[\frac{\tan\frac{R}{2}}{\tan\frac{\sjtheta}{2}+\tan\frac{R}{2}}\right]\Bigg)\nonumber\\
&\qquad+I_{\bar{n}\,\sja}(\sjtheta,R)\,.
\end{align}}%
Where the integrals $I_{~}$ are given as:
\small{\begin{align}
I_{n\,\bar{n}}(\sjtheta,R)&=-\frac{2\alpha_s}{\pi} (1-\beta ) \int_0^{u_{max}}du\frac{\text{ln}\left[1+\frac{u^2 n\cdot n_b}{\bar{n}\cdot n_b}\right]}{u}\,,\\
I_{n\,\sja}(\sjtheta,R)&=\frac{\alpha_s}{\pi}(1-\beta )\Bigg(\int_0^{1/2}du\frac{2\text{ln}\left[1+\frac{u^2 n\cdot n_b}{\bar{n}\cdot n_b}\right]}{(-1+u) u (1+u)}\nonumber\\
&+\int_{\frac{1}{2}}^{u_{max}}du\frac{ \left(u (1+u) \text{ln}\left[1+\frac{n\cdot n_b}{\bar{n}\cdot n_b}\right]-2 \text{ln}\left[1+\frac{u^2 n\cdot n_b}{\bar{n}\cdot n_b}\right]\right)}{u \left(-1+u^2\right)}\Bigg)\,,\\
I_{\bar n\,\sja}(\sjtheta,R)&=\frac{\alpha_s}{\pi}(1-\beta )\Bigg(\int_0^{1/2}du\frac{2 u \text{ln}\left[1+\frac{u^2 n\cdot n_b}{\bar{n}\cdot n_b}\right]}{(-1+u) (1+u)}\nonumber\\
& +\int_{\frac{1}{2}}^{u_{max}}du\frac{ \left((1+u) \text{ln}\left[1+\frac{n\cdot n_b}{\bar{n}\cdot n_b}\right]-2 u \text{ln}\left[1+\frac{u^2 n\cdot n_b}{\bar{n}\cdot n_b}\right]\right)}{-1+u^2}\Bigg)\\
u_{max}&=\frac{\tan\frac{R}{2}}{\tan\frac{\sjtheta}{2}}\,.
\end{align}}
The functions $B^{(1)}_{~}$ contain singular dependence on the difference between $\sjtheta$ and $R$, that is, the angle of the soft jet to the jet boundary, and are given as:
\small{\begin{align}
B_{n\,\sja}^{(1)}(\sjtheta,R)&=\frac{\alpha_s}{2\pi}(1-\beta)\Bigg( \text{ln}\Bigg[1-\frac{\tan^2\frac{\sjtheta}{2}}{\tan^2\frac{R}{2}}\Bigg] \text{ln}\left[ (\bar{n}\cdot \sja)^2\Big(1-\frac{\tan^2\frac{\sjtheta}{2}}{\tan^2\frac{R}{2}}\Big)\right]\Bigg)\,,\\
B_{\bar n\,\sja}^{(1)}(\sjtheta,R)&=\frac{\alpha_s}{2\pi}(1-\beta)\Bigg(\text{ln}\Bigg[1-\frac{\tan^2\frac{\sjtheta}{2}}{\tan^2\frac{R}{2}}\Bigg] \text{ln}\left[(\bar{n}\cdot\sja)^2\Big(1-\frac{\tan^2\frac{\sjtheta}{2}}{\tan^2\frac{R}{2}}\Big) \Big(\frac{\tan\frac{R}{2}}{\tan\frac{\sjtheta}{2}}\Big)^4\right]\Bigg)\,.
\end{align}}%
Finally, we have the constants:
\small{
\begin{align}
C_{n\bar{n}}^{(1)}&=\frac{\alpha_s}{\pi} \left(\frac{\pi ^2}{8 (1-\beta )}+(1-\beta ) \text{ln}[2]^2\right)\,,\\
C_{n\sja}^{(1)}&=\frac{\alpha_s}{4\pi(1-\beta )} \left(\pi ^2-4 \text{ln}[2]^2-8 (1-\beta ) \text{ln}[2]^2+(1-\beta )^2 \text{ln}[4] \text{ln}\left[\frac{729}{128}\right]\right)\,,\\
C_{\bar{n}\sja}^{(1)}&=\frac{\alpha_s}{8\pi(1-\beta )} \left(\pi ^2-8 (1+2 (1-\beta )) \text{ln}[2]^2\right)
\end{align}
}%
  
\subsubsection*{Boundary Soft Zero-Bin of In-Jet Soft Function}

We now calculate the boundary soft zero bin of the in-jet soft function. This is the only non-vanishing zero bin. Both constraints in the soft measurement function can be expanded in the zero bin. The jet boundary constraint can be expanded as
{\small\begin{align}
\hspace{-0.3cm}\theta\left(\tan ^2\frac{R}{2}-\frac{n\cdot k}{\bar{n}\cdot k}\right)\rightarrow \theta\left(\tan ^2\frac{R}{2}-\tan ^2\frac{\theta_{sj}}{2}+4\frac{k_{\perp_{sj}}\cdot n_{\perp_{sj}}}{(\bar{n}\cdot \sja )^2\sjabar \cdot k}\right)\,,
\end{align}}%
where we have used the expression given in \Eq{eq:fjet_constraint_expanded} for the expansion of the jet constraint. Note importantly that for the boundary soft modes, this cannot be multipole expanded, unlike for the jet modes of the soft subjet, as was discussed in \Sec{sec:ssj}. For the measurement function, we have the expansion
{\small\begin{align}
\delta\left(\eec{3}{\beta}-N_S\frac{k^0}{Q}\left[\frac{\sja\cdot k}{k^0}\frac{n\cdot k}{k^0}\right]^{\frac{\beta}{2}}\right)\rightarrow \delta   \left( \eec{3}{\beta}-2^{-1+\frac{\beta}{2}}N_S \frac{\sjabar\cdot k}{Q} \left[\frac{\sja\cdot k}{\sjabar\cdot k}\right]^{\frac{\beta}{2}}(n\cdot \sja)^{\frac{\beta}{2}}(\bar{n}\cdot\sja)^{1-\beta} \right) \,.
\end{align}}%
Furthermore, in the integrand we can make the following expansions in the zero bin
\begin{align}
\frac{n\cdot \sja }{n\cdot k\, k\cdot \sja }\rightarrow \frac{1}{\sjabar \cdot k\, k\cdot \sja }\,,\qquad 
\frac{\bar{n}\cdot \sja }{\bar{n}\cdot k\, k\cdot \sja }\rightarrow \frac{1}{\sjabar \cdot k\, k\cdot \sja }\,.
\end{align}
Performing the integration, we find the the zero bin contribution arising from the exchange between the $n$ and $\bar n$ Wilson lines vanishes
{\small\begin{align}
&\tilde{S}_{n\, \bar{n}}^{(1,\,\text{b.s.b.})}(\eeclp{3}{\beta})=0,
\end{align}}
as should be expected, since it is not related to the boundary soft modes. However, there is a non-vanishing contribution to the zero bin arising from the exchanges involving the $n_{sj}$ Wilson line, which is given by
{\small\begin{align}\label{collinear-bin-of-softs}
\tilde{S}_{\sja \, n}^{(1,\,\text{b.s.b.})}(\eeclp{3}{\beta})&=\tilde{S}_{\sja \, \bar{n}}^{(1,\,\text{b.s.b.})}(\eeclp{3}{\beta})=
\frac{\alpha_s}{2\pi}\left (     \frac{\pi^2}{6}+\frac{\pi^2}{8(1-\beta)}-\frac{\pi^2\beta}{12}     \phantom{ \left(\frac{\tan \frac{\theta_{sj}}{2}}{\anglediff}\right )^\beta   }       \right .\\ 
&\hspace{1cm}\left .+ \frac{1}{1-\beta}\ln ^2\left[ \frac{2^{-1+\frac{\beta}{2}}e^{\gamma_E}\mu\eeclp{3}{\beta}N_s(n\cdot \sja )^{\frac{\beta}{2}}}{Q}\left(\frac{\tan^2 \frac{\theta_{sj}}{2}}{\anglediff}\right )^{1-\beta}\right]
\right)\,.\nonumber
\end{align}}%
Here the superscript ``$\text{b.s.b.}$" indicates that this is the contribution from the boundary soft zero bin.

\subsubsection*{Zero Bin Subtracted In-Jet Soft Function}

We now give the expression for the in-jet soft subjet soft function after performing the zero bin subtraction. The $\tilde{S}_{n\, \bar{n}}$ terms are unaffected by the zero bin subtraction, however we include them so that we can rewrite all contributions in a similar form. After zero bin subtraction, the contributions from the three different exchanges are given by
{\small\begin{align}
\tilde{S}_{n \,\sja}^{(1,\,\text{fin})}(\eeclp{3}{\beta})&=\frac{\alpha_s}{2\pi(1-\beta)}\text{ln}[T] \left(3 \text{ln}[T]-4 (1-\beta ) \text{ln}\left[ \bar{n}\cdot\sja\,\frac{\tan\frac{\sjtheta}{2}}{\tan\frac{R}{2}}\right]\right)\nonumber\\
&\qquad+R_{n \,\sja}^{(1)}(\sjtheta,R)+\delta R_{n \,\sja}^{(1)}(\sjtheta,R)+\tilde{B}_{n\sja}^{(1)}(\sjtheta,R)+C_{n\sja}^{(1)}\,,\\
\tilde{S}_{\bar n \,\sja}^{(1,\,\text{fin})}(\eeclp{3}{\beta})&=\frac{\alpha_s}{2\pi(1-\beta)}\text{ln}[T] \Big(\text{ln}[T]-4 (1-\beta ) \text{ln}\left[\bar{n}\cdot\sja\right]\Big)\nonumber\\
&\qquad+R_{\bar n \,\sja}^{(1)}(\sjtheta,R)+\delta R_{\bar n \,\sja}^{(1)}(\sjtheta,R)+\tilde{B}_{\bar n\sja}^{(1)}(\sjtheta,R)+C_{\bar n\sja}^{(1)}\,.
\end{align}}
The $R^{(1)}$ functions and constants $C^{(1)}$ are as defined above. The boundary functions sensitive to the angle of the soft jet to the boundary are modified as:
\begin{align}
\tilde{B}_{n \,\sja}^{(1)}&=\frac{\alpha_s}{\pi}(1-\beta )\text{ln}\Big[1-\frac{\tan^2\frac{\sjtheta}{2}}{\tan^2\frac{R}{2}}\Big] \text{ln}\left[\bar{n}\cdot \sja \Big(\frac{\tan\frac{\sjtheta}{2}}{\tan\frac{R}{2}}\Big)^2\right]\,\\
\tilde{B}_{\bar n \,\sja}^{(1)}&=\frac{\alpha_s}{\pi}(1-\beta)\text{ln}\Big[1-\frac{\tan^2\frac{\sjtheta}{2}}{\tan^2\frac{R}{2}}\Big] \text{ln}\left[\bar{n}\cdot \sja\right]\,.
\end{align}
In addition, one adds the terms:
\begin{align}
\delta R_{n \,\sja}^{(1)}(\eeclp{3}{\beta})&=\frac{\alpha_s}{\pi}\Bigg(-2(1-\beta )\text{ln}^2\Big[\frac{\tan\frac{R}{2}}{\tan\frac{\sjtheta}{2}}\Big]-\frac{\pi ^2 \left(3+2 (1-\beta )+2 (1-\beta )^2\right)}{48(1-\beta )}\Bigg)\,,\\
\delta R_{\bar n \,\sja}^{(1)}(\eeclp{3}{\beta})&=\frac{\alpha_s}{\pi}\Bigg(-2(1-\beta ) \text{ln}^2\Big[\frac{\tan\frac{R}{2}}{\tan\frac{\sjtheta}{2}}\Big]-\frac{\pi ^2 \left(3+2 (1-\beta )+2 (1-\beta )^2\right)}{48(1-\beta )}\Bigg)\,.
\end{align}

We see that the potentially large logarithm of $\anglediff$, which was present in the in-jet soft function before zero bin subtraction has been removed by the boundary soft zero bin subtraction from the terms that contribute to the anomalous dimension, again emphasizing its crucial role in the factorization theorem.  We also emphasize that the presence of a non-trivial zero bin for the soft function is an interesting feature of this factorization theorem.

\subsubsection*{Out-of-Jet Contribution to Soft Function}

In this section we calculate the contribution to the soft function from out-of-jet radiation. While we have seen that for the in-jet contribution the large logarithm of $\anglediff$ was removed by the zero bin, this will not be the case for the out-of-jet radiation. When performing the calculation we will integrate over the entire out-of-jet region, except a region of radius $R_B$ around the axis of the jet in the right hemisphere. This acts as a regulator, allowing us to calculate each of the contributions to the soft function, $S_{ij}^{(1,\,\text{out})}(B)$ separately. For reference, we take the out of jet measurement to be the cumulative energy deposited, however, we have explicitly checked that using other out-of-jet measurements lead to the same dressed gluon anomolous dimension as given in \Sec{sec:pants}.

We begin by calculating the na\"ive (non zero bin subtracted) soft function, whose integrand was given in \Eq{eq:soft_out}. For the soft gluon exchanges between the three possible pairs of Wilson lines, we find
{\small\begin{align}\label{eq:sf_1}
S_{n\bar{n}}^{(1,\,\text{out})}(B)&=-\frac{\alpha_s}{\pi}   \Bigg[-\ln \left(\tan \frac{R_B}{2}\tan \frac{\theta_{sj}}{2}\right)   \ln \left(\frac{\mu\tan \frac{\theta_{sj}}{2}}{2n\cdot \sja B}\right)\nonumber\\
&\qquad+F_{n\bar{n}}\left(\frac{\tan \frac{R}{2}}{\tan \frac{\theta_{sj}}{2}}\right)-F_{n\bar{n}}\left(\frac{1}{\tan \frac{R_B}{2}\tan \frac{\theta_{sj}}{2}}\right)-2\int_{u_{min}}^{u_{max}}\frac{du}{u}\ln \Big(\sja\cdot\bar{n}+u^2\sja\cdot n\Big)\Bigg]\,,\\
\label{eq:sf_2}
S_{\sja n}^{(1,\,\text{out})}(B)&=-\frac{\alpha_s}{\pi}\left[  \ln \left(\frac{\tan ^2\frac{R}{2}(1-\tan ^2\frac{R_B}{2}\tan ^2\frac{\theta_{sj}}{2})}{\tan ^2\frac{R}{2}-\tan ^2\frac{\theta_{sj}}{2}}\right)  \ln \left(\frac{\mu\tan \frac{\theta_{sj}}{2}}{2n\cdot \sja B}\right)  \right .\nonumber\\
& \qquad+F_{\sja n}\left(\frac{\tan \frac{R}{2}}{\tan \frac{\theta_{sj}}{2}}\right)-F_{\sja  n}\left(\frac{1}{\tan \frac{R_B}{2}\tan \frac{\theta_{sj}}{2}}\right)\nonumber\\
&\left . \qquad\qquad\qquad+2\int_{u_{min}}^{u_{max}}\frac{du}{u(1-u^2)}\ln \Big(\sja\cdot\bar{n}+u^2\sja\cdot n\Big)\right] \,,\\
S_{\sja \bar{n}}^{(1,\,\text{out})}(B)&=-\frac{\alpha_s}{\pi}\left [ \ln \left(\frac{1-\tan ^2\frac{R_B}{2}\tan ^2\frac{\theta_{sj}}{2}}{\tan ^2\frac{R_B}{2}(\tan ^2\frac{R}{2}-\tan ^2\frac{\theta_{sj}}{2})}\right)\ln \left(\frac{\mu\tan \frac{\theta_{sj}}{2}}{2n\cdot \sja B}\right) \right .\nonumber\\
\label{eq:sf_3}
&\qquad+F_{\sja \bar{n}}\left(\frac{\tan \frac{R}{2}}{\tan \frac{\theta_{sj}}{2}}\right)-F_{\sja \bar{n}}\left(\frac{1}{\tan \frac{R_B}{2}\tan \frac{\theta_{sj}}{2}}\right)\nonumber\\
& \qquad+F_{\sja n}\left(\frac{\tan \frac{R}{2}}{\tan \frac{\theta_{sj}}{2}}\right)-F_{\sja  n}\left(\frac{1}{\tan \frac{R_B}{2}\tan \frac{\theta_{sj}}{2}}\right)\nonumber\\
&\left . \qquad\qquad\qquad+2\int_{u_{min}}^{u_{max}}\frac{du\,u}{1-u^2}\ln \Big(\sja\cdot\bar{n}+u^2\sja\cdot n\Big)\right ]\,.
\end{align}}%
To simplify the notation in these expressions we have defined the following functions
{\small\begin{align}
&F_{n\bar{n}}(x)=\ln ^2x\,,\\
&F_{\sja n}(x)=\frac{1}{2}\ln (x^2-1)\ln \left(\frac{x^2-1}{x^2}\right)-\frac{1}{2}\text{Li}_{2}\left(\frac{1}{x^2}\right)\,,\\
&F_{\sja \bar{n}}(x)=-\frac{1}{2}\left[2\ln ^2(x)-\ln ^2\left(\frac{1+x}{x}\right)-\ln ^2\Big(x^2-1\Big)+2\text{Li}_{2}\left(\frac{x-1}{x}\right)+2\text{Li}_{2}\left(\frac{x}{1+x}\right)\right]\,.\\
&u_{max}=\frac{1}{\text{tan}^2\frac{R_B}{2}\text{tan}^2\frac{\sjtheta}{2}}\\
&u_{min}=\frac{\text{tan}^2\frac{R}{2}}{\text{tan}^2\frac{\sjtheta}{2}}
\end{align}
}%
Again, we see the explicit appearance of $\anglediff$ in the out-of-jet contribution to the soft function.  Although $R_B$ is required as a regulator in each of the $S_{ij}^{(1,\,\text{out})}(B)$, the sum of Eqs.~\ref{eq:sf_1}, \ref{eq:sf_2}, and \ref{eq:sf_3} are non-singular as $R_B\to 0$, and all anomalous dimensions should be expanded in this limit. For a further discussion, see \cite{Ellis:2010rwa}.

\subsubsection*{Zero Bin Subtraction for Out-of-Jet Contribution to Soft Function}
Finally, we consider possible zero bin subtractions for the out-of-jet soft function. Unlike the in-jet soft function, all zero bin contributions to the out-of-jet soft functions vanish. To see this, note that we are considering the case where the scale of out-of-jet radiation, $B$, is much less than the in-jet scale. More precisely, we are considering the formal scaling
\begin{equation}
B\sim \frac{\ecf{3}{\alpha}}{\ecf{2}{\alpha}}\sim  \left(  \ecf{2}{\alpha}   \right)^2\,.
\end{equation}
In other words, $B\sim Q \,z_s^2$, with $z_s$ the energy fraction of the global soft radiation. All other modes appearing in the factorization, in particular, the boundary soft, and soft subjet modes, are parametrically more energetic. For all possible zero bins to the soft function in the out-of-jet region, we can therefore multipole expand the measurement function appearing in \Eq{eq:soft_out}. Therefore, all such zero bins give a vanishing contribution.  This implies that the dependence on $\anglediff$ is not zero bin subtracted in the out-of-jet contribution to the soft function, unlike for the in-jet contribution. Therefore, while the global soft function does depend on $\anglediff$, it comes entirely from the out-of-jet region of integration. As we will see in \App{app:bsoft}, the boundary soft function also depends on $\anglediff$. Indeed, $\anglediff$ appears in the anomalous dimensions for both these functions, but with opposite signs, as is required by the renormalization group consistency of the factorization theorem.

\subsection{Boundary Soft Function}\label{app:bsoft}

In this section we calculate the boundary soft function. In the multi-stage matching of \Sec{sec:Fact} which gave rise to the boundary soft mode, the boundary soft modes were decoupled from the soft subjet collinear modes via a BPS field redefinition. The boundary soft function therefore has the form of a global soft function, in particular it is calculated with eikonal Feynman rules, but it only has Wilson lines in the $\sja $ and $\sjabar $ directions. This is important, as it implies that the boundary soft function has the same color structure as the soft subjet jet function. In this appendix we have assumed that the soft subjet is a gluon jet. This can be understood intuitively since the boundary softs are a collinear soft mode, and hence are genuinely boosted in the $\sja $ direction, so all other Wilson lines collapse to the $\sjabar $. Thus the color structure of the boundary soft modes is simply that of the dipole formed by the soft subjet, and all other eikonal lines merged into one.

The one loop expression for the boundary soft function is given by
{\small\begin{align}
S_{\sja\sjabar}^{(1)}(\eec{3}{\beta})&=g^2\mu^{2\epsilon}C_A\int[d^dk]_+\delta\left(\eec{3}{\beta}-N_{BS}\frac{\sjabar\cdot k}{Q}\left[\frac{\sja\cdot k}{\sjabar\cdot k}\right]^{\frac{\beta}{2}}\right)\nonumber\\
&\hspace{5cm}  \Theta\left(\tan ^2\frac{R}{2}-\frac{n\cdot k_1}{\bar{n}\cdot k_1}\right)   \frac{\sja\cdot\sjabar}{\sja\cdot k\, k \cdot\sjabar}\,.
\end{align}}%
Here we have already multipole expanded away any possible out-of-jet contributions, since the boundary soft scale is higher than the out-of-jet scale. We must again take care in expanding the jet radius constraint. From \Eq{eq:fjet_constraint_expanded}, we have
\begin{equation}
\Theta\left(\tan ^2\frac{R}{2}-\frac{n\cdot k_1}{\bar{n}\cdot k_1}\right)\to  \Theta\left(\tan ^2\frac{R}{2}-\tan ^2\frac{\theta_{sj}}{2}+4\frac{k_{\perp_{sj}}\cdot n_{\perp_{sj}}}{(\bar{n}\cdot \sja )^2\sjabar \cdot k}\right)\,,
\end{equation}
which, unlike for the soft subjet jet function, cannot be further expanded.
The one loop expression for the boundary soft function is then given by
{\small\begin{align}
S_{\sja\sjabar}^{(1)}(\eec{3}{\beta})&=g^2\mu^{2\epsilon}C_A\int[d^dk]_+\delta\left(\eec{3}{\beta}-N_{BS}\frac{\sjabar\cdot k}{Q}\left[\frac{\sja\cdot k}{\sjabar\cdot k}\right]^{\frac{\beta}{2}}\right)\nonumber\\
&\hspace{2.4cm}\Theta\left(\tan ^2\frac{R}{2}-\tan ^2\frac{\theta_{sj}}{2}+4\frac{k_{\perp_{sj}}\cdot n_{\perp_{sj}}}{(\bar{n}\cdot \sja )^2\sjabar \cdot k}\right)\frac{\sja\cdot\sjabar}{\sja\cdot k\, k \cdot\sjabar}\,.
\end{align}}%
We therefore see that the boundary soft contribution is identical to the soft subjet collinear-bin of the global softs, given in \eqref{collinear-bin-of-softs}, with the substitution $\bar n \to \bar n_{sj}$, and changing the normalization of the measurement function. We can therefore immediately write down the one-loop boundary soft function
{\small\begin{align}
\tilde{S}_{\sja \, \bar{n}_{sj}}^{(1)}(\eeclp{3}{\beta})&=
\frac{\alpha_s C_A}{2\pi}\left (     \frac{\pi^2}{6}+\frac{\pi^2}{8(1-\beta)}-\frac{\pi^2\beta}{12}     \phantom{ \left(\frac{\tan \frac{\theta_{sj}}{2}}{\anglediff}\right )^\beta   }       \right .\\ 
&\hspace{2cm}\left .+ \frac{1}{1-\beta}\ln ^2\left[ \frac{e^{\gamma_E}\mu\eeclp{3}{\beta}N_{BS}}{Q}\left(\frac{\tan^2 \frac{\theta_{sj}}{2}}{\anglediff}\right )^{1-\beta}\right]
\right)\,,\nonumber
\end{align}}%
where for simplicity we have given only the finite pieces, dropping $\epsilon$-divergences. We see that the boundary soft mode carries the dependence of the soft subjet dynamics on the difference, $\anglediff$, which is completely factorized from the collinear dynamics of the soft subjet. However, importantly, the color structure of the boundary soft is determined by the color structure of the soft subjet, showing that it is indeed describing its dynamics. The difference $\anglediff$ therefore appears in both the boundary soft function, and in the out-of-jet contribution to the soft function, as seen in \App{sec:global_soft}. The fact that it appears in both these functions is required for the renormalization group consistency of the factorization theorem.

\subsection{Anomalous Dimensions}\label{app:Anom_Dim}

In this section we collect the one-loop anomalous dimensions for all the functions calculated in this appendix. The two hard functions satisfy multiplicative renormalization group equations. For the dijet production hard function, we have
{\small\begin{equation}
\mu \frac{d}{d\mu}\ln H(Q^2,\mu)=2\text{Re}\left[ \gamma_C(Q^2,\mu)\right]\,,
\end{equation}}
with
{\small\begin{equation}
\gamma_C(Q^2,\mu)=\frac{\alpha_s C_F}{4\pi}\left(  4\log\left[  \frac{-Q^2}{\mu^2} \right] -6  \right )\,.
\end{equation}}
For the soft subjet production hard function, we have
{\small\begin{equation}
\mu\frac{d}{d\mu}\ln H^{sj}_{n\bar{n}}(\sje,\sja,\mu) =-\frac{\alpha_s C_A}{\pi}\ln \Bigg[\frac{2\mu^2\bar{n}\cdot n}{Q^2_{sj}n\cdot\sja\,\sja\cdot\bar{n}}\Bigg]-\frac{\alpha_s}{\pi}\beta_0\,.
\end{equation}}
The jet, boundary soft, and global soft functions satisfy multiplicative renormalization group equations in Laplace space, which are given by
 {\small\begin{align}
\mu\frac{d}{d\mu}\ln J_{\sja }\Big(\eeclp{3}{\beta}\Big) &=-2\frac{\alpha_s C_A}{\pi(1-\beta)}\ln \Bigg[2^{-\beta/2}\eeclp{3}{\beta}e^{\gamma_E}\frac{Q_{sj}^2}{Q^2}\frac{\mu^\beta}{Q_{sj}^\beta}\Bigg]\nonumber \\
&
\hspace{0.8cm}
-2\frac{\alpha_s C_A}{\pi(1-\beta)}\ln\Big[2^{-3+\beta}\frac{Q_{hj}}{Q}(n\cdot\sja)^{\beta}\Big]+\frac{\alpha_s}{\pi}\beta_0\,,\\
\mu\frac{d}{d\mu}\ln  S_{\sja \,\sjabar }\Big(\eeclp{3}{\beta};R\Big)&=\frac{\alpha_s C_A}{\pi(1-\beta)}\ln \left[\Big(\frac{n\cdot \sja}{2}\Big)^{\beta/2}\eeclp{3}{\beta}e^{\gamma_E}\frac{\mu}{Q}\right]+\frac{\alpha_s C_A}{\pi(1-\beta)}\ln\Bigg[\frac{Q_{hj}Q_{sj}}{4Q^2}(n\cdot\sja)^{\beta/2}\Bigg]\\
&\hspace{0.8cm}-\frac{\alpha_sC_A}{2\pi}\ln \left[\frac{\bar n \cdot n_{sj}}{n\cdot n_{sj}}\tan ^4\frac{R}{2}\right]-\frac{\alpha_sC_A}{\pi}\ln \left[1-\frac{n\cdot\sja}{\bar{n}\cdot\sja\tan ^2\frac{R}{2}}\right]\,,\nonumber\\
\mu\frac{d}{d\mu}\ln  S_{\sja \,n\,\bar{n}}\Big(\eeclp{3}{\beta},B;R\Big)&=\frac{\alpha_s C_A}{\pi(1-\beta)}\ln \left[\Big(\frac{n\cdot \sja}{2}\Big)^{\beta/2}\eeclp{3}{\beta}e^{\gamma_E}\frac{\mu}{Q}\right]+\frac{\alpha_s C_A}{\pi(1-\beta)}\ln\Bigg[\frac{Q_{hj}Q_{sj}}{4Q^2}(n\cdot\sja)^{\beta/2}\Bigg]\nonumber\\
&\hspace{-2.5cm}-\frac{\alpha_sC_A}{2\pi}\ln \left[\frac{(\bar n\cdot n_{sj})(n\cdot n_{sj})^3}{\tan ^4\frac{R}{2}}\right]+\frac{\alpha_sC_A}{\pi}\ln \left[1-\frac{n\cdot\sja}{\bar{n}\cdot\sja\tan ^2\frac{R}{2}}\right]+C_F\text{ terms}
\end{align}}%
For consistency of our soft subjet factorization theorem, the sum of the anomalous dimensions listed above should cancel.  Indeed, one can explicitly check that, up to terms proportional to $C_F$,
\begin{align}
\mu\frac{d}{d\mu}\ln H^{sj}_{n\bar{n}}(\sje,\sja,\mu)&+\mu\frac{d}{d\mu}\ln J_{\sja }\Big(\eeclp{3}{\beta}\Big)\nonumber \\
&+\mu\frac{d}{d\mu}\ln  S_{\sja \,\sjabar }\Big(\eeclp{3}{\beta};R\Big)+\mu\frac{d}{d\mu}\ln  S_{\sja \,n\,\bar{n}}\Big(\eeclp{3}{\beta},B;R\Big) = 0 \,.
\end{align}
The terms in the anomalous dimension of the global soft function proportional to $C_F$ will cancel when added with the anomalous dimensions of the hard function $H(Q^2,\mu)$ and the hard jet functions $J_{n}(Q_{J},\eec{3}{\beta})$ and $J_{\bar n}(Q_{J},B)$.

We again emphasize that the contribution to the global soft radiation's anomalous dimension that is sensitive to the soft subjet's angle to the boundary comes purely from the region of integration where the soft gluon is out of the  jet. Performing the appropriate zero bin subtractions removes any dependence from the in-jet region of integration, as was discussed in detail in \App{sec:global_soft}. The terms in the anomalous dimensions involving the soft subjet's angle to the boundary cancel between the boundary soft and global soft function, as required for renormalization group consistency.  Also, for the global soft function, we have only shown the contributions proportional to $C_A$, as required for the dressed gluon approximation.

For the functions defining the dressed-gluon approximation, as presented in \Sec{sec:pants}, the one-loop renormalization group equations are 
{\small\begin{align}
\mu\frac{d}{d\mu}\ln W_{n\bar{n}}(\sje,\sja;R)&=-\frac{\alpha_sC_A}{\pi}\ln \Bigg(1-\frac{n\cdot\sja}{\bar{n}\cdot\sja\tan ^2\frac{R}{2}}\Bigg)\,,\\
\mu\frac{d}{d\mu}\ln G_{n\bar{n}\sja}(\outj;R)&=\frac{\alpha_sC_A}{\pi}\ln \Bigg(1-\frac{n\cdot\sja}{\bar{n}\cdot\sja\tan ^2\frac{R}{2}}\Bigg)\,.
\end{align}}%
Importantly, the sum of these anomalous dimensions vanishes, so that the product \newline
$W_{n\bar{n}}(\sje,\sja;R)  G_{n\bar{n}\sja}(\outj;R)$ is indeed renormalization group invariant, as stated in \Sec{sec:pants}. Furthermore, we explicitly see that there is no dependence on $\eeclp{3}{\beta}$. As discussed in \Sec{sec:anom_dressed}, we conjecture that the anomalous dimensions of the $W_{n\bar{n}}(\sje,\sja;R)$ and $G_{n\bar{n}\sja}(\outj;R)$ functions are given to all orders in perturbation theory by
{\small\begin{align}
\mu\frac{d}{d\mu}\ln W_{n\bar{n}}(\sje,\sja;R)&=-C_A\, \Gamma_{\text{cusp}}\ln \Bigg(1-\frac{n\cdot\sja}{\bar{n}\cdot\sja\tan ^2\frac{R}{2}}\Bigg)\,,\\
\mu\frac{d}{d\mu}\ln G_{n\bar{n}\sja}(\outj;R)&=C_A\, \Gamma_{\text{cusp}} \ln \Bigg(1-\frac{n\cdot\sja}{\bar{n}\cdot\sja\tan ^2\frac{R}{2}}\Bigg)\,.
\end{align}}%
It would be interesting to explicitly verify this conjecture by performing the two-loop calculation.

\section{Factorization For Two Strongly Ordered Soft Jets}\label{app:twosubjet}

We use the factorization ansatz of \Eq{eq:N-eikonal_lines_soft} to write down the factorization structure for two soft subjets added to a dijet factorization for $e^+e^-$ collisions. We write the result assuming all measurements are in their conjugate (Laplace) space form, so that we can avoid convolutions. We start with the standard dijet factorization theorem
\begin{align}\label{eq:start_dijet}
\frac{d\sigma}{d\ecflp{2}{\alpha}d\tilde \outj}&=H(Q^2)J_n(\ecflp{2}{\alpha})J_{\bar{n}}(\tilde \outj)S_{n\bar{n}}(\ecflp{2}{\alpha};\tilde \outj)\,.
\end{align}
With a single jet, $\ecf{2}{\alpha}$ is the appropriate resolution measurement. Applying \Eq{eq:N-eikonal_lines_soft}, and trading the resolution measurement for $\ecf{3}{\alpha}$, we find
\begin{align}\label{eq:dijet_1_sj}
\frac{d\sigma}{dz_p\,d\Omega_p\,d\ecflp{3}{\alpha}\,d\tilde\outj}&=H(Q^2)J_{\bar{n}}(\tilde\outj)J_n(\ecflp{3}{\alpha}) \tilde J_{p}(\ecflp{3}{\alpha};R)H_{n\bar{n}}(z_p,\Omega_p)\nonumber\\
&
\hspace{3cm}
\times\Bigg(\frac{S_{n\bar{n} p}(\ecflp{3}{\alpha};\tilde\outj)}{S_{n\bar{n}}(\ecflp{3}{\alpha};\tilde\outj)}\Bigg)S_{n\bar{n}}(\ecflp{3}{\alpha};\tilde\outj)\,.
\end{align}
Recall, that we use $p$ to label the more energetic of the soft subjets. We have used the tilde notation of \Eq{eq:ssubjetfact} to indicate that the jet function for the soft subjet must be refactorized into jet function and a boundary soft function. Once this refactorization is performed, and we cancel the $n \bar n$ soft function, \Eq{eq:dijet_1_sj} is the same as the factorization theorem given in \Eq{fact_inclusive_form_1}. 

Now we can add another soft subjet, strongly ordered with respect to the first and denoted by $q$, by performing a tree level matching. At tree level, only dipoles can contribute to the production of the soft subjet, $q$. This softest subjet can be produced from the initial $n\bar{n}$ dipole, or from either of the $na$ or $\bar{n}a$ dipoles formed from the previous soft subjet. Applying \Eq{eq:N-eikonal_lines_soft} to each of the soft functions in \Eq{eq:dijet_1_sj}, and trading for the correct resolution variable $\ecf{4}{\alpha}$ gives,
\small{\begin{align}\label{eq:dijet_2_sj}
\frac{d\sigma}{dz_pd\Omega_pdz_qd\Omega_qd\ecflp{4}{\alpha}d\tilde\outj}&=H(Q^2)J_{\bar{n}}(\tilde\outj)J_n(\ecflp{4}{\alpha})\tilde J_{p}(\ecflp{4}{\alpha};R) \tilde J_{q}(\ecflp{4}{\alpha};R)S_{n\bar{n}}(\ecflp{4}{\alpha};\tilde\outj)\nonumber\\
&\qquad H_{n\bar{n}}(z_p,\Omega_p)\Bigg(\frac{S_{n\bar{n} p}(\ecflp{4}{\alpha};\tilde\outj)}{S_{n\bar{n}}(\ecflp{4}{\alpha};\tilde\outj)}\Bigg)\Bigg\{\Big(C_F-\frac{C_A}{2}\Big)H_{n\bar{n}}(z_q,\Omega_q)\Bigg(\frac{S_{n\bar{n} q}(\ecflp{4}{\alpha};\tilde\outj)}{S_{n\bar{n}}(\ecflp{4}{\alpha};\tilde\outj)}\Bigg)\nonumber\\
&\qquad+\frac{C_A}{2}H_{np}(z_q,\Omega_q)\Bigg(\frac{S_{npq}(\ecflp{4}{\alpha};\tilde\outj)}{S_{np}(\ecflp{4}{\alpha};\tilde\outj)}\Bigg)\nonumber\\
&\qquad+\frac{C_A}{2}H_{\bar{n}p}(z_q,\Omega_q)\Bigg(\frac{S_{\bar{n}pq}(\ecflp{4}{\alpha};\tilde\outj)}{S_{\bar{n}p}(\ecflp{4}{\alpha};\tilde\outj)}\Bigg)+...\Bigg\}
\,.
\end{align}}%
With four jets, there is a non-trivial basis of possible color structures, which must be included in the factorization theorem. We have explicitly indicated the tree level matching's color factors for the soft subjet production. Again, both the $q$, and $p$ soft subjet jet functions must be refactorized into boundary soft and jet functions to achieve a complete factorization of the soft subjet dynamics. Finally, the $...$ terms in \Eq{eq:dijet_2_sj} denote terms which involve all three eikonal lines $n,\bar{n},$ and $p$ in the production of the second soft subjet. These terms do not appear in the tree-level matching.

\bibliography{ngl_factorization}{}
\bibliographystyle{jhep}

\end{document}